%% file: main.tex
\pdfoutput=1

\RequirePackage[l2tabu, orthodox]{nag}
\RequirePackage{amsmath}
\documentclass[table,xcdraw,smallextended]{svjour3}
\usepackage[utf8]{inputenc}

\makeatletter
\newsavebox\zb@x
\newcounter{z@@m}
\usepackage{calc}
\newdimen\B@r\newdimen\P@r
\newdimen\@zw\newdimen\@zh\newdimen\@zd

\newcommand{\zoombox}[2][0]{%
	\leavevmode%
	\sbox\zb@x{#2}%
	\setlength\B@r{1pt*\ratio{\wd\zb@x}{\ht\zb@x+\dp\zb@x}}%
	\setlength\P@r{1pt*\ratio{\paperwidth}{\paperheight}}%
	\ifdim\B@r>\P@r\relax%
	\setlength\@zw{\wd\zb@x}\setlength\@zh{\@zw*\ratio{\paperheight}{\paperwidth}}%
	\setlength\@zd{(\@zh-\ht\zb@x-\dp\zb@x)*\real{0.5}+\dp\zb@x}%
	\setlength\@zh{\@zh-\@zd}%
	\else%
	\setlength\@zh{\ht\zb@x+\dp\zb@x}%
	\setlength\@zw{\@zh*\ratio{\paperwidth}{\paperheight}}%
	\setlength\@zh{\ht\zb@x}\setlength\@zd{\dp\zb@x}%
	\fi%
	\makebox[0pt][l]{\makebox[\wd\zb@x][c]{\makebox[\@zw][l]{%
				\pdfdest name {zbfs\thez@@m} fitr
				width  \@zw\space
				height \@zh\space
				depth  \@zd\space
	}}}%
	\pdfdest name {zb\thez@@m} fitr
	width  \wd\zb@x\space
	height \ht\zb@x\space
	depth  \dp\zb@x\space
	\immediate\pdfannot 
	width  \wd\zb@x\space
	height \ht\zb@x\space
	depth  \dp\zb@x\space
	{%
		/Subtype/Link/H/N
		/Border [0 0 #1 [1 2]]
		/A <<
		/S/JavaScript
		/JS (
		if(typeof(zoomed)=='undefined'||!zoomed){
			var lastView=this.viewState;
			if(app.fs.isFullScreen) this.gotoNamedDest('zbfs\thez@@m');
			else this.gotoNamedDest('zb\thez@@m');
			zoomed=true;
		}else{
			this.viewState=lastView;
			zoomed=false;
		}
		)
		>>
	}%
	\usebox{\zb@x}%
	\stepcounter{z@@m}%
}   
\makeatother

\usepackage[numbers]{natbib}
\usepackage[hyphens,spaces,obeyspaces]{url}
\urldef{\bitbucketurl}\url{https://bitbucket.org/leonate/jxls/commits/607c10e84336e08b5ba9fd12a21b3f450cf733d3#chg-jxls-poi/pom.xml}

\usepackage[nounderscore]{syntax} %
\usepackage{color}
\usepackage{listings}
\usepackage{graphicx}
\usepackage{rotating}

\usepackage{diagbox}
\usepackage{floatrow}
\usepackage{wrapfig}

\usepackage{xcolor}
\usepackage{bbding}
\usepackage{pifont}

\usepackage{float}
\floatstyle{plaintop}
\restylefloat{table}

\usepackage{xcolor}
\usepackage[linesnumbered,ruled,vlined]{algorithm2e}

\usepackage{graphicx,pifont}%
\let\oldding\ding%
\renewcommand{\ding}[2][1]{\scalebox{#1}{\oldding{#2}}}

\SetCommentSty{mycommfont}
\SetKwComment{Comment}{$\triangleright$\ }{}

\SetKwInput{KwData}{Input}
\SetKwInput{KwResult}{Output}

\usepackage[noend]{algpseudocode}      
\usepackage{paralist}
\usepackage{amsmath}
\usepackage{amssymb}
\usepackage{subfigure}
\usepackage{bchart}        
\usepackage{numprint}         
\usepackage{xspace}   
\usepackage{mdwlist}   
\usepackage{enumerate}  
\usepackage[]{tabularx}  
\usepackage[framemethod=TikZ]{mdframed}
\usepackage{mathtools, cuted}
\usepackage{framed}
\usepackage{newfloat} 
\usepackage{xfrac}
\usepackage{colortbl}
\usepackage{tikz}
\usepackage[ligature, inference]{semantic}
\usepackage{paralist}
\usepackage{wrapfig}

\usepackage{parcolumns}

\usepackage{dsfont}
\usepackage[utf8]{inputenc}
\usepackage[T1]{fontenc}
\usepackage{xcolor}
\usepackage{lscape}
\usepackage{url}
\usepackage{multirow}
\usepackage{array}
\usepackage{paralist}
\usepackage{makecell}
\usepackage{adjustbox}

\usepackage{graphicx}
\usepackage{tabu}
\usepackage{booktabs}

\usepackage{adjustbox}

\usepackage[colorinlistoftodos,prependcaption]{todonotes}
\newboolean{showcomments}
\setboolean{showcomments}{true}
\ifthenelse{\boolean{showcomments}}
 { \newcommand{\mynote}[2]{
      \fbox{\bfseries\sffamily\scriptsize#1}
        {\small$\blacktriangleright$\textsf{\textcolor{red}{{\em #2}\bfseries }}$\blacktriangleleft$}}}
        { \newcommand{\mynote}[2]{}}

\usepackage[switch]{lineno}

\usepackage{hyperref}

\usepackage{booktabs}
\usepackage{threeparttable}

{\end{tabular}\par\medskip}

\definecolor{pblue}{rgb}{0.13,0.13,1}
\definecolor{pgreen}{rgb}{0,0.5,0}
\definecolor{pred}{rgb}{0.9,0,0}
\definecolor{pgrey}{rgb}{0.46,0.45,0.48}

\newcommand{\nosemic}{\renewcommand{\@endalgocfline}{\relax}}%
\newcommand{\dosemic}{\renewcommand{\@endalgocfline}{\algocf@endline}}%

\lstdefinelanguage{Pom}{
  morekeywords={<dependency>,</dependency>,<groupId>,</groupId>,<artifactId>,</artifactId>,<version>,</version>,<scope>,</scope>,<parent>,</parent>,<name>,</name>,<dependencies>,</dependencies>,<exclusion>,</exclusion>,<exclusions>,</exclusions>,<packaging>,</packaging>},
  otherkeywords={<dependency>,</dependency>,<groupId>,</groupId>,<artifactId>,</artifactId>,<version>,</version>,<scope>,</scope>,<parent>,</parent>,<name>,</name>,<dependencies>,</dependencies>,<exclusion>,</exclusion>,<exclusions>,</exclusions,<packaging>,</packaging>}
}

\definecolor{titlecolor}{RGB}{153,204,255}
\mdfdefinestyle{mpdframe}{
    frametitlebackgroundcolor   =titlecolor,
    frametitlerule              =true,
    nobreak                     =true,
    roundcorner                 =5pt,
    middlelinewidth             =0.8pt,
    innermargin                 =0.2cm,
    outermargin                 =0.2cm,
    innerleftmargin             =0.4cm,
    innerrightmargin            =0.4cm,
    innertopmargin              =0.3cm,
    innerbottommargin           =0.3cm
}

\newcommand{\nbdirect}{$44,488$\@\xspace}
\newcommand{\nbinherited}{$180,693$\@\xspace}
\newcommand{\nbtransitive}{$498,263$\@\xspace}

\newcommand{\nbartifacts}{$9,639$\@\xspace}
\newcommand{\nbedges}{$723,444$\@\xspace}

\newcommand{\nbpr}{$30$\@\xspace}
\newcommand{\nbaccrejpr}{$21$\@\xspace}
\newcommand{\nbaccpr}{$18$\@\xspace}

\newcommand{\totalremdeps}{$131$\@\xspace}

\newcommand{\totaldirectpr}{$15$\@\xspace}

\newcommand{\totaltranstpr}{$8$\@\xspace}
\newcommand{\totalacctranstpr}{$4$\@\xspace}

\newcommand{\percbloat}{$75.1\%$\@\xspace}
\newcommand{\nbbloat}{$543,610$\@\xspace}

\newcommand{\nbbd}{$19,673$\@\xspace}
\newcommand{\nbbt}{$412,288$\@\xspace}
\newcommand{\nbbi}{$111,649$\@\xspace}

\newcommand{\nbut}{$85,975$\@\xspace}

\newcommand{\percbd}{$2.7\%$\@\xspace}
\newcommand{\percbt}{$57\%$\@\xspace}
\newcommand{\percbi}{$15.4\%$\@\xspace}

\newcommand{\percud}{$3.4\%$\@\xspace}
\newcommand{\percut}{$11.9\%$\@\xspace}

\newcommand{\ie}{i.e.\@\xspace}
\newcommand{\aka}{a.k.a.\@\xspace}

\newcommand{\eg}{e.g.\@\xspace}

\newcommand{\wrt}{w.r.t.\@\xspace}

\newcommand{\mv}{Maven\@\xspace}
\newcommand{\mc}{Maven Central\@\xspace}
\newcommand{\mdg}{Maven Dependency Graph\@\xspace}
\newcommand{\mytool}{\textsc{DepClean}\@\xspace}

\newcommand{\pom}{{POM}\@\xspace}
\newcommand{\poms}{{POMs}\@\xspace}
\newcommand\jxls{{\textsc{Jxls}}\xspace}
\newcommand\jxlspoi{{\texttt{jxls-poi}}\xspace}
\newcommand\codec{{\texttt{commons-codec}}\xspace}

\usepackage{wasysym}
\mathlig{->}{\rightarrow}

\title{
A Comprehensive Study of Bloated Dependencies in\\ the Maven Ecosystem
}

\author{\footnotesize{C\'esar Soto-Valero} 
\href{https://orcid.org/0000-0003-0541-6411}{\includegraphics[scale=0.05]{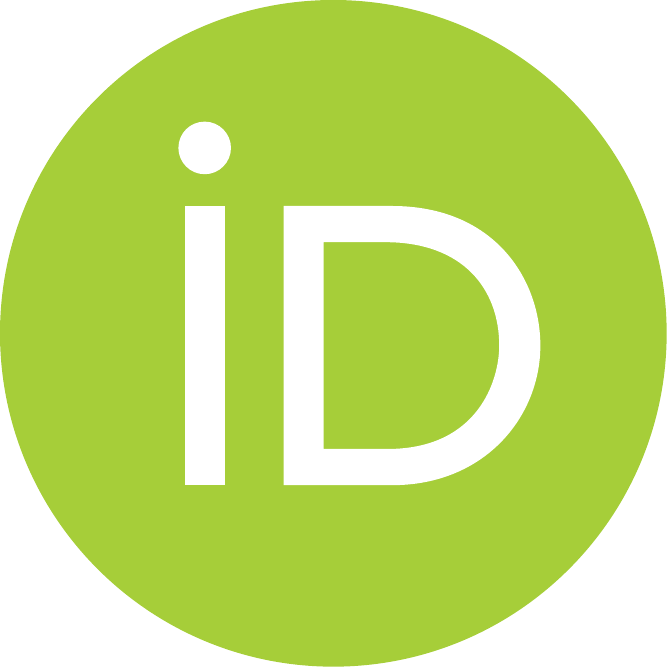}}
\and Nicolas Harrand 
\href{https://orcid.org/0000-0002-2491-2771}{\includegraphics[scale=0.05]{figures/ORCID-iD_icon-vector.pdf}}
\and Martin Monperrus 
\href{https://orcid.org/0000-0003-3505-3383}{\includegraphics[scale=0.05]{figures/ORCID-iD_icon-vector.pdf}}
\and Benoit Baudry 
\href{https://orcid.org/0000-0002-4015-4640}{\includegraphics[scale=0.05]{figures/ORCID-iD_icon-vector.pdf}}
}

\authorrunning{Soto-Valero et al.} %

\institute{
C\'esar Soto-Valero, Nicolas Harrand, Martin Monperrus, and Benoit Baudry \at KTH Royal Institute of Technology\\
Department of Software and Computer Systems\\
Stockholm, Sweden \\
\email{cesarsv@kth.se}          
}

\date{Received: date / Accepted: date}

\begin{document}
	
\maketitle

\begin{abstract}

Build automation tools and package managers have a profound influence on software development. They facilitate the reuse of third-party libraries, support a clear separation between the application's code and its external dependencies, and automate several software development tasks. However, the wide adoption of these tools introduces new challenges related to dependency management. 
In this paper, we propose an original study of one such challenge: the emergence of \emph{bloated dependencies}. 

Bloated dependencies are libraries that the build tool packages with the application's compiled code but that are actually not necessary to build and run the application. This phenomenon artificially grows the size of the built binary and increases maintenance effort. We propose a tool, called \mytool, to analyze the presence of bloated dependencies in \mv artifacts. We analyze \nbartifacts Java artifacts hosted on \mc, which include a total of \nbedges dependency relationships. Our key result is that \percbloat of the analyzed dependency relationships are bloated. In other words, it is feasible to reduce the number of dependencies of \mv artifacts up to $1/4$ of its current count. 
We also perform a qualitative study with \nbpr notable open-source projects. Our results indicate that developers pay attention to their dependencies and are willing to remove bloated dependencies: \nbaccpr/\nbaccrejpr answered pull requests were accepted and merged by developers, removing \totalremdeps dependencies in total. %
\end{abstract}


\newpage

\section{Introduction}
\label{sec:introduction}

Software reuse, a long time advocated software engineering practice~\cite{naur1969software,Krueger1992}, has boomed in the last years thanks to the widespread adoption of build automation and package managers~\cite{cox2019surviving}.
Package managers provide both a large pool of reusable packages, \aka libraries, as well as systematic ways to declare what are the packages on which an application depends. Examples of such package management systems  include Maven for Java, npm for Node.js, or Cargo for Rust. Build tools automatically fetch all the packages that are needed to compile, test, and deploy an application. 

Package managers boost software reuse by creating a clear separation between the application and its third-party dependencies. Meanwhile, they introduce new challenges for the developers of software applications, who now need to manage those third-party dependencies~\cite{cox2019surviving} to avoid entering into the so-called ``dependency hell''. These challenges relate to ensuring high quality dependencies~\cite{Salza2019}, maintaining the dependencies up-to-date~\cite{Bavota2015}, or making sure that heterogeneous licences are compatible~\cite{Wu2017}. 

Our work focuses on one specific challenge of dependency management: the existence of \emph{bloated dependencies}. This refers to packages that are declared as dependencies for an application, but that are actually not necessary to build or run it. The major issue with bloated dependencies is that the final deployed binary file includes more code than necessary: an artificially large binary is an issue when the application is sent over the network (\eg, web applications) or it is deployed on small devices (\eg, embedded systems). Bloated dependencies could also embed vulnerable code that can be exploited, while being actually useless for the application~\cite{Gkortzis2019}. Overall, bloated dependencies needlessly increase the difficulty of managing and evolving software applications.

We propose a novel, unique, and large scale analysis about bloated dependencies. 
So far, research on bloated dependencies has been only touched with a study of copy-pasted dependency declarations by McIntosh and colleagues~\cite{McIntoshicse2014}. Our previous work gives preliminary results on this topic in the context of Java \cite{Harrand2019}. To our knowledge, there is no systematic analysis of the presence of bloated dependencies nor about the importance of this problem for developers.

Our work focuses on \mv, the most popular package manager and automatic build system for Java and languages that compile to the JVM. In \mv, developers declare dependencies in a specific file, called the \pom file. In order to analyze thousands on artifacts on \mc, the largest repository of Java artifacts, manual analysis is not a feasible solution. To overcome this problem, we have designed and implemented a tool called \mytool, that performs an automatic analysis of dependencies usage in \mv projects. Given an application and its \pom file, \mytool collects the complete dependency tree (the list of dependencies declared in the \pom, as well as the transitive dependencies) and analyzes the bytecode of the artifact and all its dependencies to determine the presence of bloated dependencies. Finally, \mytool generates a variant of the \pom in which bloated dependencies are removed. 

Armed with \mytool, we structure our analysis of bloated dependencies  in two parts. First, we automatically analyse \nbartifacts artifacts and their \nbedges dependencies. We found that \percbloat of these dependencies are bloated, mostly due to transitive dependencies and the complexities of dependency management in multi-module projects. Second, we perform a user study involving \nbpr artifacts, for which the code is available as open-source on GitHub and which are actively maintained. For each project, we use \mytool to generate a \pom file without bloated dependencies and submitted the changes as a pull request to the project. Notably, our work yielded \nbaccpr merged pull requests by open-source developers and \totalremdeps bloated dependencies were removed.

To summarize, this paper makes the following contributions:
\begin{itemize}
	\item A comprehensive study of bloated dependencies in the context of the Maven package manager. We are the first to quantify the magnitude of bloat on a large scale (\nbartifacts Maven artifacts) showing that \percbloat of dependencies are bloated.
	\item A tool called \mytool to automatically analyze and remove bloated dependencies in Java applications packaged with \mv. \mytool can be used in future research on package management as well as by practitioners.
	\item A qualitative assessment of the opinion of developers regarding bloated dependencies. Through the submission of pull requests to notable open-source projects, we show that developers care about removing dependency bloat: \nbaccpr/\nbaccrejpr of answered pull requests have been merged, removing \totalremdeps bloated dependencies.
\end{itemize}

The remainder of this paper is structured as follows. \autoref{sec:background} introduces the key concepts about dependency management with \mv and presents a illustrative example. \autoref{sec:bloat} introduces the new terminology and describes the implementation of \mytool. \autoref{sec:methodology} presents the research questions that drive our study, as well as the methodology followed. \autoref{sec:results} covers our experimental results for each research question. Sections~\ref{sec:discussion} and \ref{sec:related} provide a comprehensive discussion of the results obtained and present the threats to the validity of our study. Section~\ref{sec:conclusion} concludes this paper and provides future research directions.

\section{Background}
\label{sec:background}

In this section, we provide an overview of the \mv package management system and present the essential terminology for this work. Then, we illustrate these concepts with a concrete example.

\subsection{Maven Dependency Management Terminology}

\mv is a popular package manager and build automation tool for Java projects and other languages that compile to the JVM (\eg, Scala, Kotlin, Groovy, Clojure, or JRuby). \mv supports developers in the management of the software build, from compilation to deployment~\citep{Varanasi2014}. \mv relies on a specific configuration file in XML format, known as the \pom (acronym for ``Project Object Model''), that allows customizing all the build lifecycle. This \pom file contains information about the project and all the configuration details utilized by \mv during its different building phases (\eg, \emph{compile}, \emph{test}, \emph{package}, \emph{install}, and \emph{deploy}). As in object-oriented programming, \pom files can inherit from a base \pom, known as the \mv parent \pom. 

\textbf{Maven Project.}
A \mv project is a set of source code files and configuration files. There are two ways of designing a \mv project: as a \emph{single-module}, or as a \emph{multi-module} project. The first has a single \pom file, which may include all the configurations needed to package the project as a single artifact (JAR file). The latter allows to separately build multiple artifacts in a certain specific order through a so-called aggregator POM. In multi-module projects, developers can define in a common parent \pom the dependencies used by all the modules, this facilitates the update of dependency versions at once. The \mv \emph{reactor algorithm}\footnote{\scriptsize{\url{https://maven.apache.org/guides/mini/guide-multiple-modules.html}}} ensures a deterministic build order in multi-module projects.

\textbf{Maven Artifact.} In this paper, we refer to \emph{artifacts} as compiled \mv projects that have been deployed to some external binary code repository for later reuse. In \mv, artifacts are typically packaged as JAR files, which contain Java bytecode, incl. public API members. Each artifact is uniquely identified with a triplet ($G$:$A$:$V$), where $G$ is the  \textit{groupId} that identifies the organization that develops the artifact, $A$ is the \textit{artifactId} that identifies the name of the library (or the module in the case of multi-module projects), and $V$ corresponds to its \textit{version}, which follows the \mv versioning scheme\footnote{\scriptsize{\url{https://maven.apache.org/pom.html\#Version_Order_Specification}}}. Currently, the most popular public repository to host \mv artifacts is the \mc repository. Artifacts in \mc are immutable, \ie, once deployed, they cannot be modified nor updated. Accordingly, all the versions of a given library remain permanently available in the repository~\cite{SotoValero2019}.

\textbf{Dependency Resolution.} 
The dependency resolution mechanism is a core feature of \mv. It works in two steps:
(1) it determines the set of dependencies required to build an individual artifact, and
(2) it fetches dependencies that are not present locally from external repositories such as \mc. \mv constructs a \emph{dependency tree}, which represents the dependency relationships between the \poms of the resolved dependencies. One can distinguish between three types of \mv dependencies: \textit{direct}, \textit{inherited}, and \textit{transitive}. Direct dependencies are those explicitly declared in the \pom file of the project or of the module. Inherited dependencies are those declared in the parent \pom. Transitive dependencies are those dependencies obtained from the transitive closure of direct and inherited dependencies. \mv relies on the \emph{dependency mediation algorithm}\footnote{\scriptsize{\url{https://maven.apache.org/guides/introduction/introduction-to-dependency-mechanism.html}}} to determine what version of an artifact is chosen when multiple versions of the same dependency are present in the dependency tree.

\subsection{A Brief Journey in the Dependencies of the \jxls Library} \label{sec:journey}

In this section, we illustrate the concepts introduced previously with one concrete example: 
\jxls\footnote{\scriptsize{\url{http://jxls.sourceforge.net}}}, an open source Java library for generating Excel reports. \jxls uses a special XLS template markup to define the output formatting and data layout for the reports.  It is implemented as a multi-module \mv project with a parent \pom: \texttt{jxls-project}, composed of three modules: \texttt{jxls}, \texttt{jxls-examples}, and \texttt{jxls-poi}. 

\begin{figure}[!ht]
	\centering
	\input{figures/jxls_dt.tex}
	\caption{Excerpt of the dependency tree of the multi-module \mv project \jxls. version $1.0.15$ (dependencies used for testing are not shown for the sake of simplicity).}
	\label{fig:jxls_dt}
\end{figure}
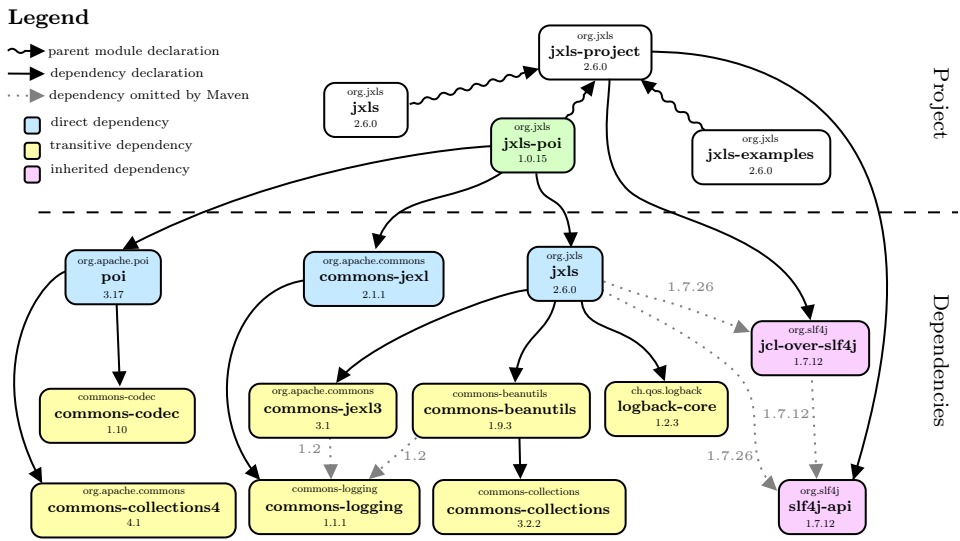

\autoref{lst:example-pom} shows an excerpt of the \pom file of the  \texttt{jxls-poi} module, version $1.0.15$.  It declares \texttt{jxls-project} as its parent \pom in lines $1-5$ and a direct dependency towards the \texttt{poi} Apache library in lines $10-14$. \autoref{fig:jxls_dt} depicts an excerpt of its \mv dependency tree (we do not show testing dependencies here, such as JUnit, to make the figure more readable). Nodes in blue, pink, and yellow represent direct, inherited, and transitive dependencies, respectively, for the \jxlspoi artifact (as reported by the \texttt{dependency:tree} \mv plugin). As we observe, it explicitly declares three direct dependencies: \texttt{commons-jexl}, \texttt{jxls}, and \texttt{poi}, while artifacts \texttt{jcl-over-slf4j} and \texttt{slf4j-api} are inherited dependencies of \jxlspoi because they are declared in its parent \pom. Since the artifact \texttt{poi} depends on Apache \codec and Apache \texttt{commons-collections4}, these artifacts are automatically added to the classpath of \jxlspoi as transitive dependencies. 

\begin{lstlisting}[basicstyle=\footnotesize\ttfamily,language=Pom, float, caption={Excerpt of the \pom file corresponding to the  module \jxlspoi of the multi-module \mv project \jxls.}, label={lst:example-pom}, numbers=left, linewidth=0.92\columnwidth]
<parent>
    <groupId>org.jxls</groupId>
    <artifactId>jxls-project</artifactId>
    <version>2.6.0</version>
</parent>
<artifactId>jxls-poi</artifactId>
<packaging>jar</packaging>
<version>1.0.15</version>
<dependencies>
    <dependency>
        <groupId>org.apache.poi</groupId>
        <artifactId>poi</artifactId>
        <version>3.17</version>
    </dependency>
    ...
</dependencies>
\end{lstlisting}

The library \texttt{jcl-over-slf4j} declares a dependency towards \texttt{slf4j-api}, version 1.7.12, which is omitted by \mv since it is already added from the \texttt{jxls-project} parent \pom. On the other hand, \jxls declares dependencies to version 1.7.26 of \texttt{jcl-over-slf4j} and \texttt{slf4j-api}, but the lower version 1.7.12 was chosen over it since it is nearer to the root in the dependency tree and, by default, \mv resolves version conflicts with a nearest-wins strategy. Once \mv finishes the dependency resolution, the classpath of \jxlspoi includes the following artifacts: \texttt{poi}, \codec, \texttt{commons-\allowbreak{}collections4}, \texttt{commons-\allowbreak{}jexl}, \texttt{commons-\allowbreak{}logging}, \texttt{jxls}, \texttt{commons-\allowbreak{}jex13}, \texttt{commons-\allowbreak{}beanutils}, \texttt{commons-\allowbreak{}collections}, \texttt{logback-\allowbreak{}core}, \texttt{jcl-\allowbreak{}over-\allowbreak{}slf4j}, and \texttt{slf4j-\allowbreak{}api}. The goal of our work is to determine if all the artifacts in the classpath of Maven projects such as \texttt{jxls-poi} are actually needed to build and run those projects.

\section{Bloated Dependencies} 
\label{sec:bloat}

In this section, we introduce  the key concept presented and studied in the rest of this paper: bloated dependencies. Then we describe our methodology to study bloated dependencies, as well as our tooling to automatically detect and remove them from \mv projects. 

\subsection{Novel Concepts}

Dependencies among \mv artifacts form a graph, according to the information declared in their \poms. The Maven Dependency Graph (MDG) is defined as follows:

\begin{definition} \textbf{\textit{(Maven Dependency Graph)}} \label{def:mdg} The MDG is a vertex-labelled graph, where vertices are \mv artifacts (uniquely identified by their $G$:$A$:$V$ coordinates), and edges represent dependency relationships among them.
	Formally, the MDG is defined as $\mathcal{G = (\mathcal{V}, \mathcal{E})}$, where:
	\begin{itemize}
		\item $\mathcal{V}$ is the set of artifacts in the \mc repository
		
		\item $\mathcal{E} \subseteq \mathcal{V} \times \mathcal{V}$ represent the set of directed edges that determine dependency relationships between each artifact $v \in \mathcal{V}$ and its dependencies
		
	\end{itemize}
\end{definition}

Each artifact in the MDG has its own Maven Dependency Tree (MDT), which is the transitive closure of all the dependencies needed to build the artifact, as resolved by \mv.

\begin{definition} \textbf{\textit{(Maven Dependency Tree)}} The MDT of an artifact $v \in \mathcal{V}$ is a directed acyclic graph of artifacts, with $v$ as the root node, and a set of edges $\mathcal{E}$ representing dependency relationships between them.
\end{definition}

In this work, we introduce the novel concept of \emph{bloated dependency} as follows:

\begin{definition} \textbf{\textit{(Bloated Dependency)}} \label{def:bloated_dep} An artifact $p$ is said to have a bloated dependency relationship $\varepsilon_b \in \mathcal{E}$ if there is a path in its MDT, between $p$ and any dependency $d$ of $p$, such as none of the elements in the API of $d$ are used, directly or indirectly, by $p$. 
\end{definition}

In order to reason about bloated dependencies, we extend the MDT to model the dependency trees of \mv artifact depending on usage relationships. To do so, we introduce edge labels to represent two types of dependency edge statuses: used or bloated. Consequently, we define the Dependency Usage Tree (DUT) as follows: 

\begin{definition} \textbf{\textit{(Dependency Usage Tree)}} \label{def:duf}
	The DUT, defined as $\mathcal{G = (\mathcal{V}, \mathcal{E}, \mathcal{r})}$, is an extension of the MDG that implements a dependency usage labelling function, which assigns one of the following six dependency usage types to the relationship between \mv artifacts and their dependencies in the MDT: $\mathcal{r} : \mathcal{E} \rightarrow \{\text{ud, ui, ut, bd, bi, bt}\}$ such that:
	
	\begin{equation*}
		\mathcal{r} (\langle p,d\rangle) = \begin{cases}
			\text{ud}, & \text{if } d \text{ is used and it is directly declared by } p\\
			\text{ui}, & \text{if } d \text{ is used and it is inherited from a parent of } p\\
			\text{ut}, & \text{if } d \text{ is used and it is resolved transitively by } p\\
			\text{bd}, & \text{if } d \text{ is bloated and it is directly declared by } p\\
			\text{bi}, & \text{if } d \text{ is bloated and it is inherited from a parent of } p\\
			\text{bt}, & \text{if } d \text{ is bloated and it is resolved transitively by } p\\
		\end{cases}
	\end{equation*}
\end{definition}

\subsection{Example}

\begin{figure}[!ht]
	\centering
	\input{figures/jxls_dt_labelled.tex}
	\caption{Dependency Usage Tree (DUT) for the example presented in \autoref{fig:jxls_dt}. Edges are labelled according to \autoref{def:duf} to reflect the usage status between \jxlspoi and each one of its dependencies.
	}
	\label{fig:jxls_dt_labelled}
\end{figure}
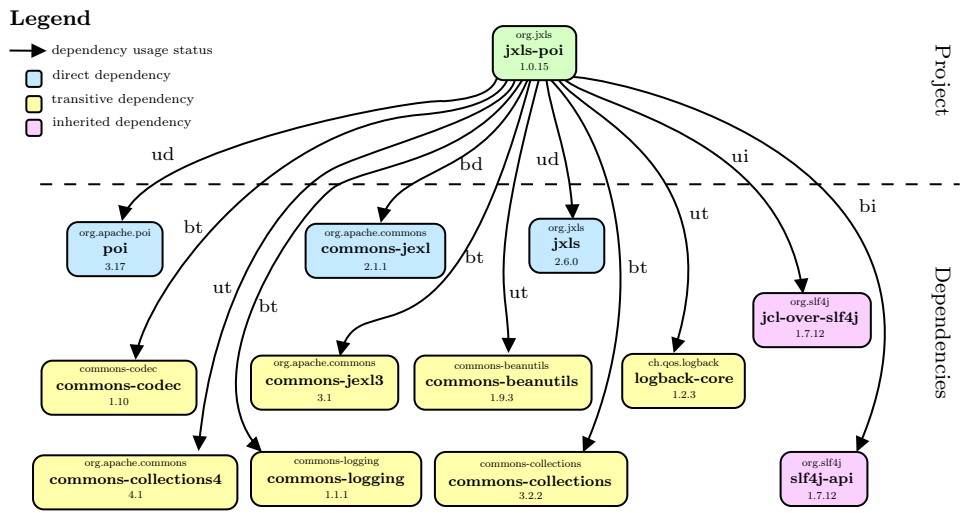

\autoref{fig:jxls_dt_labelled} illustrates how the dependency usage function creates a DUT for the example presented in \autoref{fig:jxls_dt}. If we analyze the bytecode contained in \jxlspoi, no references to any API member in the direct dependencies \texttt{commons-jexl} (explicitly declared in the \pom) and \texttt{sl4j} (inherited from its parent \pom) can be found. Therefore, according to \autoref{def:duf}, they are labelled as bloated-direct (bd) and bloated-inherited (bi) dependencies, respectively. According to the bytecode usage analysis, the direct dependency \texttt{commons-jexl} can be safely removed directly from the \pom, while the inherited dependency \texttt{sl4j} can be safely removed in the parent. Indeed, after analyzing the Git history of \jxls, we found that these changes were made in the \pom by the authors of the project in a later released version of this library\footnote{\scriptsize{\bitbucketurl}}.

Now let us consider dependencies obtained transitively through the \mv dependency resolution mechanism.
\texttt{commons-\allowbreak{}codec} and \texttt{commons-\allowbreak{}collections} are included by transitive closure from \jxlspoi, but they are never used, directly in its bytecode or indirectly through a used API member of its dependencies. Therefore, these bloated-transitive (bt) dependencies are not necessary to build \jxlspoi.

To sum up, the set of dependencies of a given \mv artifact can be partitioned into six dependency usage types, according to two criteria: (1) the dependency usage status (used or bloated), and (2) the dependency type (direct, inherited, or transitive). \autoref{tab:consolidation} summarizes this partition for \jxlspoi.

\begin{table}[htb]
	\scriptsize
	\caption{Contingency table of the different types of dependency relationships studied in this work for the example presented in \autoref{sec:journey}.}
	\vspace{-10pt}
	\begin{tabular}{l|l|l} 
		& \textbf{Used} & \textbf{Bloated} \\
		\hline
		\textbf{Direct} & poi, jxls & commons-jexl \\
		\hline
		\textbf{Inherited} & jcl-over-sl4j & sl4j-api \\
		\hline
		\multirow{2}{*}{\textbf{Transitive}} & commons-beanutils,logback-core,  & commons-logging, commons-collections, \\
		& commons-collections4 & commons-codec, commons-jexl3 \\
		\hline
	\end{tabular}
	\vspace{-10pt}
	\label{tab:consolidation}
\end{table}

\subsection{\mytool: A Tool for Detecting and Removing Bloated Dependencies}

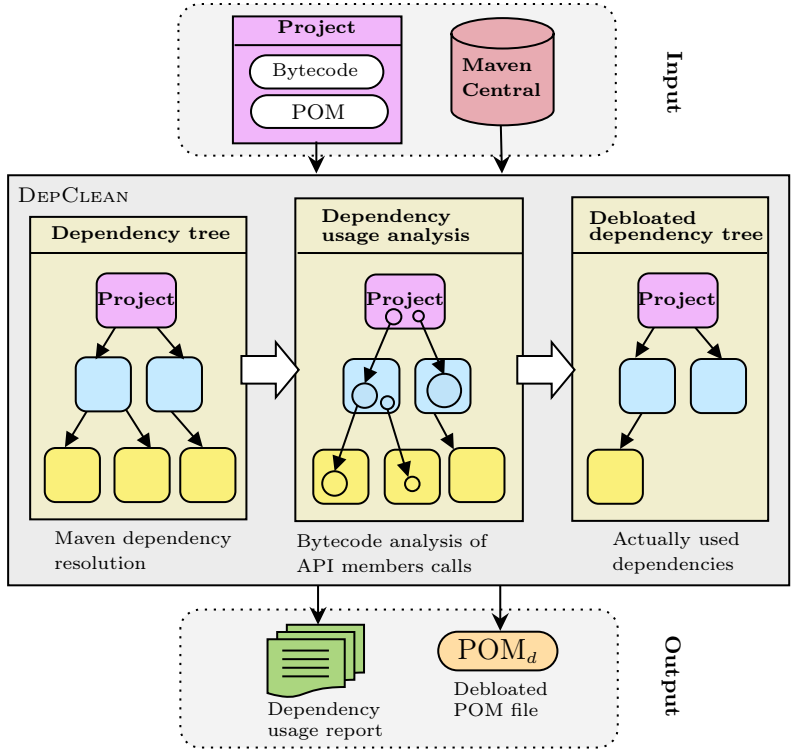
\begin{figure}
	\centering
	\input{figures/depclean.tex}
	\caption{Overview of the tool \mytool to detect and remove bloated dependencies in \mv projects. Rounded squares represent artifacts, circles inside the artifacts are API members, arrows between API members represents bytecode calls between artifacts, arrows between artifacts represent dependency relationships between them.
	}
	\label{fig:depclean}
\end{figure}

For our study, we design and implement a specific tool called \mytool. An overview of \mytool is shown in \autoref{fig:depclean}, it works as follows. It receives as inputs a built \mv project and a repository of artifacts, then it extracts the dependency tree of the projects and constructs a DUT to identify the set of dependencies that are actually used by the project. \mytool has two outputs:
(1) it returns a report with the usage status of all types of dependencies, and (2) it produces an alternative version of the \pom file ($\text{\pom}_d$) with all the bloated dependencies removed (\ie, the XML node of the bloated dependency is removed). 

Algorithm~\ref{alg:debloat} details the main procedure of \mytool. The algorithm receives as input a \mv artifact $p$ that includes a set of dependencies in its dependency tree, denoted as $DT$, and returns a report of the usage status of its dependencies and a debloated version of its \pom. Notice that \mytool computes two transitive closures: over the \mv dependency tree (line \ref{line:algo:dt}) and over the call graph of API members (line~\ref{line:algo:used}).

The algorithm first  copies the \pom file of $p$, resolving all its direct and transitive dependencies locally, and obtaining the dependency tree (lines \ref{line:algo:copy} and \ref{line:algo:dt}). If $p$ is a  module of a multi-module project, then all the dependencies declared in its parent \pom are included as direct dependencies of $p$.
Then, the algorithm proceeds to construct a set with the dependencies that are actually used by $p$ (line \ref{line:algo:used}). 

\begin{algorithm}
	\label{alg:debloat}
	\SetAlgoLined
	\SetNoFillComment
	\KwData{A \mv artifact $p$.} 
	\KwResult{A report of the bloated dependencies of $p$ and a clean POM file $f$ of $p$ without bloated dependencies.}
	
	$f$ $\leftarrow$ \textit{copyPOM}($p$)\;  \label{line:algo:copy}
	$DT$ $\leftarrow$ \textit{getDependencyTree}($p$)\; \label{line:algo:dt}
	$UD$ $\leftarrow$ \textit{getUsedDependencies}($p$, $DT$)$\text{\;}$ \label{line:algo:used} \tcp{refer to Algorithm~\ref{alg:isUsed}}
	
	\label{line:algo:resolve}
	\ForEach{$d \in DT$}{
		\If{$d \in UD$}{\label{line:algo:api} 
			\textbf{continue}$\text{\;}$ \tcp{do nothing, the dependency is actually used by $p$}
		}
		\eIf{\textit{isDeclared}($d, f$)}{ 
			report $d$ as bloated-direct and \textit{remove}($d$, $f$)$\text{\;}$ \label{line:algo:remove}
		}{
			report $d$ as bloated-inherited or bloated-transitive, and \textit{exclude}($d$, $f$)$\text{\;}$ \label{line:algo:exclude}
		}
	}
	\Return $f$\;
	\caption{Main procedure to detect and remove bloated dependencies.}
\end{algorithm}

\begin{algorithm}
	\SetAlgoLined
	\SetNoFillComment
	\KwData{A \mv artifact $p$ and its dependency tree $DT$.} 
	\KwResult{A set of dependencies $\text{\textit{UD}}$ actually used by $p$.}
	$UD$ $\leftarrow$ $\emptyset$\;
	\ForEach{$d \in DT$}{
		\eIf{\textit{$(p,d) \text{ is a direct or inherited dependency}$}}{ 
			$part_d$ $\leftarrow$ \textit{extractMembers}($p, d$)$\text{\;}$           \tcp{extract the subpart of $d$ used by $p$}
			\eIf{$part_d \neq \emptyset$}{ 
				add($d$, $UD$)\;
			}{
				\textbf{continue}$\text{\;}$
			}
		}{ 
			find the path $l = [p, ..., d]$ connecting $p$ and $d$ in $DT$\;
			$a$ $\leftarrow$ $p$\;
			\ForEach{$b \in l_{2, ...,n}$}{
				$part_b$ $\leftarrow$ \textit{extractMembers}($a, b$)\;
				\eIf{$part_b = \emptyset$}{
					\textbf{break}$\text{\;}$
				}{
					$a$ $\leftarrow$ $part_b$\;
				}
				
				\If{$b = d$}{
					add($d$, $UD$)\;
				}
			}
		}
	}
	\caption{Procedure to obtain all the dependencies used, directly or indirectly, by a \mv artifact.}
	\label{alg:isUsed}
\end{algorithm}

Algorithm~\ref{alg:isUsed} explains the bytecode analysis. The detection component statically analyzes the bytecode of $p$ and all its dependencies to check which API members are being referenced by the artifact, either directly or indirectly. Notice that it behaves differently if the included artifact is a direct, inherited, or a transitive dependency. If none of the API members of a dependency $d \in DT$ are called, even indirectly via transitive dependencies, then $d$ is considered to be bloated and we proceed to remove it. 

In \mv, we remove bloated dependencies in two different ways: (1) if the bloated dependency is explicitly declared in the \pom, then we remove its declaration clause directly (line \ref{line:algo:remove} in Algorithm~\ref{alg:debloat}), 
or (2) if the bloated dependency is inherited from a parent \pom or induced transitively, then we excluded it in the \pom (line \ref{line:algo:exclude} in Algorithm~\ref{alg:debloat}). This exclusion consists in adding an \texttt{<exclusion>} clause inside a direct dependency declaration, with the \emph{groupId} and \emph{artifactId} of the transitive dependency to be excluded. Excluded dependencies are not added to the classpath of the artifact by way of the dependency in which the exclusion was declared.

\mytool is implemented in Java as a Maven plugin that extends the \texttt{maven-dependency-analyzer}\footnote{\scriptsize{\url{ http://maven.apache.org/shared/maven-dependency-analyzer}}} tool maintained by the Apache foundation. For the construction of the dependency tree, \mytool relies on the \texttt{maven-dependency-plugin} with the \texttt{copy-dependencies} and \texttt{tree} goals. Internally, \mytool addresses some of the key aspects of static analysis in Java. It relies on the ASM\footnote{\scriptsize{
		\url{https://asm.ow2.io}}} library to visit all the \texttt{.class} files in order to register bytecode calls towards classes, methods, fields, and annotations among \mv artifacts. 
To do so, we define a customized parser that reads entries in the constant pool table of \texttt{.class} files directly, in case it contains special references that ASM does not support. This allows the plugin to statically capture reflection calls that are based on string literals and concatenations.
\mytool adds unique features to the \texttt{maven-dependency-analyzer} to detect and report transitive and inherited bloated dependencies, and to produce a debloated version of the \pom file. \mytool is open-source and reusable from \mc, the source code is available at \url{https://github.com/castor-software/depclean}. 

\section{Experimental Methodology}
\label{sec:methodology}

In this section, we present the research questions that articulate our study. We also describe the experimental protocols used to select and analyze \mv artifacts for an assessment of the impact of bloated dependencies in this ecosystem.

\subsection{Research Questions}

\newcommand{\RQone}{\textbf{RQ1}: \textit{How frequently do bloated dependencies occur?}} 

\newcommand{\RQtwo}{\textbf{RQ2}: \textit{How do the reuse practices affect bloated dependencies?}}

\newcommand{\RQthree}{\textbf{RQ3}: \textit{To what extent are developers willing to remove bloated-direct dependencies?}}

\newcommand{\RQfour}{\textbf{RQ4}: \textit{To what extent are developers willing to exclude bloated-transitive dependencies?}}

Our investigation of bloated dependencies in the Maven ecosystem is composed of four research questions grouped in two parts. In the first part, we perform a large scale quantitative study to answer the following research questions:

\begin{itemize}
	\item \RQone~With this research question, we aim at quantifying the amount of bloated dependencies among \nbartifacts Maven artifacts. We measure direct, inherited and transitive dependencies to provide an in-depth assessment of the  dependency bloat  in the \mv ecosystem. 
	
	\item \RQtwo~In this research question, we analyze bloated dependencies with respect to two distinctive aspects of reuse in the \mv ecosystem: the additional complexity of the \mv dependency tree caused by transitive dependencies, and the choice of a multi-module architecture.
\end{itemize}

The second part of our study focuses on \nbpr notable \mv projects and presents the qualitative feedback about how developers react to bloated dependencies, and to the solutions provided by \mytool. It is guided by the following research questions:

\begin{itemize}
	\item \RQthree~Direct dependencies are those that are explicitly declared in the \pom. Hence, those dependencies are easy to remove since it only requires the modification of a \pom that developers can easily change. In this research question, we use \mytool to detect and fix bloated-direct dependencies. Then, we communicate the results to the developers. We report on their feedback.
	
	\item \RQfour~Transitive dependencies are those not explicitly declared in the \pom but induced from other dependencies. We exchange with developers about such cases. This gives unique insights about how developers react to excluding transitive dependencies from their projects.
\end{itemize}

\subsection{Experimental Protocols}

\subsubsection{Protocol of the Quantitative Study (RQ1 \& RQ2)}

\begin{figure}[htb]
	\centering
	\input{figures/exp_framework.tex}
	\caption{Experimental framework used to collect artifacts and analyze bloated dependencies in the \mv ecosystem.}
	\label{fig:exp_framework}
	\vspace{-10pt}
\end{figure}
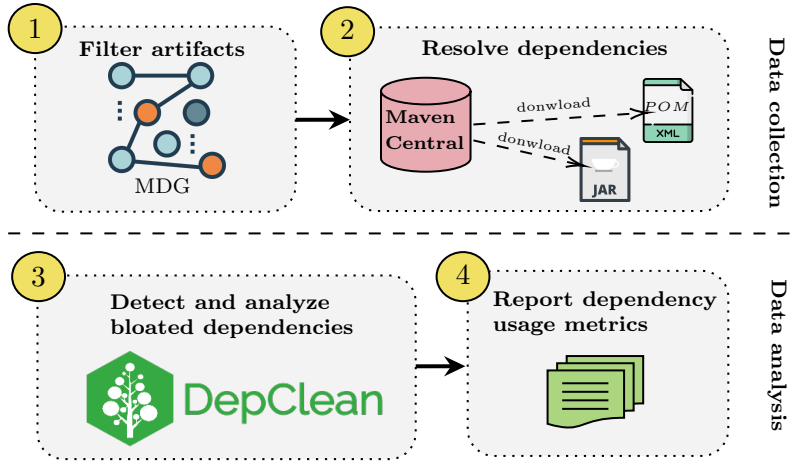

\autoref{fig:exp_framework} shows our process to build a dataset of \mv artifacts in order to answer RQ1 and RQ2. Steps \ding[1.2]{172} and \ding[1.2]{173} focus on the collection of a representative set of \mv artifacts:  we sample our study subjects from the whole \mv Dependency Graph, then we resolve the dependencies of each study subject. In steps  \ding[1.2]{174} and \ding[1.2]{175}, we analyze dependency usages with \mytool and compute the set of metrics to answer RQ1 and RQ2. 

\ding[1.2]{172} \textbf{Filter artifacts.}
In the first step, we leverage the \mdg (MDG) from previous research~\citep{Benelallam2019}, a graph database that captures the complete dependency relationships between artifacts in \mc at a given point in time. Then, we randomly select a representative sample of $14,699$~\mv artifacts. Representativeness is achieved by sampling over the probability distribution of the number of dependencies per artifact in the MDG, per the recommendation of Shull ~\citep[Chapter~8.3.1]{Shull2007}. From the sampled artifacts, we select as study subjects all the  artifacts that meet the following additional criteria. This results in a dataset of \nbartifacts artifacts.

\begin{itemize}
	\item Public API: The subjects must contain at least one \texttt{.class} file with one or more public methods, \ie, can be reused via external calls to their API.
	
	\item Diverse: The subjects all have different \emph{groupId} and \emph{artifactId}, \ie, they belong to different \mv projects.
	
	\item Reused: The subjects are used by at least one client via direct declaration.
	
	\item Complex: The subjects  have at least one direct dependency with \emph{compile} scope, \ie, we can analyze the dependency tree and the reused artifacts.
	
	\item Latest: The subjects are the latest released version at the time of the experiment (October, 2019).
\end{itemize}

\ding[1.2]{173} \textbf{Resolve dependencies.}
In the second step, we download the binaries of all the selected artifacts and their \poms from \mc and we resolve all their direct and transitive dependencies to a local repository. To ensure the consistency of our analysis, we discard the artifacts that depend on libraries hosted in external repositories. In case of any other error when downloading some dependency, we exclude the artifact from our analysis.

\autoref{tab:subjects} shows the descriptive statistics about \nbartifacts artifacts in our dataset for RQ1 and RQ2.  The dataset includes \nbdirect direct, \nbinherited inherited, and \nbtransitive transitive dependency relationships (\nbedges in total). We report about their dependencies with \emph{compile} scope, since those dependencies are necessary to build the artifacts. Columns \#C, \#M, and \#F give the distribution of the number of classes, methods, and fields per artifact (we count both the public and private API members). The size of artifacts varies, from small artifacts with one single class (\eg, \texttt{org.elasticsearch.client:\allowbreak{}transport:\allowbreak{}6.2.4}), to large libraries with thousands of classes (\eg, \texttt{org.apache.hive:\allowbreak{}hive-exec:\allowbreak{}3.1.0)}. In total, we analyze the bytecode of more than $30$ millions of API members. Columns \#D, \#I, and \#T accounts for the distributions of direct, inherited, and transitive dependencies, respectively. \texttt{com.bbossgroups.pdp:\allowbreak{}pdp-system:\allowbreak{}5.0.3.9} is the artifact with the largest number of declared dependencies in our dataset, with $148$ dependency declarations in its \pom file, while   \texttt{be.atbash.json:\allowbreak{}octopus-accessors-smart:\allowbreak{}0.9.1} has the maximum number of transitive dependencies: $1,776$. Interestingly, the distributions of direct and transitive dependencies are notably different, the total number of transitive dependencies is an order of magnitude larger than direct dependencies, with mean of $20$ and $4$, respectively. 

\begin{table}[htb]
	\centering
	\scriptsize
	\caption{Descriptive statistics of the \nbartifacts~\mv artifacts selected to conduct our quantitative study of bloated dependencies (RQ1 \& RQ2).}
	\vspace{-10pt}
	\begin{tabular}{l|ccc|ccc} 
		\toprule
		& \multicolumn{3}{c|}{API Members} & \multicolumn{3}{c}{Dependencies} \\ 
		\cline{2-7}
		& \#C & \#M & \#F & \#D & \#I & \#T \\ 
		\hline
		
		\rowcolor[HTML]{EEEEEE}
		Min.  & $1$  & $1$  & $0$  &  $1$  & $0$  &  $0$ \\
		1st-Q  & $10$  & $63$  & $21$  & $2$  & $0$  & $6$   \\
		
		\rowcolor[HTML]{EEEEEE}
		Median & $32$  & $231$  & $75$  &  $4$  & $2$  & $20$  \\
		
		3rd-Q  & $111$  & $891$  & $279$  &  $7$  & $18$  &  $59$ \\
		
		\rowcolor[HTML]{EEEEEE}
		Max.  & $47,241$  & $435,695$  & $129,441$  &  $148$  & $453$  & $1,776$  \\ 
		\hline
		Total  & $2,397,879$  & $22,633,743$  & $6,510,584$  &   \nbdirect  &  \nbinherited  & \nbtransitive   \\
		\bottomrule
	\end{tabular}
	\vspace{-10pt}
	\label{tab:subjects}
\end{table}

\ding[1.2]{174} \textbf{Dependency usage analysis.}
This is the first step to answer RQ1 and RQ2: collect the status of all dependency relationships for each artifact in our dataset.
For each artifact, we first unpack its JAR file, as well as its dependencies. Then, for each JAR file, we  analyze all the bytecode calls to API class members using  \mytool. This provides a Dependency Usage Tree (DUT) for each artifact, on which each dependency relationship is labeled with one of the six categories as we illustrated in \autoref{tab:consolidation}: bloated-direct (bd), bloated-inherited (bi), bloated-transitive (bt), used-direct (ud), used-inherited (ui), or used-transitive (ut).

\ding[1.2]{175} \textbf{Collect dependency usage metrics.}
This last step consists of collecting a set of metrics about the global presence of bloated dependencies. We focus on the presence of bloat with respect to the complexity of the Maven dependency trees, and with respect to the multi-module Maven architecture. 

In RQ1, we analyze our dataset as a whole, looking at the usage status of dependency relationships from two perspectives.

\emph{Global distribution of dependency usage.} This is the normalized distribution of each category of dependency usage, over each dependency relationship of each of the \nbartifacts artifacts in our dataset.

\emph{Distribution of dependency usage type, per artifact.} For each of the six types of dependency usage, we compute the normalized ratio over the total number of dependency relationships for each artifact in our dataset.

In RQ2, we analyze how the specific reuse strategies of \mv relate to the presence of bloated dependencies. First, we relate bloated dependencies to the complexity of the \mv dependency tree of each artifact, according to the following metrics. 

\emph{Bloated dependencies \wrt the number of transitive dependencies.} For each artifact that has at least one transitive dependency, we determine the relation between the ratio of transitive dependencies and the ratio of bloated dependencies.

\emph{Bloated-transitive dependencies \wrt to the height of the dependency tree.} Given an artifact and its \mv dependency tree, the height of the tree is the longest path between the root and its leaves. To compute this metric, we group our artifacts according to the height of their tree. The maximum dependency tree height that we observed is 14. However, there are only 58 artifacts with a tree higher than 10. So, we group all artifacts with height $\geq 10$. For each subset of artifact with the same height, we compute the size of the subset and the distribution of bloated-transitive dependencies of each artifact in the subset.

Second, we distinguish the presence of bloated dependencies between single and multi-module \mv projects, according to the following metrics: 

\emph{Global distribution of dependency usage in a single or multi-module project.} We present two plots that measure the distribution of each type of dependency usage in the set of single and multi-module projects. It is to be noted that the plot for single-module projects does not include bloated-inherited (bi) and used-inherited (ui) dependencies since they have no inherited dependencies.

\emph{Distribution of dependency usage type, per artifact, in a single or multi-module project.} We present two plots that provide six distributions each: the distribution of each type of dependency usage type for artifacts that are in a single-module or multi-module project.

\subsubsection{Protocol of the Qualitative Study (RQ3 \& RQ4)} 
\label{sec:methodology_qualitative}

In RQ3 and RQ4, we perform a qualitative assessment about the relevance of bloated dependencies for the developers of open-source projects. We systematically select \nbpr notable open-source projects hosted on  GitHub to conduct this  analysis. We query the GitHub API to list all the Java projects ordered by their number of stars. Then, we randomly select a set of projects that fulfil all the following criteria: ($1)$ we can build them successfully with \mv, ($2)$ the last commit was at the latest in October 2019, ($3)$ they declare at least one dependency in the \pom, ($4)$ they have a description in the README about how to contribute through pull requests, and ($5)$ they have more than $100$ stars on GitHub.

\autoref{tab:projects_subjects} shows the selected \nbpr projects per those criteria, to which we submitted at least one pull request. They are listed in decreasing order  of stars on GitHub. The first column shows the name of the project as declared on GitHub, followed by the name of the targeted module if the project is multi-module. Notice that in the case of \texttt{jenkins} we submitted two pull requests targeting two distinct modules: \texttt{core} and \texttt{cli}. Columns two to four describe the projects according to its category as assigned to the corresponding released artifact in \mc, the number of commits in the master branch in October 2019, and the number of starts at the moment of conducting this study. Columns five to seven report about the total number of direct, inherited, and transitive dependencies included in the dependency tree of each considered project. 

\begin{table}
	\centering
	\tiny
	\caption{\mv projects selected to conduct our qualitative study of bloated dependencies (RQ3 \& RQ4).}
	\vspace{-10pt}
	\begin{threeparttable}
		\begin{tabular}{l|lcc|ccc} 
			\toprule
			\multirow{2}{*}{Project} & \multicolumn{3}{c|}{Description} & \multicolumn{3}{c}{Dependencies} \\ 
			\cline{2-7}
			& \multicolumn{1}{l}{Category} & \#Commits & \#Stars & \multicolumn{1}{c}{\#D} & \multicolumn{1}{c}{\#I} & \multicolumn{1}{c}{\#T}  \\ 
			\hline
			\rowcolor[HTML]{EEEEEE}
			
			\texttt{jenkins [core]} & Automation Server & 29,040 & 14,578 & \multicolumn{1}{c}{51} & \multicolumn{1}{c}{2} & \multicolumn{1}{c}{87} \\
			\rowcolor[HTML]{EEEEEE}
			\texttt{jenkins [cli]} & Automation Server & 29,040 & 14,578 & \multicolumn{1}{c}{17} & \multicolumn{1}{c}{2} & \multicolumn{1}{c}{0} \\
			\texttt{mybatis-3 [mybatis]} & Relational Mapping & 3,145 & 12,196 & \multicolumn{1}{c}{23} & \multicolumn{1}{c}{0} & \multicolumn{1}{c}{51} \\
			
			\rowcolor[HTML]{EEEEEE}
			\texttt{flink [core]} & Streaming & 19,789 & 11,260 & \multicolumn{1}{c}{14} & \multicolumn{1}{c}{10} & \multicolumn{1}{c}{34} \\
			\texttt{checkstyle [checkstyle]} & Code Analysis & 8,897 & 8,897 & \multicolumn{1}{c}{18} & \multicolumn{1}{c}{0} & \multicolumn{1}{c}{36} \\
			
			\rowcolor[HTML]{EEEEEE}
			\texttt{auto [common]} & Meta-programming & 1,081 & 8,331 & \multicolumn{1}{c}{8} & \multicolumn{1}{c}{0} & \multicolumn{1}{c}{24} \\
			\texttt{neo4j [collections]} & Graph Database & 66,602 & 7,069 & \multicolumn{1}{c}{8} & \multicolumn{1}{c}{2} & \multicolumn{1}{c}{21}  \\
			
			\rowcolor[HTML]{EEEEEE}
			\texttt{CoreNLP} & NLP & 15,544 & 6,812 & \multicolumn{1}{c}{23} & \multicolumn{1}{c}{0} & \multicolumn{1}{c}{45} \\
			\texttt{moshi [moshi-kotlin]} & JSON Library & 793 & 5,731 & \multicolumn{1}{c}{14} & \multicolumn{1}{c}{0} & \multicolumn{1}{c}{21} \\
			
			\rowcolor[HTML]{EEEEEE}
			\texttt{async-http-client [http-client]} & HTTP Client & 4,034 & 5,233 & \multicolumn{1}{c}{29} & \multicolumn{1}{c}{16} & \multicolumn{1}{c}{130} \\
			\texttt{error-prone [core]} & Defects Detection & 4,015 & 4,915 & \multicolumn{1}{c}{44} & \multicolumn{1}{c}{0} & \multicolumn{1}{c}{35} \\
			
			\rowcolor[HTML]{EEEEEE}
			\texttt{alluxio [core-transport]} & Database & 30,544 & 4,442 & \multicolumn{1}{c}{6} & \multicolumn{1}{c}{14} & \multicolumn{1}{c}{73} \\
			\texttt{javaparser [symbol-solver-logic]} & Code Analysis & 6,110 & 2,784 & \multicolumn{1}{c}{3} & \multicolumn{1}{c}{0} & \multicolumn{1}{c}{8} \\
			
			\rowcolor[HTML]{EEEEEE}
			\texttt{undertow [benchmarks]} & Web Server & 4,687 & 2,538 & \multicolumn{1}{c}{10} & \multicolumn{1}{c}{0} & \multicolumn{1}{c}{19} \\
			\texttt{wc-capture [driver-openimaj]} & Webcam & 629 & 1,618 & \multicolumn{1}{c}{3} & \multicolumn{1}{c}{0} & \multicolumn{1}{c}{84} \\
			
			\rowcolor[HTML]{EEEEEE}
			\texttt{teavm [core]} & Compiler & 2,334 & 1,354 & \multicolumn{1}{c}{9} & \multicolumn{1}{c}{0} & \multicolumn{1}{c}{9}  \\
			\texttt{handlebars [markdown]} & Templates & 916 & 1,102 & \multicolumn{1}{c}{6} & \multicolumn{1}{c}{0} & \multicolumn{1}{c}{13} \\
			
			\rowcolor[HTML]{EEEEEE}
			\texttt{jooby [jooby]} & Web Framework & 2,462 & 1,083 & \multicolumn{1}{c}{23} & \multicolumn{1}{c}{0} & \multicolumn{1}{c}{68} \\
			\texttt{tika [parsers]} & Parsing library & 4,650 & 929 & \multicolumn{1}{c}{81} & \multicolumn{1}{c}{0} & \multicolumn{1}{c}{67} \\
			
			\rowcolor[HTML]{EEEEEE}
			\texttt{orika [eclipse-tools]} & Object Mapping & 970 & 864 &  \multicolumn{1}{c}{3} &  \multicolumn{1}{c}{0} &  \multicolumn{1}{c}{3}  \\
			\texttt{spoon [core]} & Meta-programming & 2,971 & 840 & \multicolumn{1}{c}{16} & \multicolumn{1}{c}{2} & \multicolumn{1}{c}{59}  \\
			
			\rowcolor[HTML]{EEEEEE}
			\texttt{accumulo [core]} & Database & 10,314 & 763 & \multicolumn{1}{c}{26} & \multicolumn{1}{c}{1} & \multicolumn{1}{c}{51} \\
			\texttt{couchdb-lucene} & Text Search & 1,121 & 752 & \multicolumn{1}{c}{25} & \multicolumn{1}{c}{0} & \multicolumn{1}{c}{112} \\
			
			\rowcolor[HTML]{EEEEEE}
			\texttt{jHiccup} & Profiling & 215 & 519 & \multicolumn{1}{c}{4} & \multicolumn{1}{c}{0} & \multicolumn{1}{c}{1}  \\
			\texttt{subzero [server]} & Cryptocurrency & 158 & 499 & \multicolumn{1}{c}{6} & \multicolumn{1}{c}{0} & \multicolumn{1}{c}{100}  \\
			
			\rowcolor[HTML]{EEEEEE}
			\texttt{vulnerability-tool [shared]} & Security & 1,051 & 324 & \multicolumn{1}{c}{6} & \multicolumn{1}{c}{4} & \multicolumn{1}{c}{2} \\
			\texttt{para [core]} & Cloud Framework & 1,270 & 310 & \multicolumn{1}{c}{47} & \multicolumn{1}{c}{2} & \multicolumn{1}{c}{112} \\
			
			\rowcolor[HTML]{EEEEEE}
			\texttt{launch4j-maven-plugin} & Deployment Tool & 316 & 194 & \multicolumn{1}{c}{7} & \multicolumn{1}{c}{0} & \multicolumn{1}{c}{61} \\
			\texttt{jacop} & CP Solver & 1,158 & 155 & \multicolumn{1}{c}{7} & \multicolumn{1}{c}{0} & \multicolumn{1}{c}{9} \\
			
			\rowcolor[HTML]{EEEEEE}
			\texttt{selenese-runner-java} & Interpreter & 1,688 & 117 & \multicolumn{1}{c}{23} & \multicolumn{1}{c}{0} & \multicolumn{1}{c}{148} \\
			\texttt{commons-configuration} & Config library & 3,159 & 100 & \multicolumn{1}{c}{31} & \multicolumn{1}{c}{0} & \multicolumn{1}{c}{49}  \\
			\bottomrule
			
		\end{tabular}
		
	\end{threeparttable}
	\vspace{-10pt}
	\label{tab:projects_subjects}
\end{table}

\begin{figure}[!ht]
	\centering
	\frame{\includegraphics[width=0.8\columnwidth]{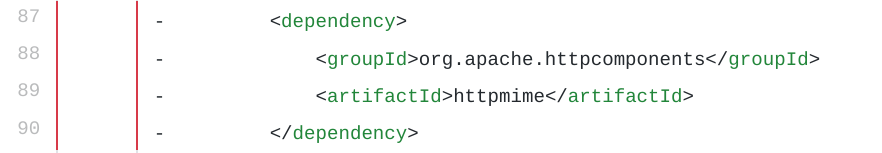}}
	\caption{Example of commit removing the bloated-direct dependency \texttt{org.apache.httpcomponents:\allowbreak{}httpmime} in the project Undertow.}
	\label{fig:removed_dep_example-2}
	\vspace{-10pt}
\end{figure}

We answer RQ3 according to the following protocol: first, we run \mytool and try to build the artifact with the debloated \pom file; second, we analyze the history of the project to propose a relevant change to the developers; third, we propose a change in the \pom file in the form of a pull request; four, we discuss the pull request through GitHub. \autoref{fig:removed_dep_example-2} shows an excerpt of the diff of such a change. Since our submitted pull requests represent a small modification in the \pom, they have a good chance to be reviewed by developers and discussed before being merged or rejected.

In the first step of the protocol, we use \mytool to obtain a report about the usage  of dependencies. We analyze dependencies with both compile and test scope. Once a bloated-direct dependency is found, we remove it directly in the \pom and proceed to build the project. If the project builds successfully after the removal (all the tests pass), we submit the pull request with the corresponding change. If after the removal of the dependency the build fails, then we consider the dependency as used dynamically and do not suggest to remove it. %
In the case of multi-module projects, with bloated dependencies in several modules, we submitted a single pull request per module.

\begin{figure}[!ht]
	\centering
	\frame{\includegraphics[width=0.8\columnwidth]{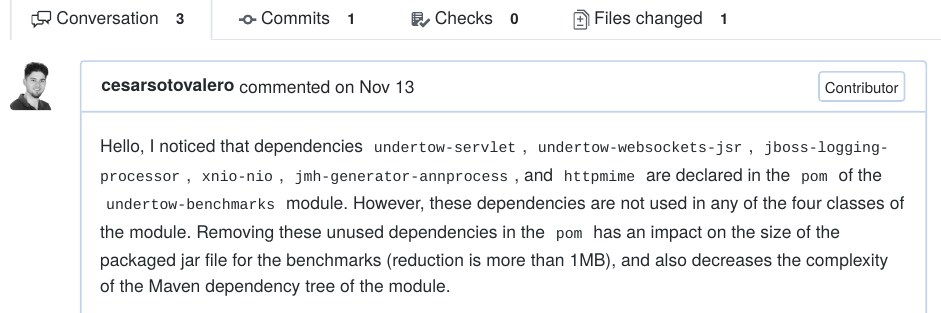}}
	\caption{Example of message of a pull request sent to the project Undertow on GitHub.}
	\label{fig:pr_undertow}
	\vspace{-10pt}
\end{figure}

For each pull request, we analyze the Git history of the \pom file to determine when the bloated dependency was declared or modified. Our objective is to collect information in order to understand how the dependencies of the projects change during their evolution. This allows us to prepare a more informative  pull request message and to support our discussion with developers. We also report on the benefits of tackling these bloated dependencies by describing the differences between the original and the debloated packaged artifact of the project in terms of the size of the bundle and the complexity of its dependency tree, when the difference was significant. 
Each pull request includes an explanatory message. \autoref{fig:pr_undertow} shows an example of the pull request message  submitted to the project Undertow\footnote{\scriptsize{\url{https://github.com/undertow-io/undertow}}}. The message explains the motivations of the proposed change, as well as the negative impact of keeping these bloated dependencies in the project. 

\begin{figure}[!ht]
	\centering
	\frame{\includegraphics[width=0.8\columnwidth]{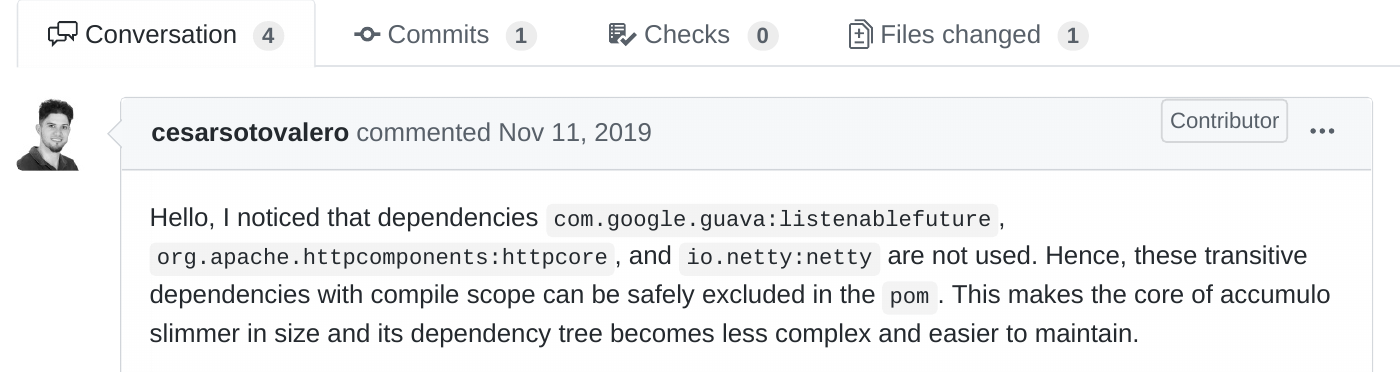}}
	\caption{Example of message in a pull request sent to the project Apache Accumulo on GitHub.}
	\label{fig:exclude_dep_message}
	\vspace{-10pt}
\end{figure}

\begin{figure}[!ht]
	\centering
	\frame{\includegraphics[width=0.8\columnwidth]{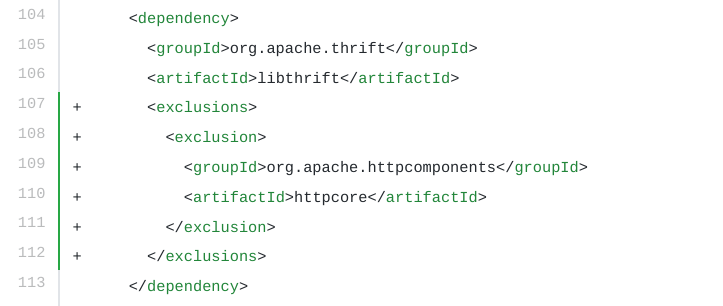}}
	\caption{Example of commit excluding the bloated-transitive dependency \texttt{org.apache.httpcomponents:\allowbreak{}httpcore} in the project Apache Accumulo.}
	\label{fig:excluded_dep_example}
	\vspace{-10pt}
\end{figure}

To answer RQ4, we follow the same pull request submission protocol as for RQ3. We use \mytool to detect bloated-transitive dependencies and submit pull requests suggesting the addition of the corresponding exclusion clauses in each project \pom. \autoref{fig:exclude_dep_message} shows an example of a pull request message submitted to the project Apache Accumulo\footnote{\scriptsize{\url{https://github.com/apache/accumulo}}}, while \autoref{fig:excluded_dep_example} shows an excerpt of the commit proposing the exclusion of the transitive dependency \texttt{org.apache.httpcomponents:\allowbreak{}httpcore} from the direct dependency \texttt{org.apache.thrift:libthrift} in its \pom.

Additional information related to the selected projects and the research methodology employed is publicly available as part of our replication package at \url{https://github.com/castor-software/depclean-experiments}. 

\section{Experimental Results}
\label{sec:results}

We now present the results of our in-depth analysis of bloated dependencies in the \mv ecosystem.

\subsection{\textbf{\RQone}}

In this first research question, we investigate the status of all the dependency relationships of the \nbartifacts \mv artifacts under study.

\begin{figure}[!ht]
	\centering
	\includegraphics[width=0.9\columnwidth]{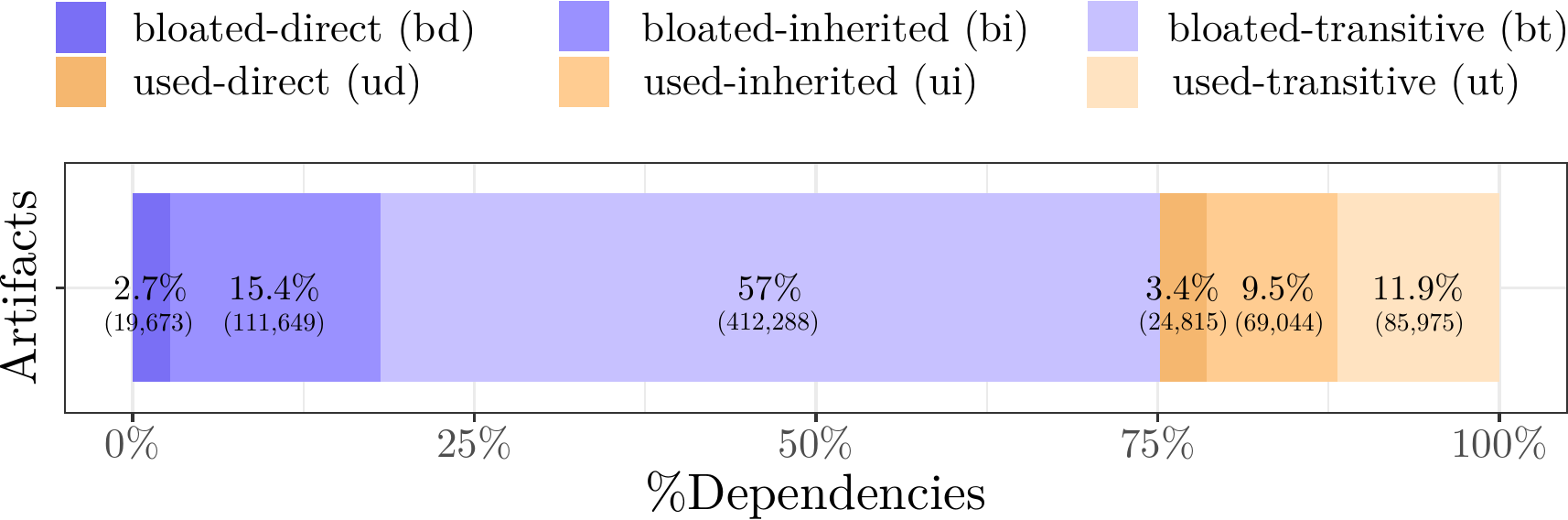}
	\caption{Ratio per usage status of the \nbedges dependency relationships analyzed. Raw counts are inside parenthesis below each percentage.}
	\label{fig:filtered_type_bloat_barplot}
	\vspace{-10pt}
\end{figure}

\autoref{fig:filtered_type_bloat_barplot} shows the overall status of the \nbedges dependency relationships in our dataset. The x-axis represents the percentages, per usage type, of all the dependencies considered in the studied artifacts. The first observation is that the bloat phenomenon is massive: \nbbloat (\percbloat) of all dependencies are bloated, they are not needed to compile and run the code. This bloat is divided into three separate categories: \nbbd (\percbd) are bloated-direct dependency relationships (explicitly declared in the \poms); \nbbi (\percbi) are bloated-inherited dependency relationships from parent module(s); and \nbbt (\percbt) are bloated-transitive dependencies. 

Consequently, \autoref{fig:filtered_type_bloat_barplot} shows that \percbloat of the relationships (edges in the dependency usage tree) are bloated dependencies. Note that this observation does not mean that \nbedges artifacts are unnecessary and can be removed from Maven Central. The same artifact can occur in several DUTs and can be part of a bloated dependency relationship in some DUTs and part of a used relationships in the other DUTs.

In the following, we discuss in more detail and give examples to illustrate the occurrence of the three types of bloated dependency relationships analyzed.

\textbf{Bloated-direct.} We found that $2.7\%$ of the dependencies declared in the \pom file of the studied artifacts are not used at all via bytecode calls. As an example, the Apache Ignite\footnote{\scriptsize{\url{https://github.com/apache/ignite}}} project has deployed an artifact: \texttt{org.apache.ignite:ignite-zookeeper:2.4.0}, which contains only one class in its bytecode: \texttt{TcpDiscoveryZookeeperIpFinder}, and it declares a direct dependency in the \pom towards \texttt{slf4j}, a widely used Java logging library. However, if we analyze the bytecode of \texttt{ignite-zookeeper}, no call to any API member of \texttt{sl4j} exist and therefore it is a bloated-direct dependency. After a manual inspection of the commit history of the  \pom, we found that \texttt{sl4j} was extensively used across all the modules of Apache Ignite at the early stages of the project, but it was later replaced by a dedicated logger, and its declaration remained intact in the \pom.

\textbf{Bloated-inherited.} 
In our dataset, a total of $4,963$ artifacts are part of multi-module Maven projects. Each of these artifacts declares a set of dependencies in its \pom file, and also inherits a set of dependencies from a parent \pom. 
\mytool marks those inherited dependencies are either bloated-inherited or used-inherited. Our dataset includes a total of \nbbi dependency relationships labeled as bloated-inherited, which represents \percbi of all dependencies under study and 61.8\% of the total of inherited dependencies. For example, the artifact \texttt{org.apache.drill:\allowbreak{}drill-protocol:\allowbreak{}1.14.0} inherits dependencies \texttt{commons-codec} and \texttt{commons-io} from its parent \pom \texttt{org.apache.drill:\allowbreak{}drill-root:\allowbreak{}1.14.0}, however, those dependencies are not used in this module, and therefore they are bloated-inherited dependencies. 

\textbf{Bloated-transitive.} In our dataset, bloated-transitive dependencies represent the majority of the bloated dependency relationships: \nbbt (\percbt). This type of bloat is a natural consequence of the \mv dependency resolution mechanism, which automatically resolves all the dependencies whether they are explicitly declared in the \pom file of the project or not. Transitive dependencies are the most common type of dependency relationships, having a direct impact on the growth of the dependency trees. This type of bloat is the most common in the \mv ecosystem. For example, the artifact \texttt{org.eclipse.milo:sdk-client:0.2.1} ships the transitive dependency \texttt{gson} in its MDT, induced from its direct dependency towards \texttt{bsd-parser-core}. However, the part of \texttt{bsd-parser-core} used by \texttt{sdk-client} does not calls any API member of \texttt{gson}, and therefore it is a bloated-transitive dependency. %

In the following, we discuss the dependencies that are actually used. We observe that direct dependencies represent only \percud of the total of dependencies in our dataset. This means that the majority of the dependencies that are necessary to build Maven artifacts are not declared explicitly in the \pom files of these artifacts.

It is interesting to note that \nbut of the dependencies used by the artifacts under study are transitive dependencies. This kind of dependency usage occurs in two different scenarios: (1) the artifact uses API members of some transitive dependencies, without declaring them in its own \pom file; or (2) the transitive dependency is a necessary provider for another artifact in the dependency tree. 

We now discuss an example of the first scenario based on the \texttt{org.apache.streams:streams-filters:0.6.0} artifact from the Apache Streams\footnote{\scriptsize{\url{https://streams.apache.org}}} project. It contains two classes: \texttt{VerbDefinitionDropFilter} and \texttt{VerbDefinitionKeepFilter}. \autoref{lst:used_undeclared} shows part of the source code of the class \texttt{VerbDefinitionDropFilter}, which imports the class \texttt{PreCondition} from library \texttt{guava} (line \ref{line:import}) and uses its static method \texttt{checkArgument} in line \ref{line:usage} of method \texttt{process}. However, if we inspect the \pom of  \texttt{streams-filters}, we  notice that there is no dependency declaration towards \texttt{guava}. It declares a dependency towards \texttt{streams-core}, which in turn depends on the \texttt{streams-utils} artifact that has a direct dependency towards \texttt{guava}. Hence, \texttt{guava} is a used-transitive dependency of \texttt{streams-filters}, called from its source code.

\begin{lstlisting}[basicstyle=\footnotesize\ttfamily, language=java, float, numbers=left, linewidth=0.92\columnwidth, escapeinside={!}{!}, caption={Code snippet of the class \texttt{VerbDefinitionDropFilter} present in the artifact \texttt{org.apache.streams:\allowbreak{}streams-filters:\allowbreak{}0.6.0}. The library \texttt{com.google.guava:\allowbreak{}guava:\allowbreak{}20.0} is included in its classpath via transitive dependency and called from the source code, but no dependency towards \texttt{guava} is declared in its \pom.}, label={lst:used_undeclared}]
package org.apache.streams.filters;
!\label{line:import}!import com.google.common.base.Preconditions;
public class VerbDefinitionDropFilter implements StreamsProcessor {
   ...
   !\label{line:method}!@Override
   public List<StreamsDatum> process(StreamsDatum entry) {
      ...
      !\label{line:usage}!Preconditions.checkArgument(entry.getDocument() instanceof Activity);
      return result;
   }
   ...
}
\end{lstlisting}

Let us now present an example of the second scenario. \autoref{lst:used_undeclared_indirect} shows an excerpt of the class \texttt{AuditTask} included in the artifact \texttt{org.duracloud:\allowbreak{}auditor:\allowbreak{}4.4.3}, from the project DuraCloud\footnote{\scriptsize{\url{https://duraspace.org}}}. In line \ref{line:usage_json}, the method \texttt{getPropsSerializer} instantiates the \texttt{JaxbJsonSerializer} object that belongs to the direct dependency \texttt{org.duracloud:\allowbreak{}common-json:\allowbreak{}4.4.3}. This object in turns creates an \texttt{ObjectMapper} from the transitive dependency \texttt{jackson-mapper-asl}. Hence, \texttt{jackson-mapper-asl} is a necessary, transitive provider for \texttt{org.duracloud:\allowbreak{}auditor:\allowbreak{}4.4.3}.

\begin{lstlisting}[basicstyle=\footnotesize\ttfamily, language=java, float, numbers=left, linewidth=0.92\columnwidth, escapeinside={!}{!}, caption={Code snippet of the class \texttt{AuditTask} present in the artifact \texttt{org.duracloud:\allowbreak{}auditor:\allowbreak{}4.4.3}. The library \texttt{org.codehaus.jackson:\allowbreak{}jackson-mapper-asl:\allowbreak{}1.6.2} is used indirectly through the direct dependency \texttt{org.duracloud:\allowbreak{}common-json:\allowbreak{}4.4.3}.}, label={lst:used_undeclared_indirect}]
package org.duracloud.audit.task;
import org.duracloud.common.json.JaxbJsonSerializer;
public class AuditTask extends TypedTask {
   ...
   private static JaxbJsonSerializer<Map<String, String>> getPropsSerializer() {
        !\label{line:usage_json}!return new JaxbJsonSerializer<>((Class<Map<String, String>>) (Object) new HashMap<String, String>().getClass());
   }
   ...
}
\end{lstlisting}

\begin{figure}
	\centering
	\includegraphics[width=0.9\columnwidth]{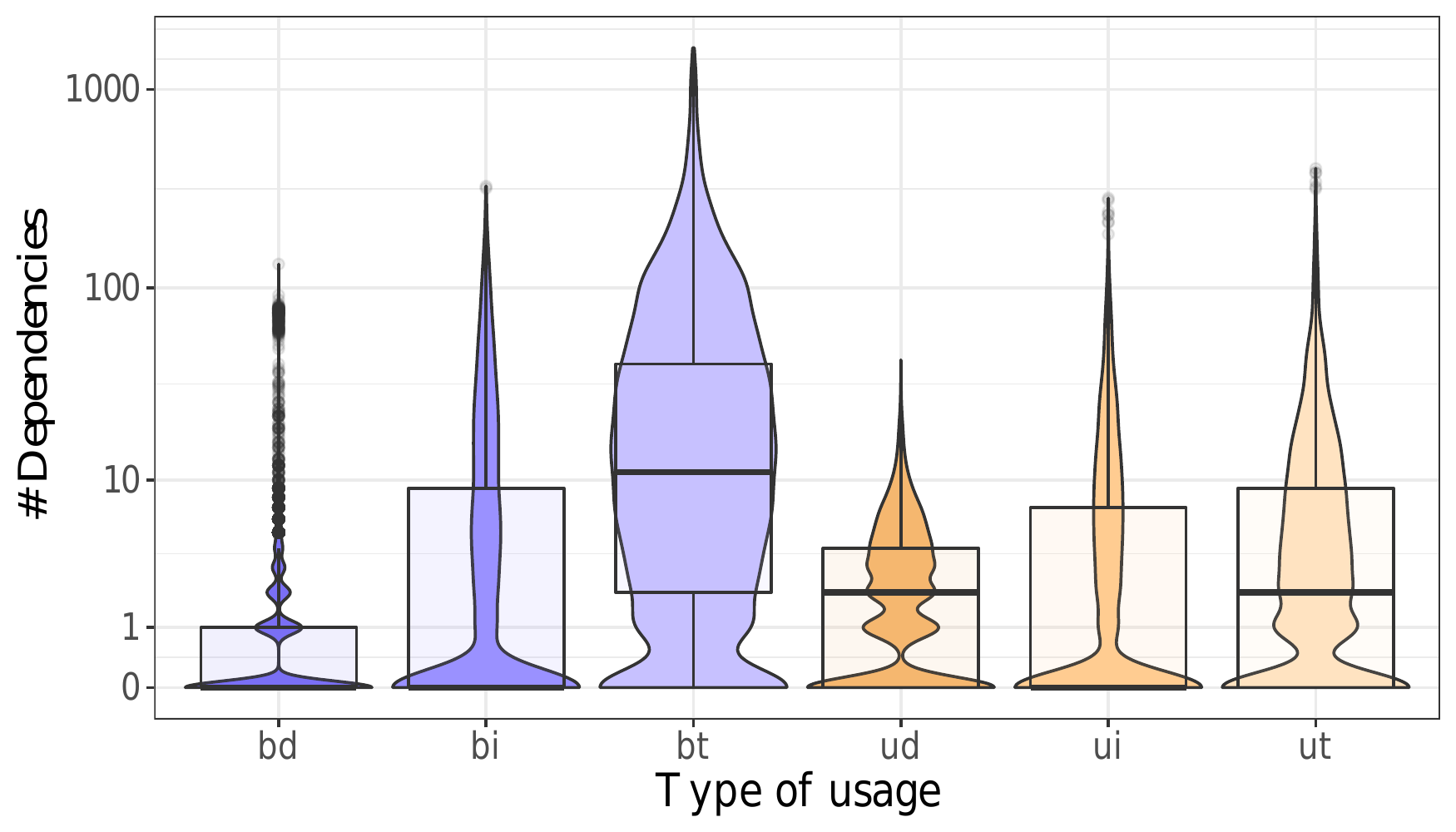}
	\caption{Distributions of the six types of dependency usage relationships for the studied artifacts. The thicker areas on each curve represent concentrations of artifacts per type of usage.}
	\label{fig:filtered_type_bloat_boxplot}
	\vspace{-10pt}
\end{figure}

\autoref{fig:filtered_type_bloat_boxplot} shows the distributions of dependency usage types per artifact. The figure presents superimposed log-scaled box-plots and violin-plots of the number of dependency relationships corresponding to the six usage types studied. Box-plots indicate the standard statistics of the distribution (\ie, lower/upper inter-quartile range, max/min values, and outliers), while violin plots indicate the entire distribution of the data. 

We observe that the distributions of the bloated-direct (bd) and bloated-transitive (bt) dependencies vary greatly.
Bloated-direct dependencies are distributed between $0$ and $1$ (1st-Q and 3rd-Q), with a median of $0$; whereas the second ranges between $2$ and $41$ (1st-Q and 3rd-Q), with a median of $11$. These values are in line with the statistics presented in \autoref{tab:subjects}, since the number of direct and transitive dependencies in general differ approximately by one order of magnitude. 
Overall, from the \nbartifacts \mv artifacts studied, $3,472$ ($36\%$) have at least one bloated-direct dependency, while $8,305$ ($86.2\%$) have at least one  bloated-transitive. 

On the other hand, the inter-quartile range of bloated-direct (bd) dependencies is more compact than the used-direct (ud). In other words, the dependencies declared in the \pom are mostly used. This result is expected, since developers have more control over the edition (adding/removing dependencies) of the \pom file of their artifact.

The median number of used-transitive (ut) dependencies is significantly lower than the median number of bloated-transitive (bt) dependencies ($2$, vs. $11$). This suggests that the the default dependency resolution mechanism of \mv is suboptimal.

The number of outliers in the box-plots differs for each usage type. Notably, the bloated-direct dependencies have more outliers (in total, $25$ artifacts have more than $100$ bloated-direct dependencies). In particular, the artifact \texttt{com.bbossgroups.pdp:pdp-system:5.0.3.9} has the maximum number of bloated-direct dependencies: $133$, out of the $147$ declared in its \pom. The total number of artifacts with at least one bloated-direct dependency in our dataset is $2,298$, which represents $23.8\%$ of the \nbartifacts studied artifacts.\\

\begin{mdframed}[style=mpdframe]
	\textbf{Answer to RQ1:} The analysis of the \nbedges analysed  dependency relationships in our dataset reveals that $543,610$ (\percbloat) of them are bloated. Most of the bloated dependencies are transitive \nbbt (\percbt). Overall, $36\%$ of the artifacts have at least one bloated dependency that is declared in their \pom file. To our knowledge, this is the first scientific observation of this phenomenon. \\
	\noindent\textbf{Implications:} Since developers have more control over direct dependencies, up to $17,673$ (\percbd) of dependencies can be removed directly from the \pom of \mv artifacts, in order to obtain smaller binaries and reduced attack surface. RQ3 will explore the willingness of developers to do so.
\end{mdframed}

\subsection{\textbf{\RQtwo}}

Most of the bloated dependencies in our dataset are either transitive (\percbt) or inherited (\percbi). In this research question, we investigate how the reuse practices that lead to those types of dependencies relate to the type of bloated dependency that emerges in \mv artifacts.

\begin{figure}[!ht]
	\centering
	\includegraphics[width=0.8\columnwidth]{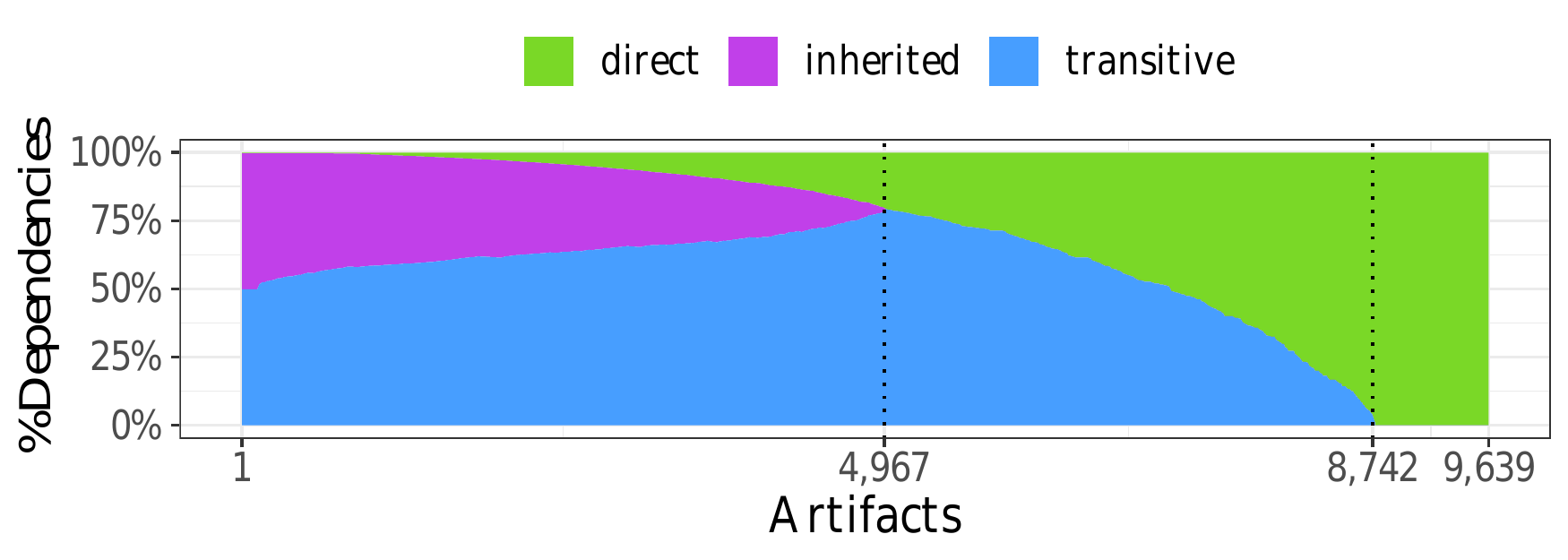}
	\caption{Distribution of the percentages of direct, inherited, and transitive dependencies for the \nbartifacts artifacts considered in this study}
	\label{fig:nb_direct_and_transitive_deps_area_3}
	\vspace{-10pt}
\end{figure}

\autoref{fig:nb_direct_and_transitive_deps_area_3} shows the distributions, in percentages, of the direct, inherited, and transitive dependencies for the \nbartifacts studied artifacts. The  artifacts are sorted, from left to right, in increasing order  according to the ratio of direct dependencies. The y-axis indicates the ratio of each type of dependency for a give artifact. First, we observe that $4,967$ artifacts belong to multi-module projects. Among these artifacts,  the extreme case (far left of the plot) is \texttt{org.janusgraph:\allowbreak{}janusgraph-berkeleyje:\allowbreak{}0.4.0}, which declares $1.4\%$ of its dependencies in its \pom, while the $48.6\%$ of dependencies are inherited from parent \pom files, and $50\%$ are transitive. Second, we observe that the ratio of transitive dependencies is not equally distributed. On the right side of the plot,  $879$ ($9.1\%$) artifacts have no transitive dependency (they have $100\%$ direct dependencies). Meanwhile, $5,561$ ($57.7\%$) artifacts have more than $50\%$ transitive dependencies. The extreme case is  \texttt{org.apereo.cas:\allowbreak{}cas-server-core-api-validation:\allowbreak{}6.1.0}, with $77.6\%$ transitive dependencies. 

\subsubsection{Transitive dependencies}
\label{sec:transitive}

\begin{figure}[!ht]
	\centering
	\includegraphics[width=0.9\columnwidth]{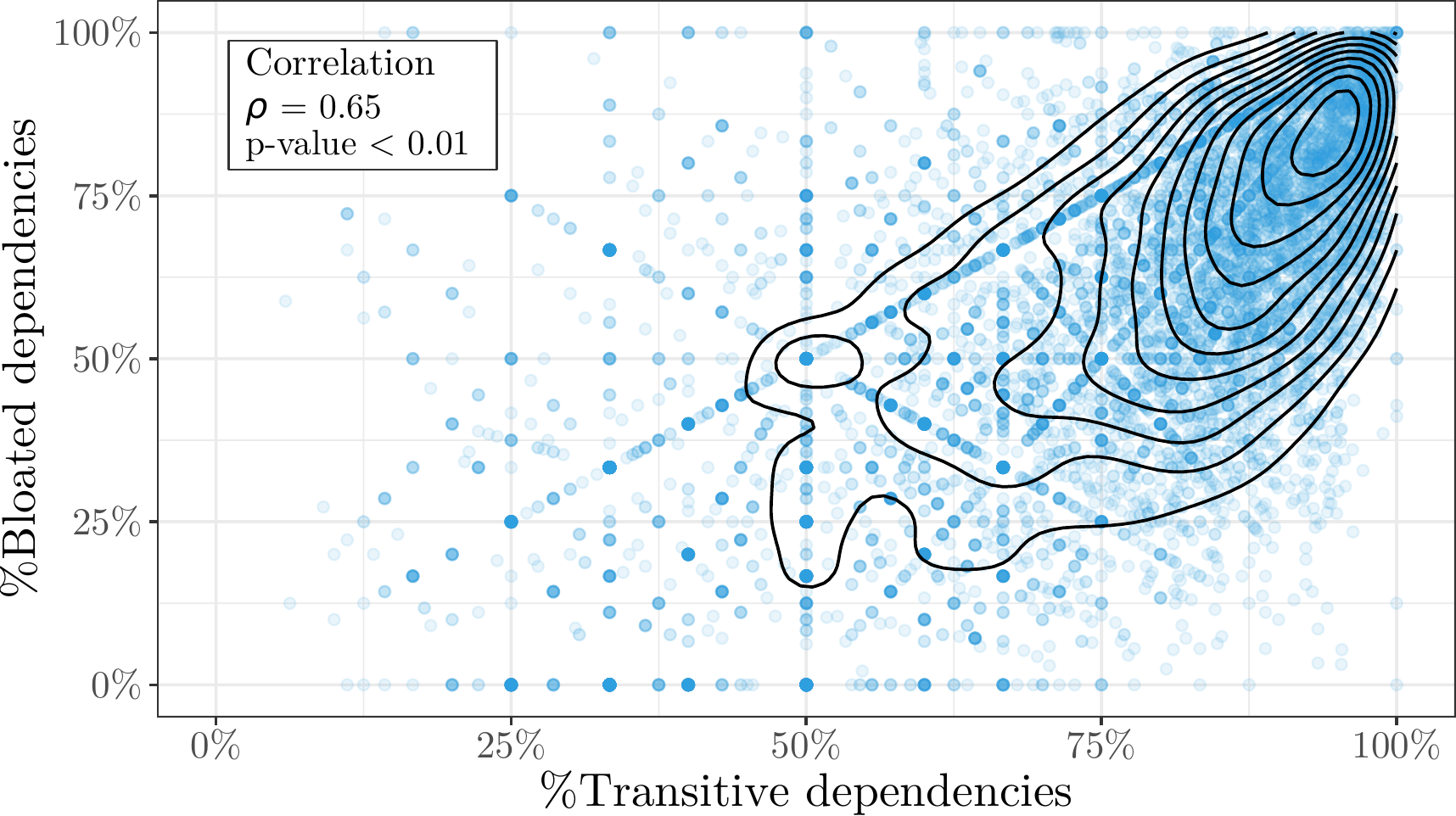}
	\caption{Relation between the percentages of transitive dependencies and the percentage of bloated dependencies in the \nbartifacts studied libraries.}
	\label{fig:transitive_bloated}
	\vspace{-10pt}
\end{figure}

\autoref{fig:transitive_bloated} plots the relation between the ratio of transitive dependencies and the ratio of bloated dependencies. The key insight in this figure is that the larger concentration of artifacts is skewed to the top right corner. This indicates that artifacts with a high percentage of transitive dependencies also tend to exhibit higher percentages of bloated dependencies. Indeed, both variables are positively correlated, according to the Spearman's rank correlation test ($\rho$ = 0.65, p-value < $0.01$).

\begin{figure}[!ht]
	\centering
	\includegraphics[width=0.9\columnwidth]{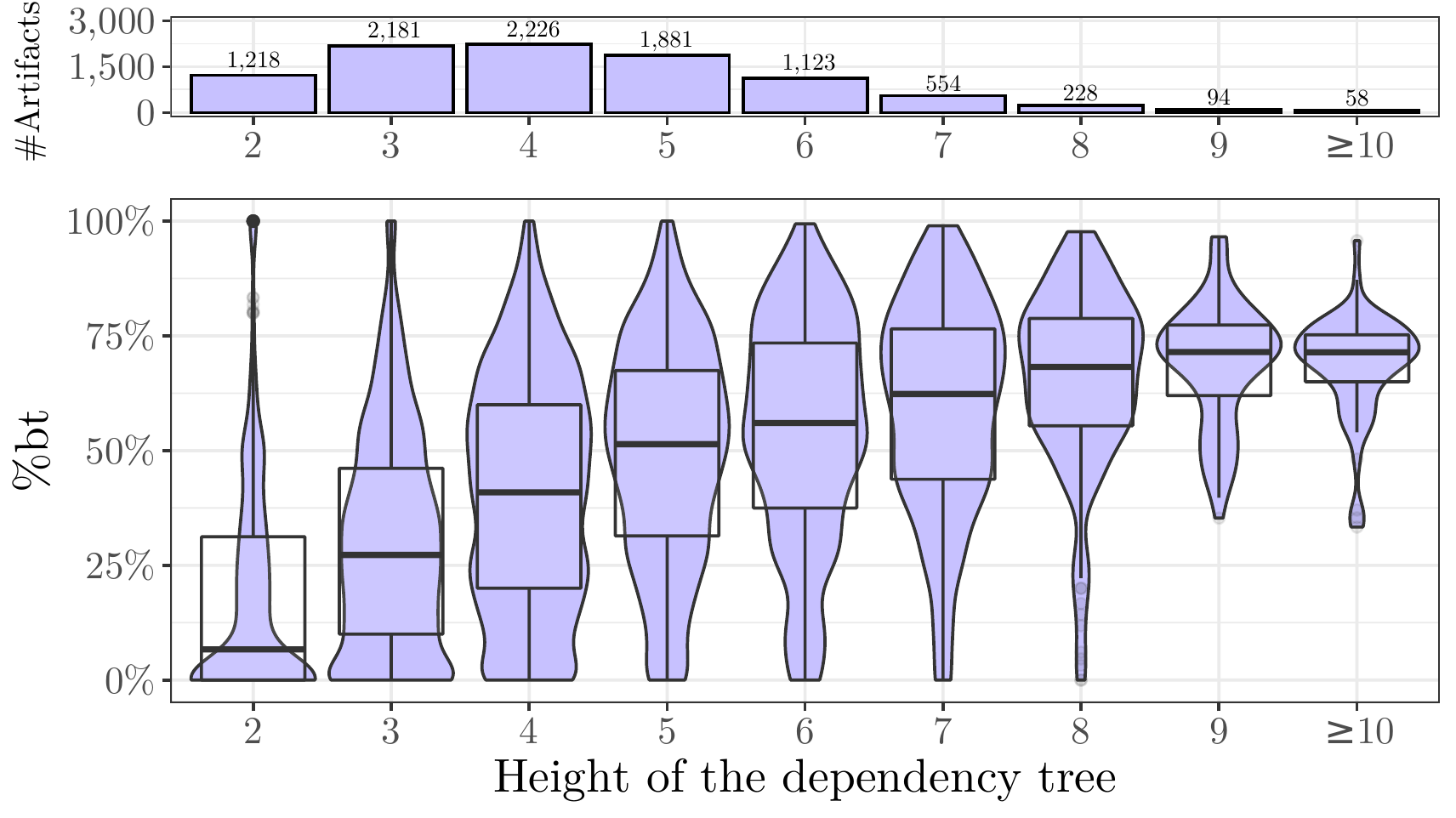}
	\caption{Distribution of the percentages of bloated-transitive dependencies for our study subjects with respect to the height of the dependency trees. Height values greater than $10$ are aggregated. The bar plot at the top represents the number of study subjects for each height.}
	\label{fig:dt_height_comb}
	\vspace{-10pt}
\end{figure}

\autoref{fig:dt_height_comb} shows the distribution of the ratio of transitive bloated dependencies according to the height of the dependency tree. The artifact in our dataset with the largest height is \texttt{top.wboost:\allowbreak{}common-base-spring-boot-starter:\allowbreak{}3.0.RELEASE}, with a height of $14$. 

The bar plot on top of \autoref{fig:dt_height_comb} indicates the number of artifacts that have the same height. We observe that most of the artifacts have a height of $4$: $2,226$ artifacts in total. Considering the number of dependencies, this suggests that the dependency trees tend to be wider than deep. This is because any dependency that already appears at a level closer to the root will be omitted by \mv if it is referred to at a deeper level.  

Looking at the $58$ artifacts with height $\geq 10$, we notice that most of them belong to multi-module projects, and declare other modules in the same project as their direct dependencies. This is a regular practice of multi-module projects, which allows to release each module as an independent artifact. Meanwhile, this drastically increase the complexity of dependency trees. For example, the artifact \texttt{org.wso2.carbon.devicemgt:\allowbreak{}org.wso2.carbon.apimgt.handlers:\allowbreak{}3.0.192} is the extreme case of this practice in our dataset, with a dependency tree of height $11$ and two direct dependencies towards other modules of the same project that in turn depends on other modules of this project. As a result, this artifact has $342$ bloated-transitive and $87$ bloated-inherited dependencies, and is part of a multi-module project with a total of $79$ modules released in \mc.

The plot in \autoref{fig:dt_height_comb} shows a clear increasing trend of bloated-transitive dependencies as the height of the dependency tree increases. For artifacts with a dependency tree of height greater than $9$, at least $28\%$ of their transitive dependencies are bloated, while the median of the percentages of bloated-transitive dependencies for artifacts with height larger than $5$ is more than $50\%$. This finding confirms and complements the results of \autoref{fig:transitive_bloated}, showing that the complexity of the dependency tree is directly related to the occurrence of bloat.

In order to validate the graphical finding, we perform a Spearman's rank correlation test between the number of bloated-transitive dependencies and the size of the dependency tree, \ie, the number of nodes in each tree. We found that there is a significant positive correlation between both variables ($\rho$ =0.67, p-value < $0.01$). 
This confirms that the actual usage of transitive dependencies decreases with the increasing complexity of the dependency tree. This result is aligned with our previous study that suggest that most of the public API members of transitive dependencies are not used by its clients~\citep{Harrand2019}. 

In summary, our results point to the excess of transitive dependencies as one of the fundamental causes of the existence of bloated dependencies in the \mv ecosystem.

\subsubsection{Single-module vs. multi-module}
\label{sec:inherited}

\begin{figure}[!ht]
	\centering
	\includegraphics[width=0.9\columnwidth]{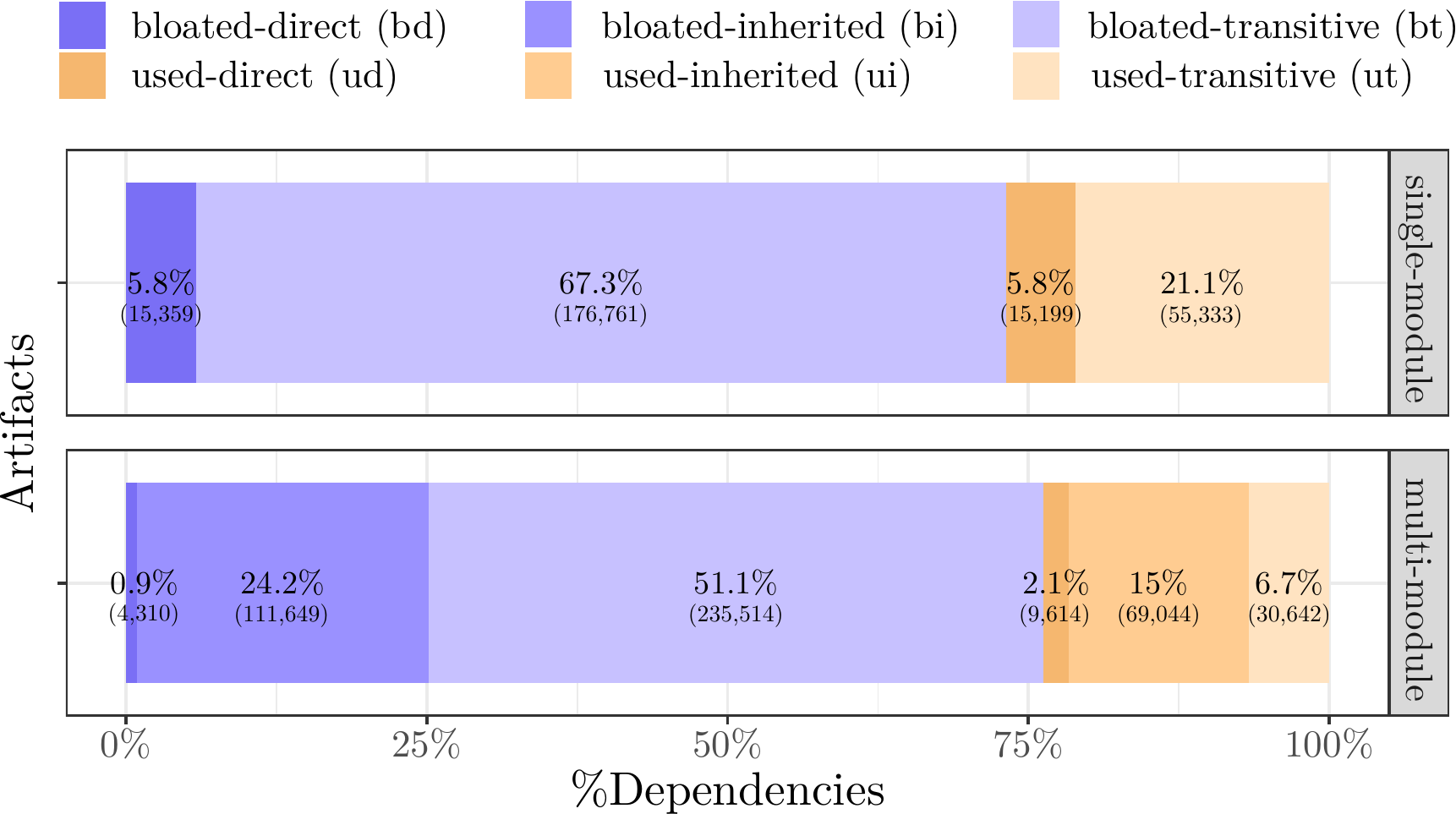}
	\caption{Comparison between multi-module and single-module artifacts according to the percentage status of their dependency relationships. Raw counts are inside parenthesis below each percentage.}
	\label{fig:multimod_barplot}
	\vspace{-10pt}
\end{figure}

Let us investigate on the differences between single and multi-module architectures with respect to the presence of bloated dependencies. \autoref{fig:multimod_barplot} compares the distributions  of bloated and used dependencies between multi-module and single-module artifacts in our dataset. We notice that, in general, multi-module artifacts have slightly more bloat than single-module, precisely $3.1\%$ more (the percentage of bloat in single-module is $5.8\% + 67.3\% = 73.1\%$ vs. $0.9\% + 24.2\% + 51.1\% = 76.2\%$ in multi-module). More interestingly, we observe that a majority of the inherited dependencies are bloated: $24.2\%$ of the dependencies among multi-module project are bloated-inherited (bi), while only $15\%$ are used-inherited (ui). This suggest that most of the dependencies inherited by \mv artifacts that belong to multi-module artifacts are not used by these modules.

We observe that the percentage of bloated-direct dependencies in multi-module artifacts is very small ($0.9\%$) in comparison with single-module ($5.8\%$). Meanwhile, the percentage of bloated-transitive dependencies in single-module ($67.3\%$) is larger than in multi-module ($51.1\%$). This is due to the ``shift''~of a part of  direct and transitive dependencies  into inherited dependencies when using a parent \pom. Indeed, the ``shift'' ~from direct to inherited is the main motivation for having a parent \pom: to have one single declaration of dependencies for many artifacts instead of letting each artifact manage their own dependencies. 

\begin{figure}[!ht]
	\centering
	\includegraphics[width=0.9\columnwidth]{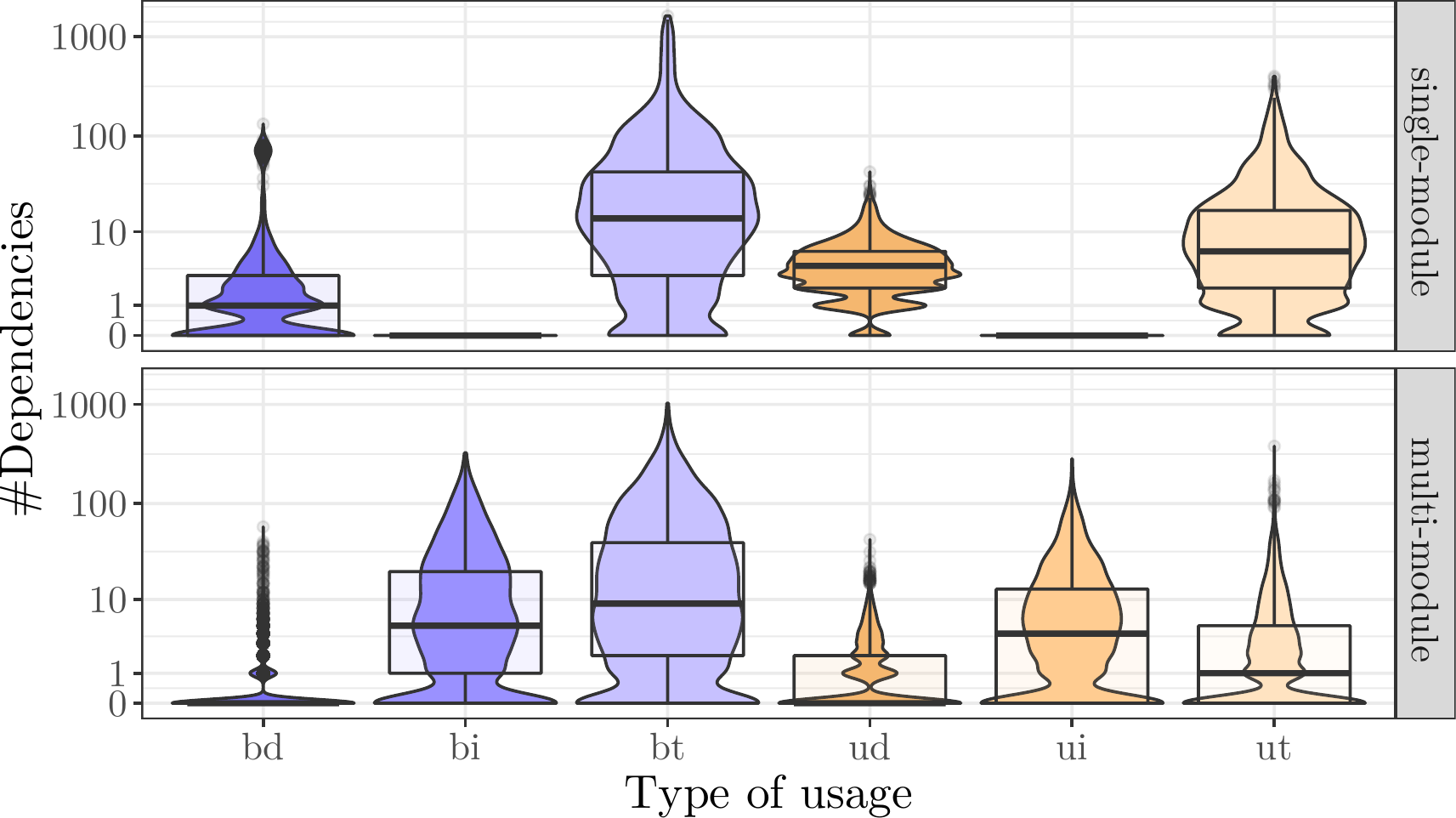}
	\caption{Comparison between multi-module and single-module projects according to their distributions of dependency usage relationships.}
	\label{fig:ndeps_singlemod_multimod}
	\vspace{-10pt}
\end{figure}

This ``shift\textquotedblright ~in the nature of dependencies between single and multi-module artifacts is further emphasized in \autoref{fig:ndeps_singlemod_multimod}. This plot shows superimposed log scaled box-plots and violin-plots comparing the distributions of the number of distinct dependency usage types per artifact, for single-module (top part of the figure) and multi-module (bottom part). 

We observe that multi-module artifacts have less bloated-direct (1st-Q = 0, median = 0, 3rd-Q = 0) and less bloated-transitive (1st-Q = 2, median = 9, 3rd-Q = 40), compared to single-modules, as shown in \autoref{fig:ndeps_singlemod_multimod}. However, multi-module artifacts have a considerably larger number of bloated-inherited dependencies instead (1st-Q = 1, median = 5, 3rd-Q = 20). The extreme case in our dataset is the artifact \texttt{co.cask.cdap:\allowbreak{}cdap-standalone:\allowbreak{}4.3.4}, with $326$ bloated-inherited dependencies in total. 

In summary, the multi-module architecture in \mv projects  contributes to limit redundant dependencies and facilitates the consistent versioning of dependencies in large projects.
However, it introduces two challenges for developers. 
First, it leads to the emergence of bloated-inherited dependencies because of the friction of maintaining a common parent \pom file: it is more difficult to remove dependencies from a parent \pom than from an artifact's own \pom.
Second, it is more difficult for developers to be aware of and understand the dependencies that are inherited from the parent \pom. This calls for better tooling and user interfaces to help developer grasp their inherited dependencies, which is virtually absent in the current \mv tooling ecosystem.\\

\begin{mdframed}[style=mpdframe]
	\textbf{Answer to RQ2:} Reuse practices through to complex dependency trees and multi-module architecture are correlated with the presence of bloated dependencies: the higher a dependency tree, the more bloated-transitive dependencies; bloat is more pervasive in multi-module than single-module artifacts. \\
	\noindent\textbf{Implications:} Developers should carefully consider reusing artifacts with several dependencies because they introduce bloat. They should also contemplate the risks of having bloated dependencies when considering adopting a multi-module architecture.
\end{mdframed}

\subsection{\textbf{\RQthree}} \label{sec:rq3}

In this research question, our goal is to see how developers react when made aware of bloated-direct dependencies in their projects. We do this by proposing the removal of  bloated-direct dependencies to lead developers of mature open-source projects, as described in \autoref{sec:methodology_qualitative}.

\autoref{tab:pr_direct} shows the list of $18$ pull requests submitted. Each pull request proposes the removal of at least one bloated-direct dependency in the \pom. We received response from developers for \totaldirectpr pull request. The first and second columns in the table show the name of the project and the URL to the pull request on GitHub. Columns three and four represent the number of bloated dependencies removed in the \pom and the total number of dependencies removed from the dependency tree with the proposed change, incl. transitive ones. The last column shows the status of the pull request (\textcolor{teal}{\ding{51}}~accepted, \textcolor{teal}{\ding{51}$^\star$}~accepted with changes, \textcolor{red}{\ding{55}} rejected, or \textcolor{blue}{\ding{84}}~pending). The last row represent the acceptance rate calculated with respect to the projects with response, \ie, the total number of dependencies removed divided by the number of proposed removals. For example, for project \texttt{undertow} we propose the removal of $6$ bloated dependencies in its module \texttt{benchmarks}. As a result of this change, $17$ transitive dependencies were removed from the dependency tree the module.

Overall, from the pull requests with responses from developers, $14/15$ were accepted and merged. In total, $68$ dependencies were removed from the dependency trees of the projects. This result demonstrates the relevance of handling bloated-direct dependencies for developers, and the practical usefulness of \mytool.

Let us now summarize the developer feedback. 
First, all developers agreed about the importance of refining the projects' \poms. This is reflected in the positive comments received. Second, their quick responses suggest that it is easy for them to understand the issues associated with the presence of bloated-direct dependencies in their projects. In $8/15$ projects, the response time was less than $24$ hours, which is an evidence that developers consider this type of improvement as a priority. 

Our results also evidence the fact that we, as external contributors to those projects, were able to identify the problem and propose a solution using \mytool. In the following, we discuss four cases of pull requests that are particularly interesting and the feedback provided by developers.

\begin{table}[htb]
	\tiny
	\centering
	\caption{List of pull requests proposing the removal of bloated-direct dependencies created for our experiments.}
	\begin{flushleft}
		\vspace{-10pt}
		
		\begin{threeparttable}
			\begin{tabular}{l|l|cc|c} 
				\toprule
				\multirow{2}{*}{Project} & \multirow{2}{*}{Pull-request URL} & \multicolumn{2}{c|}{\begin{tabular}[c]{@{}c@{}}Removed\\Dependencies \end{tabular}} & \multirow{2}{*}{PR*} \\ 
				\cline{3-4}
				&  & \#D & Total &  \\ 
				\hline
				
				\rowcolor[HTML]{EEEEEE}
				\texttt{jenkins [core, cli]} & \url{https://github.com/jenkinsci/jenkins/pull/4378} & $2$ & $2$ & \textcolor{teal}{\ding{51}$^\star$}\\
				\texttt{mybatis-3 [mybatis]} & \url{https://github.com/mybatis/mybatis-3/pull/1735} & $2$ & $4$  & \textcolor{teal}{\ding{51}}\\
				
				\rowcolor[HTML]{EEEEEE}
				\texttt{flink [core]} & \url{https://github.com/apache/flink/pull/10386} & $1$ &  $1$ & \textcolor{teal}{\ding{51}}\\
				\texttt{checkstyle} & \url{https://github.com/checkstyle/checkstyle/issues/7307} & $1$ &   $4$ & \textcolor{red}{\ding{55}}\\
				
				\rowcolor[HTML]{EEEEEE}
				\texttt{neo4j [collections]} & \url{https://github.com/neo4j/neo4j/pull/12339} & $1$ &  $2$ & \textcolor{teal}{\ding{51}}\\
				\texttt{async-http-client [http-client]} & \url{https://github.com/AsyncHttpClient/async-http-client/pull/1675} & $1$ &  $12$ & \textcolor{teal}{\ding{51}}\\
				
				\rowcolor[HTML]{EEEEEE}
				\texttt{error-prone [core]} & \url{https://github.com/google/error-prone/pull/1409} & $1$ &  $1$ & \textcolor{teal}{\ding{51}}\\
				\texttt{alluxio [core-transport}] & \url{https://github.com/Alluxio/alluxio/pull/10567} & $1$ &  $11$ & \textcolor{teal}{\ding{51}}\\
				
				\rowcolor[HTML]{EEEEEE}
				\texttt{javaparser [symbol-solver-logic]} & \url{https://github.com/javaparser/javaparser/pull/2403} & $2$ &  $9$ & \textcolor{teal}{\ding{51}} \\
				\texttt{undertow [benchmarks]} & \url{https://github.com/undertow-io/undertow/pull/824} & $6$ &  $17$ & \textcolor{teal}{\ding{51}}\\
				
				\rowcolor[HTML]{EEEEEE}
				\texttt{handlebars [markdown]} & \url{https://github.com/jknack/handlebars.java/pull/719} & $1$ & $1$ & \textcolor{blue}{\ding{84}} \\
				\texttt{jooby} & \url{https://github.com/jooby-project/jooby/pull/1412} & $1$ &  $1$ & \textcolor{teal}{\ding{51}}\\
				
				\rowcolor[HTML]{EEEEEE}
				\texttt{couchdb-lucene} & \url{https://github.com/rnewson/couchdb-lucene/pull/279} & $3$ &  $3$ & \textcolor{blue}{\ding{84}}\\
				\texttt{jHiccup} & \url{https://github.com/giltene/jHiccup/pull/42} & $1$ &  $1$ & \textcolor{teal}{\ding{51}}\\
				
				\rowcolor[HTML]{EEEEEE}
				\texttt{subzero [server]} & \url{https://github.com/square/subzero/pull/122} & $1$ &  $4$ & \textcolor{blue}{\ding{84}}\\
				\texttt{vulnerability-tool [shared]} & \url{https://github.com/SAP/vulnerability-assessment-tool/pull/290} & $1$ &  $1$ & \textcolor{teal}{\ding{51}}\\
				
				\rowcolor[HTML]{EEEEEE}
				\texttt{launch4j-maven-plugin} & \url{https://github.com/lukaszlenart/launch4j-maven-plugin/pull/113} & $2$ &  $4$ &  \textcolor{teal}{\ding{51}}\\
				\texttt{jacop} & \url{https://github.com/radsz/jacop/pull/35} & $2$ &  $4$ & \textcolor{teal}{\ding{51}} \\
				
				\hline
				Acceptance rate & \multicolumn{1}{c|}{--} & $23/25$  & $68/73$ & $14/15$ \\
				
				\bottomrule
			\end{tabular}
			
			\begin{tablenotes}[para,flushleft]
				*Status of the pull request: \textcolor{teal}{\ding{51}} Accepted. \textcolor{teal}{\ding{51}$^\star$} Accepted with changes. \textcolor{red}{\ding{55}} Rejected. 
				\textcolor{blue}{\ding{84}} Pending. 
			\end{tablenotes}
			
		\end{threeparttable}
		
		\vspace{-10pt}
		\label{tab:pr_direct}
	\end{flushleft}
	
\end{table}

\subsubsection{jenkins}

\mytool detects that \texttt{jtidy} and \texttt{commons-codec} are bloated-direct dependencies present in the modules \texttt{core} and \texttt{cli} of \texttt{jenkins}. \texttt{jtidy} is a HTML syntax checker and pretty printer. \texttt{commons-codec} is an Apache library that provides an API to encode/decode from various formats such as Base64 and Hexadecimal.

Developers were reluctant to remove \texttt{jtidy} due to their concerns of affecting the users of \texttt{jenkins}, which could be potential consumers of this dependency.
After further inspection, they found that the class \texttt{HTMLParser} of the  \texttt{nis-notification-lamp-plugin}\footnote{\scriptsize{\url{https://github.com/jenkinsci/nis-notification-lamp-plugin}}} project relies on \texttt{jtidy} transitively for performing HTML parsing.

Developers also pointed out the fact that there is no classloader isolation in \texttt{jenkins}, and hence all dependencies in its \texttt{core} module automatically become part of its public API. A developer also referred to issues related to past experiences removing unused dependencies. He argued that external projects have depended on that inclusion and their builds were broken by such a removal. For example, the Git client plugin mistakenly included Java classes from certain Apache contrib authentication classes. When they removed the dependency, some downstream consumers of the library were affected, and they had to include the dependency directly. 

Consequently, we received the following feedback from an experienced developer of \texttt{jenkins}: 

\emph{\textquotedblleft We're not precluded from removing an unused dependency, but I think that the project values compatibility more than removal of unused dependencies, so we need to be careful that removal of an unused dependency does not cause a more severe problem than it solves.\textquotedblright}

After some discussions, developers agreed with the removal of \texttt{commons-codec} in module \texttt{cli}. Our pull request was edited by the developers and merged to the master branch one month after.

\subsubsection{checkstyle}

\mytool identifies the direct dependency  \texttt{junit-jupiter-engine} as bloated. This is a test scope dependency that was added to the \pom of \texttt{checkstyle} when migrating integration tests to JUnit 5. The inclusion of this dependency was necessary due to the deprecation of \texttt{junit-platform-surefire-provider} in the Surefire \mv plugin. However, the report of \mytool about this bloated-direct dependency was a false positive. The reason for this output occurs because \texttt{junit-jupiter-engine} is commonly used through reflective calls that cannot be captured at the bytecode level. 

Although this pull request was rejected, developers expressed interest in \mytool, which is encouraging. They also proposed a list of features for the improvement of our tool. For example, the addition of an exclusion list in the configuration of \mytool for dependencies that are known to be used dynamically, improvements on the readability of the generated report, and the possibility of causing the build process to fail in case of detecting the presence of any bloated dependency. We implemented each of the requested functionalities in \mytool. As a result, developers opened and issue to integrate \mytool in the Continuous Integration (CI) pipeline of \texttt{checkstyle}, in order to automatically manage their bloated dependencies\footnote{\scriptsize{\url{https://github.com/checkstyle/checkstyle/issues/7307}}}.

\subsubsection{alluxio}

\begin{figure}[!ht]
	\centering
	\frame{\includegraphics[width=1\columnwidth]{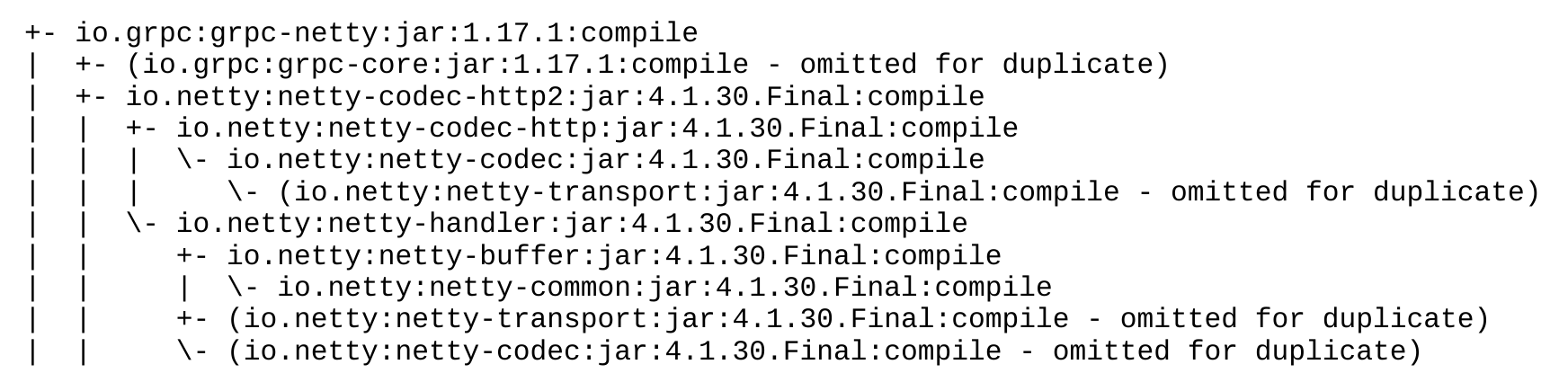}}
	\caption{Transitive dependencies induced by the bloated-direct dependency \texttt{grpc-netty} in the dependency tree of module \texttt{alluxio-core-transport}. The tree is obtained with the \texttt{dependency:tree} \mv goal.}
	\label{fig:dt_alluxio-2}
	\vspace{-10pt}
\end{figure}

\mytool detects that the direct dependency \texttt{grpc-netty}, declared in the module \texttt{alluxio-core-transport} is bloated. \autoref{fig:dt_alluxio-2} shows that this dependency also induces a total of $10$ transitive dependencies that are not used ($4$ of them are omitted by \mv due to their duplication in the dependency tree). Developers accepted our pull request and also manifested their interest on using \mytool for managing unused dependencies in the future.

\subsubsection{undertow}

\mytool detects a total of 6 bloated-direct dependencies in the \texttt{benchmarks} module of the project \texttt{undertow}: \texttt{undertow-\allowbreak{}servlet}, \texttt{undertow-\allowbreak{}websockets-\allowbreak{}jsr}, \texttt{jboss-\allowbreak{}logging-\allowbreak{}processor}, \texttt{xnio-\allowbreak{}nio}, \texttt{jmh-\allowbreak{}generator-\allowbreak{}annprocess}, and \texttt{httpmime}. In this case, we received a rapid positive response from the developers two days after the submission of the pull request. Removing the suggested bloated-direct dependencies has a significant impact on the size of the packaged JAR artifact of the \texttt{undertow-benchmarks} module. We compare the sizes of the bundled JAR before and after the removal of those dependencies: the binary size reduction represents more than 1MB. It is worth mentioning that this change also reduced the complexity of the  dependency tree of the module.\\

\begin{mdframed}[style=mpdframe]
	\textbf{Summary of RQ3:} We use \mytool to propose $18$ pull requests removing bloated-direct dependencies, from which 15/18 were answered. $14/15$ pull requests with response were accepted and merged by open-source developers ($68$ dependencies were removed from the dependency tree of $14$ projects).\\
	\noindent\textbf{Implications:} Removing bloated-direct dependencies is relevant for developers and it is perceived as a valuable contribution. This type of change in the \pom files are small, and they can have a significant impact on the dependency tree of \mv projects.
\end{mdframed}

\subsection{\textbf{\RQfour}}

In this research question, our goal is to see how developers react when made aware of bloated-transitive dependencies. We do this by proposing the exclusion of bloated-transitive dependencies to them, as described in \autoref{sec:methodology_qualitative}.

\autoref{tab:pr_transitive} shows the list of $13$ pull requests submitted. Each pull request proposes the exclusion of at least one transitive dependency in the \pom. We received response from developers for \totaltranstpr pull requests.  The first and second columns show the name of the project and the URL to the pull request on GitHub. Columns three and four represent the number of bloated-transitive dependencies explicitly excluded and the total number of dependencies removed in the dependency tree as resulting from the exclusion. The last column shows the status of the pull request (\textcolor{teal}{\ding{51}}~accepted, \textcolor{red}{\ding{55}} rejected, or \textcolor{blue}{\ding{84}}~pending). The last row represents the acceptance rate with respect to the projects with response. For example, for the project \texttt{spoon} we propose the exclusion of four-bloated transitive dependencies in its \texttt{core} module. As a result of this change, $31$ transitive dependencies were removed from the dependency tree of this module.

Overall, from the pull requests with responses from developers, \totalacctranstpr were accepted and $4$ were rejected. In total, $63$ bloated dependencies were removed from the dependency trees of $4$ projects. We notice that the accepted pull requests involve those projects for which the exclusion of transitive dependencies also represents the removal of a large number of other dependencies from the dependency tree. This result suggests that developers are more careful concerning this type of contributions. 

As in RQ3, we obtained valuable feedback from developers about the pros and cons of excluding bloated-transitive dependencies. In the following, we provide unique qualitative insights about the most interesting cases and explain the feedback obtained from developers to the research community.

\begin{table}[htb]
	\tiny
	\caption{List of pull requests proposing the exclusion of bloated-transitive dependencies.}
	\begin{flushleft}
		\vspace{-10pt}
		
		\begin{center}
			\begin{threeparttable}
				
				\begin{tabular}{l|l|cc|c} 
					\toprule
					\multirow{2}{*}{Project} & \multirow{2}{*}{Pull-request URL} & \multicolumn{2}{c|}{\begin{tabular}[c]{@{}c@{}}Excluded\\Dependencies \end{tabular}} & \multirow{2}{*}{PR*} \\ 
					\cline{3-4}
					&  & \#T & Total  &  \\ 
					\hline
					
					\rowcolor[HTML]{EEEEEE}
					\texttt{jenkins [core]} & \url{https://github.com/jenkinsci/jenkins/pull/4378} & $2$ & $2$ &  \textcolor{red}{\ding{55}} \\
					\texttt{auto [common]} & \url{https://github.com/google/auto/pull/789} & $2$ & $2$ &  \textcolor{red}{\ding{55}} \\
					
					\rowcolor[HTML]{EEEEEE}
					\texttt{moshi [moshi-kotlin]} & \url{https://github.com/square/moshi/pull/1034} & $3$ & $3$ & \textcolor{red}{\ding{55}} \\
					\texttt{spoon [core]} & \url{https://github.com/INRIA/spoon/pull/3167} & $4$ & $31$ &  \textcolor{teal}{\ding{51}} \\
					
					\rowcolor[HTML]{EEEEEE}
					\texttt{moshi [moshi-kotlin]} & \url{https://github.com/square/moshi/pull/1034} & $3$ & $3$ & \textcolor{red}{\ding{55}} \\
					\texttt{wc-capture [driver-openimaj]} & \url{https://github.com/sarxos/webcam-capture/pull/750} & $1$ & $1$ & \textcolor{blue}{\ding{84}} \\
					
					\rowcolor[HTML]{EEEEEE}
					\texttt{teavm [core]} & \url{https://github.com/konsoletyper/teavm/pull/439} & $1$ & $2$ & \textcolor{blue}{\ding{84}} \\
					\texttt{tika [parsers]} & \url{https://github.com/apache/tika/pull/299} & $1$ & $2$ & \textcolor{blue}{\ding{84}} \\
					
					\rowcolor[HTML]{EEEEEE}
					\texttt{orika [eclipse-tools]} & \url{https://github.com/orika-mapper/orika/pull/328} & $1$ & $2$ & \textcolor{blue}{\ding{84}} \\
					\texttt{accumulo [core]} & \url{https://github.com/apache/accumulo/pull/1421} & $3$ & $3$ &   \textcolor{teal}{\ding{51}} \\
					
					\rowcolor[HTML]{EEEEEE}
					\texttt{para [core]} & \url{https://github.com/Erudika/para/pull/69} & $1$ & $20$ &  \textcolor{teal}{\ding{51}} \\
					\texttt{\texttt{selenese-runner-java}} & \url{https://github.com/vmi/selenese-runner-java/pull/313} & $2$ & $9$ &  \textcolor{teal}{\ding{51}} \\
					
					\rowcolor[HTML]{EEEEEE}
					\texttt{commons-configuration} & \url{https://github.com/apache/commons-configuration/pull/40} & $2$ & $9$ & \textcolor{blue}{\ding{84}} \\
					
					\hline
					Acceptance rate & \multicolumn{1}{c|}{--}   & $10/17$ & $63/70$ & $4/8$ \\
					
					\bottomrule
				\end{tabular}
				
				\begin{tablenotes}[para,flushleft]
					*Status of the pull request: \textcolor{teal}{\ding{51}} Accepted. \textcolor{red}{\ding{55}} Rejected. 
					\textcolor{blue}{\ding{84}} Pending. 
				\end{tablenotes}
				
			\end{threeparttable}
		\end{center}
		
		\vspace{-10pt}
		\label{tab:pr_transitive}
	\end{flushleft}
\end{table}

\subsubsection{jenkins}

\mytool detects the bloated-transitive dependencies \texttt{constant-\allowbreak{}pool-\allowbreak{}scanner} and \texttt{eddsa} in the  module \texttt{core} of \texttt{jenkins}. These bloated dependencies were induced through the direct dependencies \texttt{remoting} and \texttt{cli}, respectively. In the message of the pull request, we explain how their exclusion contributes to make the \texttt{core} of \texttt{jenkins} slimmer and the dependency tree clearer.

Although both dependencies were confirmed as unused in the \texttt{core} module of \texttt{jenkins}, developers rejected our pull request. They argue that excluding such dependencies has no valuable repercussion for the project and might actually affect its clients, which is correct. For example, \texttt{constant-\allowbreak{}pool-\allowbreak{}scanner} is used by external components, \eg, the class \texttt{RemoteClassLoader} in the \texttt{remoting}\footnote{\scriptsize{\url{https://github.com/jenkinsci/remoting}}} project relies on this library to inspect the bytecode of remote dependencies. 

As shown in the following quote from an experienced developer of Jenkins, there is a general consensus on the usefulness of removing bloated dependencies, but developers need strong facts to support the removal of transitive dependencies:

\emph{\textquotedblleft Dependency removals and exclusions are really useful, and my recommendation would be to avoid them if there is no substantial gain.\textquotedblright}

\subsubsection{auto}

\mytool reports on the bloated-transitive dependencies \texttt{listenablefuture} and \texttt{auto-value-annotations} in module \texttt{auto-common} of the Google \texttt{auto} project. We proposed the exclusion of these dependencies and submitted a pull request with the \pom change.

Developers express several concerns related to the exclusion of these dependencies. For example, a developer believes that it is not worth  maintaining exclusion lists for dependencies that cause no problem. They point out that although \texttt{listenableFuture} is a single class file dependency, its presence in the dependency tree is vital to the  project, since it overrides the version of the \texttt{guava} library that have many classes. Therefore, the inclusion of this dependency is a strategy followed by \texttt{guava} to narrow the access to the interface \texttt{ListenableFuture} and not to the whole library\footnote{\scriptsize{\url{https://groups.google.com/forum/\#!topic/guava-announce/Km82fZG68Sw/discussion}}}. 

On the other hand, developers agree that \texttt{auto-value-annotations} is bloated. However, they keep it, arguing that it is a test-only dependency, and they prefer to keep annotation-only dependencies and let end users exclude them when desired.

The response from developers suggests that bloated dependencies with test scope are perceived as less harmful. This is reasonable since test dependencies are only available during the test, compilation, and execution phases and are not shipped transitively in the JAR of the artifact. However, we believe that although it is a developers' decision whether they keep this type of bloated dependency or not, the removal of testing dependencies is regularly a desirable refactoring improvement.   

\subsubsection{moshi}

\mytool detects that the bloated-transitive dependency \texttt{kotlin-\allowbreak{}stdlib-\allowbreak{}common} is present in the dependency tree of modules \texttt{moshi-\allowbreak{}kotlin}, \texttt{moshi-\allowbreak{}kotlin-\allowbreak{}codegen}, and \texttt{moshi-\allowbreak{}kotlin-\allowbreak{}tests} of project \texttt{moshi}. This dependency is induced from a common dependency of these modules: \texttt{kotlin-\allowbreak{}stdlib}. 

Developers rejected our pull requests, arguing that excluding such transitive dependency prevents the artifacts from participating in the proper dependency resolution of their clients. They suggest that clients interested in reducing the size of their projects can use specialized shrinking tools, such as ProGuard\footnote{\scriptsize{\url{https://www.guardsquare.com/en/products/proguard}}}, for this purpose. 

Although the argument of developers is valid, we believe that delegating the task of bloat removal to their library clients imposes an unnecessary burden on them. On the other hand, recent studies reveal that library clients do not widely adopt the usage of dependency analysis tools for quality analysis purposes~\cite{Nguyen2020}.  

\subsubsection{spoon}

\mytool detects that the transitive dependencies \texttt{org.\allowbreak{}eclipse.\allowbreak{}core.\allowbreak{}resources}, \texttt{org.\allowbreak{}eclipse.\allowbreak{}core.\allowbreak{}runtime}, \texttt{org.\allowbreak{}eclipse.\allowbreak{}core.\allowbreak{}filesystem}, and \texttt{org.\allowbreak{}eclipse.\allowbreak{}text} \texttt{org.\allowbreak{}eclipse.\allowbreak{}jdt.\allowbreak{}core} are bloated. All of these transitive dependencies were induced by the inclusion of the direct dependency \texttt{org.eclipse.jdt.core}, declared in the \pom of \texttt{core} module of the \texttt{spoon} library.

\autoref{tab:spoon} shows how the exclusion of these bloated-transitive dependencies has a positive impact on the size and the number of classes of the library. As we can see, by excluding these dependencies the size of the \texttt{jar-with-dependencies} of the \texttt{core} module of \texttt{spoon} is trimmed from $16.2$MB to $12.7$MB, which represents a significant reduction in size of $27.6\%$. After considering this improvements, the developers confirmed the relevance of this change and merged our pull request into the master branch of the project.

\begin{table}[htb]
	\scriptsize
	\caption{Comparison of the size and number of classes in the bundled JAR of the \texttt{core} module of \texttt{spoon}, before and after the exclusion of bloated-transitive dependencies.}
	\vspace{-10pt}
	\begin{tabular}{l|c|c} 
		& \textbf{JAR Size(MB)} & \textbf{\#Classes} \\
		\hline
		\textbf{Before} & $16.2$ & $7,425$ \\
		\hline
		\textbf{After} & $12.7$ & $5,593$ \\
		\hline
		\textbf{Reduction(\%)} & $27.6\%$ & $24.7\%$ \\
		\hline
	\end{tabular}
	\vspace{-10pt}
	\label{tab:spoon}
\end{table}

\subsubsection{accumulo}

\mytool detects the bloated-transitive dependencies \texttt{listenablefuture}, \texttt{httpcore} and \texttt{netty} in the  \texttt{core} module of Apache \texttt{accumulo}. These dependencies were confirmed as bloated by the developers. However, they manifested their concerns regarding their exclusion, as expressed in the following comment:

\emph{\textquotedblleft I'm not sure I want us to take on the task of maintaining an exclusion set of transitive dependencies from all our deps POMs, because those can change over time, and we can't always know which transitive dependencies are needed by our dependencies.\textquotedblright}.

After the discussion, developers decided to accept and merge the pull request. Overall, developers considered that the proposal is a good idea. They suggest that it would be better to approach the communities of each of the direct dependencies that they use, and encourage them to mark those dependencies as \emph{optional}, thus they would not be automatically inherited by their users.

\subsubsection{para}

\mytool detects the bloated-transitive dependency \texttt{flexmark-\allowbreak{}jira-\allowbreak{}converter}. This dependency is induced through the direct dependency \texttt{flexmark-\allowbreak{}ext-\allowbreak{}emoji}, declared in the \texttt{core} module of the \texttt{para} project. Our further investigation on the \mv dependency tree of this module revealed that this bloated dependency adds a total of $19$ additional dependencies to the dependency tree of the project, of which $15$ are detected as duplicated by \mv.

Because of this large number (19) of bloated-transitive dependencies removed, developers accepted the pull request and merged the change into the master branch of the project the same day of the pull request submission.\\

\begin{mdframed}[style=mpdframe]
	\textbf{Answer to RQ4:} We use \mytool to propose $13$
	pull requests excluding bloated-transitive dependencies. \totalacctranstpr out of $8$ pull requests with response were accepted and merged by developers ($63$ dependencies were removed from the dependency tree of \totalacctranstpr projects). \\
	\noindent\textbf{Implications:} The handling of bloated-transitive dependencies is a topic with no clear consensus among developers. Developers consider this kind of bloat as relevant, but they are concerned about the maintenance of a list of exclusion directives. Some developers agree to remove them based on practical facts (e.g., JAR size reduction) while other developers prefer to keep bloated-transitive dependencies in order to avoid the potential negative impact on their clients.
\end{mdframed}

\section{Discussion}
\label{sec:discussion}

In this section, we discuss the implications of our findings and the threats to the validity of the results obtained. 

\subsection{Implications of Results}

Our study of bloat in \pom files complements previous studies about the challenges and risks of dependency management for security. In this context, bloated dependencies represent an unnecessary additional source of  vulnerabilities, opening the door to potential attackers which may exploit code segments that could not have been reached otherwise~\cite{Mitropoulos2014, Gkortzis2019}. The  complexity of dependency management systems is  a critical engineering concern since dependency issues can  scale through entire ecosystems. For example, in 2027, the Equifax data  breach\footnote{\scriptsize{\url{https://www.equifaxsecurity2017.com}}}  caused by a  vulnerability in a Java library impacted the personal information of  147 million American citizens. Various initiatives have emerged to cope with this issue. GitHub tracks dependency vulnerability reports from various sources\footnote{\scriptsize{\url{https://help.github.com/en/articles/about-security-alerts-for-vulnerable-dependencies}}} and provides security alerts to affected repositories; dependabot\footnote{\scriptsize{\url{https://dependabot.com/}}} automatically submits pull requests to suggest dependency updates.  \mytool can complement these security scanning solutions in order to determine if a dependency can be removed rather than upgraded. 

Our results indicate that most of the dependency bloat is due to transitive dependencies and the \mv dependency heritage mechanism. The official \mv dependency management guidelines\footnote{\scriptsize{\url{https://maven.apache.org/guides/introduction/introduction-to-dependency-mechanism.html}}} encourage developers to take control over the dependency resolution process via their explicit declaration in the \pom file. This is a good practice to provide better documentation for the project and to keep one artifact's dependencies independent from the choices of other libraries down the dependency tree. Dependencies declared in this way have priority over the \mv mediation mechanism, allowing developers to have a clear knowledge about which library version they are expecting to be used through transitive dependencies. However, since backward compatibility is not always guaranteed, having fixed transitive dependency versions, and therefore non-declared dependencies, still remains as a widely extended practice. In this context, the introduction of the module construct in Java 9 provides a higher level of aggregation above packages. This new language element, if largely adopted, may help to reduce the transitive explosion of dependencies. Indeed, this mechanism forces developers to declare explicitly what other modules are required to use in a given module. This leads to two benefits: (1) it enables reuse declaration at a finer grain than dependencies, and (2) it makes the debloat techniques described in this work safer as it constrains reflection to white-listed modules.

Our results show that even notable open-source projects, which are maintained by development communities with strict development rules, are affected by dependency bloat. Developers confirmed and removed most of the reported bloated-direct dependencies detected by \mytool. However, they are more careful about excluding bloated-transitive dependencies. The addition of exclusion clauses to the \pom files is perceived by some developers as an unnecessary maintainability burden. Interestingly, our quantitative results indicate that bloated-transitive dependency relationships represent the largest portion of bloated dependencies, yet, our qualitative study reveals that these bloated relationships are also the ones that developers find the most challenging to handle and reason about. Overall, this work opens the door to new research opportunities on debloating \poms and other build files.

\subsection{Threats to Validity}

In the following, we discuss about construct, internal and external threats to the validity of our study.

\textbf{Construct validity.} The threats to construct validity are related to the novel concept of bloated dependencies and the metrics utilized for its measurement. For example, the DUT constructed by \mytool could be incomplete due to issues during the resolution of the dependencies. We mitigate this threat by building \mytool on top of \mv plugins to collect the information about the dependency relationships. We also exclude from the study those artifacts for which we were unable to retrieve the full dependency usage information. 

It is possible that developers repackage a library as a bundle JAR file along with its dependencies, or copy the source code of dependencies directly into their source code, in order to avoid dependency related issues. Consequently, \mytool will miss such dependencies, as they are not explicitly declared in the \pom file. Thus, the analysis of dependencies can underestimate the part of bloated dependencies. However, considering the size of our dataset and the feedback obtained from actively maintained projects, we believe that these corner cases do not affect our main results. 

\textbf{Internal validity.} The threats to internal validity are related to the effectiveness of \mytool to detect bloated dependencies. The dynamic features of the Java programming language, \eg, reflection or dynamic class loading present particular challenges for any automatic analysis of Java source code~\citep{Landman2017,Lindholm2014}. Since \mytool statically analyzes bytecode, anything that does not get into the bytecode is not detected (\eg, constants, annotations with source-only retention police, links in Javadocs), which can lead to false positives. To mitigate this threat, \mytool can detect classes or class members that are created or invoked dynamically using basic constructs such as \texttt{class.forName("someClass")} or \texttt{class.getMethod("someMethod", null)}. We also use an exclusion list of dependencies that are known to be used only dynamically.

\textbf{External validity.} The relevance of our findings in  other software ecosystems is one threat to external validity. Our observations about bloated dependencies are based on Java and the \mv ecosystem and our findings are restricted to this scope. More studies on other dependency management systems are needed to figure out whether our findings can be generalized. Another external threat relates to the representativeness of the  projects considered for the qualitative study. To mitigate this threat, we submitted pull requests to a set of diverse, mature, and popular open-source Java projects that belong to distinct communities and cover various application domains. This means that we contributed to improving the dependency management of projects that are arguably among the best of the open-source Java world, which aims to get as strong external validity as possible.

\section{Related Work}
\label{sec:related}

In this work, we propose the first systematic large-scale analysis of bloat in the \mv ecosystem. Here, we discuss the related works in the areas of software debloating and dependency management.

\subsection{Analysis and Mitigation of Software Bloat}

Previous studies have shown that software tends to grow over time, whether or not there is a need for it~\cite{Holzmann2015, Quach2017}. Consequently, software bloat appears as a result of the natural increase of software complexity, \eg, the addition of non-essential features to programs~\citep{Brooks1987}. This phenomenon comes with several risks: it makes software harder to understand and maintain, increases the attack surface, and degrades the overall performance. 
Our paper contributes to the analysis and mitigation of a novel type of software bloat: \emph{bloated dependencies}. 

Celik et al.~\citep{Celik2016} presented \textsc{Molly}, a build system to lazily retrieve dependencies in CI environments and reduce build time. For the studied projects, the build time speed-up reaches $45\%$ on average compared to \mv. \mytool operates differently than \textsc{Molly}: it is not an  alternative to \mv as \textsc{Molly} is, but a static analysis tool that allows \mv users to have a better understanding and control about their dependencies.

Yu et al. \citep{Yu2003} investigated the presence of unnecessary dependencies in header files of large C projects. Their goal was to reduce build time. They proposed a graph-based algorithm to statically remove unused code from applications. Their results show a reduction of build time of $89.70\%$ for incremental builds, and of $26.38\%$ for fresh builds. Our work does not focus on build performance, we analyze the pervasiveness of dependency bloat across a vast and modern ecosystem of \mv packages.

In recent years, there has been a notable interest in the development of debloating techniques for program specialization. The aim is to produce a smaller, specialized version of programs that consume fewer resources while hardening security~\cite{Babak2019}. They range from debloating command line programs written in C~\citep{Sharif2018}, to the specialization of JavaScript frameworks~\cite{Vazquez2019} and
fully fledged containerized platforms~\citep{Rastogi2017}. Most debloating techniques are built upon static analysis and are conservative in the sense that they focus on trimming unreachable code~\cite{Jiang2016}, others are more aggressive and utilize advanced dynamic analysis techniques to remove potentially reachable code~\citep{Heath2019}. Our work addresses the same challenges at a coarser granularity. \mytool removes unused dependencies, which is, according to our empirical results, a significant cause of program bloat.

Qiu et al.~\cite{Qiu2016} empirically  shows evidence that a considerable proportion of API members are not widely used, i.e., many classes, methods, and fields of popular libraries are not used in practice. Nguyen et al.~\cite{Pham2016} implement a bytecode based analysis tool to learn about the actual API usage of Android frameworks. 
L{\"a}mmel et al.~\cite{Lammel2011} perform a large-scale study on API usage based on the migration of AST code segments. Other studies have focused on understanding how developers use APIs on the daily basis~\cite{Roover2013, Bauer2014}. Some of the motivations include improving API design~\cite{Myers2016,Harrand2019} and increasing developers productivity~\cite{Lim1994}. 
All these studies hint at the presence of bloat in APIs. Yet, none of the aforementioned studies investigated the reaction of developers to bloated dependencies. In this paper, we complement the quantitative study of dependency bloat in \mc with practical insights gathered from our contributions to recognized open-source projects.

\subsection{Dependency Management in Software Ecosystems}

Library reuse and dependency management has become mainstream in software development. McIntosh et al.~\citep{McIntosh2012} analyze the evolution of automatic build systems for Java (ANT and \mv). They found that Java build systems follow linear or exponential evolution patterns in terms of size and complexity. In this context, we interpret bloated dependencies as a consequence of the tendency of build automation systems of evolving towards open-ended complexity over time.

Decan et al.~\citep{Decan2019, Decan2017} studied the fragility of packaging ecosystems caused by the increasing number of transitive dependencies. Their findings corroborate our results, showing that most clients have few direct dependencies but a high number of transitive dependencies. They also found that popular libraries tend to have  larger dependency trees. However, their work focuses primarily on the relation between the library users and their direct providers and does not take into account the inherited or transitive dependencies of those providers. We are the first, to the best of our knowledge, to conduct an empirical analysis of bloated dependencies in the Maven ecosystem considering both, users and providers, as potential sources of software bloat.

Bezemer et al.~\citep{Bezemer2017} performed a study of  unspecified dependencies, \ie, dependencies that are not explicitly declared in the build systems. They found that these unspecified dependencies are subtle and difficult to detect in make-based build systems. Seo~et al.~\citep{Seo2014} analyzed over 26 millions builds in Google to investigate the causes, types of errors made, and resolution efforts to fix the failing builds. Their results indicate that, independent of the programming language, dependency errors  are the most common cause of failures, representing more than two thirds of fails for Java. Based on our results, we hypothesize that removing dependency bloat would reduce spurious CI errors related to dependencies. 

Jezek et al.~\cite{Jezek2014} describe, with practical examples, the issues caused by transitive dependencies in \mv. They propose a static analysis approach for finding missing, redundant, incompatible, and conflicting API members in dependencies. Their experiments, based on a dataset of $29$ \mv projects, show that  problems related to transitive dependency are common in practice. They identify the use of wrong dependency scopes as a primary cause of redundancy. Our quantitative study extends this work to the scale of the \mc ecosystem, and provides additional evidence about the persistence of the dependency redundancy problem.

Callo et al.~\citep{CalloArias2011} performed a systematic review about dependency analysis solutions in software-intensive systems. Bavota et al. \citep{Bavota2015} studied performed an empirical study on the evolution of declared dependencies in the Apache community. They found that build system  specifications tend to grow over time unless explicit effort is put into refactoring them. Our qualitative results complement  previous studies that present empirical evidence that developers do not systematically update their dependency configuration files~\cite{McIntoshicse2014, Kula2018}. 

\section{Conclusion}
\label{sec:conclusion}

In this work, we presented a novel conceptual analysis of a phenomenon originated from the  practice of software reuse, which we coined as \emph{bloated dependencies}. This type of dependency relationship between software artifacts is two-side intriguing: from the perspective of the dependency management systems that are unable to avoid it, and from the standpoint of developers who declare dependencies but do not use them in their applications. 

We performed a quantitative and qualitative study of bloated dependencies in the \mv ecosystem. To do so, we implemented a tool, \mytool, which analyzes the bytecode of an artifact and all its dependencies that are resolved by \mv. As a result of the analysis, \mytool provides a report of the bloated dependencies, as well as a new version of its \pom file which removes the bloat. We use \mytool to analyze the \nbedges dependency relationships of \nbartifacts artifacts in \mc. Our results reveal that \percbd of them are bloated (\percbd are direct dependencies, \percbi are inherited from parent \poms, and \percut are transitive dependencies). Based on these results, we distilled two possible causes: the cascade of unwanted transitive dependencies induced by direct dependencies, and the dependency heritage mechanism of multi-module \mv projects. 

We complement our quantitative study of bloated dependencies with an in-depth qualitative analysis of \nbaccrejpr mature Java projects. We use \mytool to analyze these projects and submit the results obtained as pull request on GitHub. Our results indicate that developers are willing to remove bloated-direct dependencies: $14$ out of $15$ answered pull requests were accepted and merged by the developers in their code base. On the other hand, we found that developers tend to be skeptical regarding the exclusion of bloated-transitive dependencies: $4$ out of $8$ answered pull requests were accepted. Overall, the feedback from developers reveals that the removal of bloated dependencies clearly worth the additional analysis and effort.

Our study stresses the need to engineer, i.e., analyze, maintain, test \pom files. The feedback from developers shows interest in \mytool to address this challenge. While the tool is robust enough to analyze a variety of real-world projects, developers now ask questions related to the methodology for dependency debloating, \eg, when to analyze bloat? (in every build, in every release, after every \pom change), who is responsible for debloat of direct or transitive dependencies? (the lead developers, any external contributor), how to properly managing complex dependency trees to avoid dependency conflicts? These methodological questions are part of the future work to further consolidate \mytool.

\footnotesize
\bibliographystyle{abbrvnat}
\bibliography{references}

\end{document}

%% file: figures/jxls_dt.tex
\tikzset{every picture/.style={line width=0.75pt}} %

\begin{tikzpicture}[x=0.75pt,y=0.75pt,yscale=-0.85,xscale=0.85]

\draw  [dash pattern={on 4.5pt off 4.5pt}]  (44.93,138) -- (584.93,138) ;

\draw [color={rgb, 255:red, 128; green, 128; blue, 128 }  ,draw opacity=1 ][line width=0.75]  [dash pattern={on 0.84pt off 2.51pt}]  (278.73,261.03) -- (241.19,293.89) ;
\draw [shift={(238.93,295.87)}, rotate = 318.81] [fill={rgb, 255:red, 128; green, 128; blue, 128 }  ,fill opacity=1 ][line width=0.08]  [draw opacity=0] (8.93,-4.29) -- (0,0) -- (8.93,4.29) -- cycle    ;

\draw [color={rgb, 255:red, 128; green, 128; blue, 128 }  ,draw opacity=1 ][line width=0.75]  [dash pattern={on 0.84pt off 2.51pt}]  (215.7,255.4) -- (216.66,293.03) ;
\draw [shift={(216.73,296.03)}, rotate = 268.54] [fill={rgb, 255:red, 128; green, 128; blue, 128 }  ,fill opacity=1 ][line width=0.08]  [draw opacity=0] (8.93,-4.29) -- (0,0) -- (8.93,4.29) -- cycle    ;

\draw [color={rgb, 255:red, 128; green, 128; blue, 128 }  ,draw opacity=1 ] [dash pattern={on 0.84pt off 2.51pt}]  (376.73,178.83) .. controls (420.29,195.66) and (392.31,181.1) .. (460.63,207.97) ;
\draw [shift={(462.73,208.8)}, rotate = 201.52] [fill={rgb, 255:red, 128; green, 128; blue, 128 }  ,fill opacity=1 ][line width=0.08]  [draw opacity=0] (8.93,-4.29) -- (0,0) -- (8.93,4.29) -- cycle    ;

\draw [color={rgb, 255:red, 128; green, 128; blue, 128 }  ,draw opacity=1 ] [dash pattern={on 0.84pt off 2.51pt}]  (375.73,183.83) .. controls (381.23,188.33) and (423.28,205.92) .. (450.8,229.4) .. controls (477.64,252.29) and (454.49,276.78) .. (476.92,300.94) ;
\draw [shift={(478.73,302.8)}, rotate = 224.2] [fill={rgb, 255:red, 128; green, 128; blue, 128 }  ,fill opacity=1 ][line width=0.08]  [draw opacity=0] (8.93,-4.29) -- (0,0) -- (8.93,4.29) -- cycle    ;

\draw [color={rgb, 255:red, 128; green, 128; blue, 128 }  ,draw opacity=1 ][line width=0.75]  [dash pattern={on 0.84pt off 2.51pt}]  (498.6,226.93) -- (501.79,292.87) ;
\draw [shift={(501.93,295.87)}, rotate = 267.23] [fill={rgb, 255:red, 128; green, 128; blue, 128 }  ,fill opacity=1 ][line width=0.08]  [draw opacity=0] (8.93,-4.29) -- (0,0) -- (8.93,4.29) -- cycle    ;

\draw    (404.93,42.87) .. controls (412.93,43.87) and (604.27,32.93) .. (523.33,296.4) ;
\draw [shift={(523.33,296.4)}, rotate = 287.08] [fill={rgb, 255:red, 0; green, 0; blue, 0 }  ][line width=0.08]  [draw opacity=0] (8.93,-4.29) -- (0,0) -- (8.93,4.29) -- cycle    ;

\draw    (380.27,58.93) .. controls (380.27,73.93) and (372.27,106.93) .. (391.27,124.93) .. controls (409.98,142.66) and (485.94,149.72) .. (497.77,199.62) ;
\draw [shift={(498.27,201.93)}, rotate = 259.11] [fill={rgb, 255:red, 0; green, 0; blue, 0 }  ][line width=0.08]  [draw opacity=0] (8.93,-4.29) -- (0,0) -- (8.93,4.29) -- cycle    ;

\draw    (363.73,187.83) .. controls (363.73,213.31) and (397.35,199.61) .. (411.87,235.76) ;
\draw [shift={(412.73,238.03)}, rotate = 250.25] [fill={rgb, 255:red, 0; green, 0; blue, 0 }  ][line width=0.08]  [draw opacity=0] (8.93,-4.29) -- (0,0) -- (8.93,4.29) -- cycle    ;

\draw    (90.73,190.03) -- (92.62,238.04) ;
\draw [shift={(92.73,241.03)}, rotate = 267.75] [fill={rgb, 255:red, 0; green, 0; blue, 0 }  ][line width=0.08]  [draw opacity=0] (8.93,-4.29) -- (0,0) -- (8.93,4.29) -- cycle    ;

\draw    (205.92,177.83) .. controls (157.65,179.8) and (138.51,251.66) .. (173.5,294.47) ;
\draw [shift={(175.13,296.4)}, rotate = 228.82999999999998] [fill={rgb, 255:red, 0; green, 0; blue, 0 }  ][line width=0.08]  [draw opacity=0] (8.93,-4.29) -- (0,0) -- (8.93,4.29) -- cycle    ;

\draw    (337.27,114.93) .. controls (339.35,138.85) and (355.43,120.88) .. (357.59,156.02) ;
\draw [shift={(357.73,158.83)}, rotate = 267.84000000000003] [fill={rgb, 255:red, 0; green, 0; blue, 0 }  ][line width=0.08]  [draw opacity=0] (8.93,-4.29) -- (0,0) -- (8.93,4.29) -- cycle    ;

\draw    (317.13,114.4) .. controls (286.91,134.87) and (253.01,124.49) .. (243.59,157.23) ;
\draw [shift={(242.92,159.83)}, rotate = 283.09000000000003] [fill={rgb, 255:red, 0; green, 0; blue, 0 }  ][line width=0.08]  [draw opacity=0] (8.93,-4.29) -- (0,0) -- (8.93,4.29) -- cycle    ;

\draw    (332.2,184) .. controls (317.84,184.39) and (242.45,212.28) .. (221.52,236.77) ;
\draw [shift={(219.73,239.03)}, rotate = 305.75] [fill={rgb, 255:red, 0; green, 0; blue, 0 }  ][line width=0.08]  [draw opacity=0] (8.93,-4.29) -- (0,0) -- (8.93,4.29) -- cycle    ;

\draw    (348.73,188.83) .. controls (344.79,211.37) and (327.86,208.97) .. (325.02,236.19) ;
\draw [shift={(324.8,238.8)}, rotate = 273.81] [fill={rgb, 255:red, 0; green, 0; blue, 0 }  ][line width=0.08]  [draw opacity=0] (8.93,-4.29) -- (0,0) -- (8.93,4.29) -- cycle    ;

\draw    (327.27,266.93) -- (327.86,292.87) ;
\draw [shift={(327.93,295.87)}, rotate = 268.68] [fill={rgb, 255:red, 0; green, 0; blue, 0 }  ][line width=0.08]  [draw opacity=0] (8.93,-4.29) -- (0,0) -- (8.93,4.29) -- cycle    ;

\draw    (310.93,98.87) .. controls (296.21,99.26) and (190,108.9) .. (95.69,159.17) ;
\draw [shift={(94.27,159.93)}, rotate = 331.72] [fill={rgb, 255:red, 0; green, 0; blue, 0 }  ][line width=0.08]  [draw opacity=0] (8.93,-4.29) -- (0,0) -- (8.93,4.29) -- cycle    ;

\draw    (67.27,173.93) .. controls (57.37,162.05) and (5.32,219.76) .. (45.88,296.08) ;
\draw [shift={(47.13,298.4)}, rotate = 241.04] [fill={rgb, 255:red, 0; green, 0; blue, 0 }  ][line width=0.08]  [draw opacity=0] (8.93,-4.29) -- (0,0) -- (8.93,4.29) -- cycle    ;

\draw  [fill={rgb, 255:red, 213; green, 255; blue, 197 }  ,fill opacity=1 ] (310.73,88.8) .. controls (310.73,85.27) and (313.6,82.4) .. (317.13,82.4) -- (353.33,82.4) .. controls (356.87,82.4) and (359.73,85.27) .. (359.73,88.8) -- (359.73,108) .. controls (359.73,111.53) and (356.87,114.4) .. (353.33,114.4) -- (317.13,114.4) .. controls (313.6,114.4) and (310.73,111.53) .. (310.73,108) -- cycle ;
\draw  [color={rgb, 255:red, 0; green, 0; blue, 0 }  ,draw opacity=1 ][fill={rgb, 255:red, 197; green, 234; blue, 255 }  ,fill opacity=1 ] (332.2,164.8) .. controls (332.2,161.27) and (335.07,158.4) .. (338.6,158.4) -- (369.8,158.4) .. controls (373.33,158.4) and (376.2,161.27) .. (376.2,164.8) -- (376.2,184) .. controls (376.2,187.53) and (373.33,190.4) .. (369.8,190.4) -- (338.6,190.4) .. controls (335.07,190.4) and (332.2,187.53) .. (332.2,184) -- cycle ;
\draw  [color={rgb, 255:red, 0; green, 0; blue, 0 }  ,draw opacity=1 ][fill={rgb, 255:red, 197; green, 234; blue, 255 }  ,fill opacity=1 ] (200.8,167.8) .. controls (200.8,164.27) and (203.67,161.4) .. (207.2,161.4) -- (276.4,161.4) .. controls (279.93,161.4) and (282.8,164.27) .. (282.8,167.8) -- (282.8,187) .. controls (282.8,190.53) and (279.93,193.4) .. (276.4,193.4) -- (207.2,193.4) .. controls (203.67,193.4) and (200.8,190.53) .. (200.8,187) -- cycle ;
\draw  [fill={rgb, 255:red, 255; green, 254; blue, 171 }  ,fill opacity=1 ] (168.73,245.8) .. controls (168.73,242.27) and (171.6,239.4) .. (175.13,239.4) -- (248.8,239.4) .. controls (252.33,239.4) and (255.2,242.27) .. (255.2,245.8) -- (255.2,265) .. controls (255.2,268.53) and (252.33,271.4) .. (248.8,271.4) -- (175.13,271.4) .. controls (171.6,271.4) and (168.73,268.53) .. (168.73,265) -- cycle ;
\draw  [fill={rgb, 255:red, 255; green, 254; blue, 171 }  ,fill opacity=1 ] (264.73,245.8) .. controls (264.73,242.27) and (267.6,239.4) .. (271.13,239.4) -- (362.33,239.4) .. controls (365.87,239.4) and (368.73,242.27) .. (368.73,245.8) -- (368.73,265) .. controls (368.73,268.53) and (365.87,271.4) .. (362.33,271.4) -- (271.13,271.4) .. controls (267.6,271.4) and (264.73,268.53) .. (264.73,265) -- cycle ;
\draw  [fill={rgb, 255:red, 255; green, 254; blue, 171 }  ,fill opacity=1 ] (377.73,244.8) .. controls (377.73,241.27) and (380.6,238.4) .. (384.13,238.4) -- (443.33,238.4) .. controls (446.87,238.4) and (449.73,241.27) .. (449.73,244.8) -- (449.73,264) .. controls (449.73,267.53) and (446.87,270.4) .. (443.33,270.4) -- (384.13,270.4) .. controls (380.6,270.4) and (377.73,267.53) .. (377.73,264) -- cycle ;
\draw  [color={rgb, 255:red, 0; green, 0; blue, 0 }  ,draw opacity=1 ][fill={rgb, 255:red, 250; green, 208; blue, 255 }  ,fill opacity=1 ] (463.73,208.8) .. controls (463.73,205.27) and (466.6,202.4) .. (470.13,202.4) -- (526.33,202.4) .. controls (529.87,202.4) and (532.73,205.27) .. (532.73,208.8) -- (532.73,228) .. controls (532.73,231.53) and (529.87,234.4) .. (526.33,234.4) -- (470.13,234.4) .. controls (466.6,234.4) and (463.73,231.53) .. (463.73,228) -- cycle ;
\draw  [color={rgb, 255:red, 0; green, 0; blue, 0 }  ,draw opacity=1 ][fill={rgb, 255:red, 197; green, 234; blue, 255 }  ,fill opacity=1 ] (60.8,166.8) .. controls (60.8,163.27) and (63.67,160.4) .. (67.2,160.4) -- (110.4,160.4) .. controls (113.93,160.4) and (116.8,163.27) .. (116.8,166.8) -- (116.8,186) .. controls (116.8,189.53) and (113.93,192.4) .. (110.4,192.4) -- (67.2,192.4) .. controls (63.67,192.4) and (60.8,189.53) .. (60.8,186) -- cycle ;
\draw  [fill={rgb, 255:red, 255; green, 254; blue, 171 }  ,fill opacity=1 ] (45.73,248.8) .. controls (45.73,245.27) and (48.6,242.4) .. (52.13,242.4) -- (130.33,242.4) .. controls (133.87,242.4) and (136.73,245.27) .. (136.73,248.8) -- (136.73,268) .. controls (136.73,271.53) and (133.87,274.4) .. (130.33,274.4) -- (52.13,274.4) .. controls (48.6,274.4) and (45.73,271.53) .. (45.73,268) -- cycle ;
\draw  [fill={rgb, 255:red, 255; green, 254; blue, 171 }  ,fill opacity=1 ] (40.73,304.8) .. controls (40.73,301.27) and (43.6,298.4) .. (47.13,298.4) -- (154.33,298.4) .. controls (157.87,298.4) and (160.73,301.27) .. (160.73,304.8) -- (160.73,324) .. controls (160.73,327.53) and (157.87,330.4) .. (154.33,330.4) -- (47.13,330.4) .. controls (43.6,330.4) and (40.73,327.53) .. (40.73,324) -- cycle ;
\draw  [fill={rgb, 255:red, 255; green, 254; blue, 171 }  ,fill opacity=1 ] (168.73,302.8) .. controls (168.73,299.27) and (171.6,296.4) .. (175.13,296.4) -- (262.33,296.4) .. controls (265.87,296.4) and (268.73,299.27) .. (268.73,302.8) -- (268.73,322) .. controls (268.73,325.53) and (265.87,328.4) .. (262.33,328.4) -- (175.13,328.4) .. controls (171.6,328.4) and (168.73,325.53) .. (168.73,322) -- cycle ;
\draw  [fill={rgb, 255:red, 255; green, 254; blue, 171 }  ,fill opacity=1 ] (276.73,302.8) .. controls (276.73,299.27) and (279.6,296.4) .. (283.13,296.4) -- (383.33,296.4) .. controls (386.87,296.4) and (389.73,299.27) .. (389.73,302.8) -- (389.73,322) .. controls (389.73,325.53) and (386.87,328.4) .. (383.33,328.4) -- (283.13,328.4) .. controls (279.6,328.4) and (276.73,325.53) .. (276.73,322) -- cycle ;
\draw  [color={rgb, 255:red, 0; green, 0; blue, 0 }  ,draw opacity=1 ][fill={rgb, 255:red, 250; green, 208; blue, 255 }  ,fill opacity=1 ] (479.73,302.8) .. controls (479.73,299.27) and (482.6,296.4) .. (486.13,296.4) -- (524.33,296.4) .. controls (527.87,296.4) and (530.73,299.27) .. (530.73,302.8) -- (530.73,322) .. controls (530.73,325.53) and (527.87,328.4) .. (524.33,328.4) -- (486.13,328.4) .. controls (482.6,328.4) and (479.73,325.53) .. (479.73,322) -- cycle ;
\draw [color={rgb, 255:red, 0; green, 0; blue, 0 }  ,draw opacity=1 ][line width=0.75]    (26.93,55.87) -- (45.73,56.01) ;
\draw [shift={(48.73,56.03)}, rotate = 180.44] [fill={rgb, 255:red, 0; green, 0; blue, 0 }  ,fill opacity=1 ][line width=0.08]  [draw opacity=0] (8.93,-4.29) -- (0,0) -- (8.93,4.29) -- cycle    ;

\draw  [color={rgb, 255:red, 0; green, 0; blue, 0 }  ,draw opacity=1 ][fill={rgb, 255:red, 197; green, 234; blue, 255 }  ,fill opacity=1 ] (36.8,83.62) .. controls (36.8,82.63) and (37.6,81.83) .. (38.59,81.83) -- (43.95,81.83) .. controls (44.93,81.83) and (45.73,82.63) .. (45.73,83.62) -- (45.73,89.61) .. controls (45.73,90.6) and (44.93,91.4) .. (43.95,91.4) -- (38.59,91.4) .. controls (37.6,91.4) and (36.8,90.6) .. (36.8,89.61) -- cycle ;
\draw  [color={rgb, 255:red, 0; green, 0; blue, 0 }  ,draw opacity=1 ][fill={rgb, 255:red, 255; green, 254; blue, 171 }  ,fill opacity=1 ] (36.8,98.62) .. controls (36.8,97.63) and (37.6,96.83) .. (38.59,96.83) -- (43.95,96.83) .. controls (44.93,96.83) and (45.73,97.63) .. (45.73,98.62) -- (45.73,104.61) .. controls (45.73,105.6) and (44.93,106.4) .. (43.95,106.4) -- (38.59,106.4) .. controls (37.6,106.4) and (36.8,105.6) .. (36.8,104.61) -- cycle ;
\draw   (212.73,67.8) .. controls (212.73,64.27) and (215.6,61.4) .. (219.13,61.4) -- (255.33,61.4) .. controls (258.87,61.4) and (261.73,64.27) .. (261.73,67.8) -- (261.73,87) .. controls (261.73,90.53) and (258.87,93.4) .. (255.33,93.4) -- (219.13,93.4) .. controls (215.6,93.4) and (212.73,90.53) .. (212.73,87) -- cycle ;
\draw   (428.92,95.8) .. controls (428.92,92.27) and (431.78,89.4) .. (435.32,89.4) -- (503.52,89.4) .. controls (507.05,89.4) and (509.92,92.27) .. (509.92,95.8) -- (509.92,115) .. controls (509.92,118.53) and (507.05,121.4) .. (503.52,121.4) -- (435.32,121.4) .. controls (431.78,121.4) and (428.92,118.53) .. (428.92,115) -- cycle ;
\draw   (338.92,33.8) .. controls (338.92,30.27) and (341.78,27.4) .. (345.32,27.4) -- (398.52,27.4) .. controls (402.05,27.4) and (404.92,30.27) .. (404.92,33.8) -- (404.92,53) .. controls (404.92,56.53) and (402.05,59.4) .. (398.52,59.4) -- (345.32,59.4) .. controls (341.78,59.4) and (338.92,56.53) .. (338.92,53) -- cycle ;
\draw    (336.02,53.8) -- (328.31,55.92) .. controls (327.15,57.97) and (325.54,58.41) .. (323.49,57.25) .. controls (321.44,56.08) and (319.84,56.52) .. (318.67,58.57) .. controls (317.5,60.62) and (315.9,61.06) .. (313.85,59.9) .. controls (311.8,58.74) and (310.2,59.18) .. (309.03,61.23) .. controls (307.86,63.28) and (306.26,63.72) .. (304.21,62.55) .. controls (302.16,61.39) and (300.56,61.83) .. (299.39,63.88) .. controls (298.22,65.93) and (296.62,66.37) .. (294.57,65.21) .. controls (292.52,64.04) and (290.91,64.48) .. (289.74,66.53) .. controls (288.57,68.58) and (286.97,69.02) .. (284.92,67.86) .. controls (282.87,66.7) and (281.27,67.14) .. (280.1,69.19) .. controls (278.93,71.24) and (277.33,71.68) .. (275.28,70.51) .. controls (273.23,69.35) and (271.63,69.79) .. (270.46,71.84) .. controls (269.29,73.89) and (267.69,74.33) .. (265.64,73.17) -- (262.92,73.92) -- (262.92,73.92) ;

\draw [shift={(338.92,53)}, rotate = 164.61] [fill={rgb, 255:red, 0; green, 0; blue, 0 }  ][line width=0.08]  [draw opacity=0] (8.93,-4.29) -- (0,0) -- (8.93,4.29) -- cycle    ;
\draw    (370.07,62.28) -- (365.14,68.58) .. controls (365.43,70.92) and (364.4,72.23) .. (362.06,72.52) .. controls (359.72,72.81) and (358.69,74.12) .. (358.98,76.46) .. controls (359.27,78.8) and (358.24,80.11) .. (355.9,80.4) -- (354.33,82.4) -- (354.33,82.4) ;

\draw [shift={(371.92,59.92)}, rotate = 128.03] [fill={rgb, 255:red, 0; green, 0; blue, 0 }  ][line width=0.08]  [draw opacity=0] (8.93,-4.29) -- (0,0) -- (8.93,4.29) -- cycle    ;
\draw    (400.87,61.26) -- (407.13,66.24) .. controls (409.48,65.97) and (410.78,67.01) .. (411.05,69.35) .. controls (411.32,71.69) and (412.63,72.72) .. (414.97,72.45) .. controls (417.31,72.18) and (418.61,73.22) .. (418.88,75.56) .. controls (419.15,77.9) and (420.46,78.94) .. (422.8,78.67) .. controls (425.14,78.4) and (426.45,79.44) .. (426.72,81.78) .. controls (426.99,84.12) and (428.29,85.16) .. (430.63,84.89) .. controls (432.97,84.62) and (434.28,85.66) .. (434.55,88) -- (436.32,89.4) -- (436.32,89.4) ;

\draw [shift={(398.52,59.4)}, rotate = 38.44] [fill={rgb, 255:red, 0; green, 0; blue, 0 }  ][line width=0.08]  [draw opacity=0] (8.93,-4.29) -- (0,0) -- (8.93,4.29) -- cycle    ;
\draw    (26.96,42.42) .. controls (28.64,40.76) and (30.31,40.77) .. (31.96,42.45) .. controls (33.62,44.12) and (35.29,44.13) .. (36.96,42.48) -- (38,42.48) -- (46,42.53) ;
\draw [shift={(49,42.55)}, rotate = 180.35] [fill={rgb, 255:red, 0; green, 0; blue, 0 }  ][line width=0.08]  [draw opacity=0] (8.93,-4.29) -- (0,0) -- (8.93,4.29) -- cycle    ;

\draw  [color={rgb, 255:red, 0; green, 0; blue, 0 }  ,draw opacity=1 ][fill={rgb, 255:red, 250; green, 208; blue, 255 }  ,fill opacity=1 ] (36.8,112.62) .. controls (36.8,111.63) and (37.6,110.83) .. (38.59,110.83) -- (43.95,110.83) .. controls (44.93,110.83) and (45.73,111.63) .. (45.73,112.62) -- (45.73,118.61) .. controls (45.73,119.6) and (44.93,120.4) .. (43.95,120.4) -- (38.59,120.4) .. controls (37.6,120.4) and (36.8,119.6) .. (36.8,118.61) -- cycle ;
\draw [color={rgb, 255:red, 128; green, 128; blue, 128 }  ,draw opacity=1 ][line width=0.75]  [dash pattern={on 0.84pt off 2.51pt}]  (26.93,68.87) -- (45.73,69.01) ;
\draw [shift={(48.73,69.03)}, rotate = 180.44] [fill={rgb, 255:red, 128; green, 128; blue, 128 }  ,fill opacity=1 ][line width=0.08]  [draw opacity=0] (8.93,-4.29) -- (0,0) -- (8.93,4.29) -- cycle    ;

\draw (335,88) node [scale=0.5] [align=left] {org.jxls};
\draw (335.07,97.47) node [scale=0.7] [align=left] {\textbf{jxls-poi}};
\draw (335.07,106.47) node [scale=0.5] [align=left] {1.0.15};
\draw (354,164) node [scale=0.5] [align=left] {org.jxls};
\draw (354.07,173.47) node [scale=0.7] [align=left] {\textbf{jxls}};
\draw (354.07,183.47) node [scale=0.5] [align=left] {2.6.0};
\draw (242,167) node [scale=0.5] [align=left] {org.apache.commons};
\draw (242.07,176.47) node [scale=0.7] [align=left] {\textbf{commons-jexl}};
\draw (242.07,186.47) node [scale=0.5] [align=left] {2.1.1};
\draw (212,244) node [scale=0.5] [align=left] {org.apache.commons};
\draw (212.07,254.47) node [scale=0.7] [align=left] {\textbf{commons-jexl3}};
\draw (212.07,264.47) node [scale=0.5] [align=left] {3.1};
\draw (316,245) node [scale=0.5] [align=left] {commons-beanutils};
\draw (316.07,254.47) node [scale=0.7] [align=left] {\textbf{commons-beanutils}};
\draw (316.07,264.47) node [scale=0.5] [align=left] {1.9.3};
\draw (414,244) node [scale=0.5] [align=left] {ch.qos.logback};
\draw (414.07,253.47) node [scale=0.7] [align=left] {\textbf{logback-core}};
\draw (414.07,262.47) node [scale=0.5] [align=left] {1.2.3};
\draw (497,208) node [scale=0.5] [align=left] {org.slf4j};
\draw (497.07,217.47) node [scale=0.7] [align=left] {\textbf{jcl-over-slf4j}};
\draw (497.07,226.47) node [scale=0.5] [align=left] {1.7.12};
\draw (89,167) node [scale=0.5] [align=left] {org.apache.poi};
\draw (89.07,176.47) node [scale=0.7] [align=left] {\textbf{poi}};
\draw (89.07,186.47) node [scale=0.5] [align=left] {3.17};
\draw (91,247) node [scale=0.5] [align=left] {commons-codec};
\draw (91.07,256.47) node [scale=0.7] [align=left] {\textbf{commons-codec}};
\draw (91.07,266.47) node [scale=0.5] [align=left] {1.10};
\draw (101,303) node [scale=0.5] [align=left] {org.apache.commons};
\draw (101.07,312.47) node [scale=0.7] [align=left] {\textbf{commons-collections4}};
\draw (101.07,321.47) node [scale=0.5] [align=left] {4.1};
\draw (219,302) node [scale=0.5] [align=left] {commons-logging};
\draw (218.7,312.4) node [scale=0.7] [align=left] {\textbf{commons-logging}};
\draw (219.07,321.47) node [scale=0.5] [align=left] {1.1.1};
\draw (333,303) node [scale=0.5] [align=left] {commons-collections};
\draw (332.7,313.4) node [scale=0.7] [align=left] {\textbf{commons-collections}};
\draw (333.07,322.47) node [scale=0.5] [align=left] {3.2.2};
\draw (504,303) node [scale=0.5] [align=left] {org.slf4j};
\draw (504.07,312.47) node [scale=0.7] [align=left] {\textbf{slf4j-api}};
\draw (504.07,322.47) node [scale=0.5] [align=left] {1.7.12};
\draw (97,55.9) node [scale=0.7] [align=left] {{\scriptsize dependency declaration}};
\draw (87,84.9) node [scale=0.7] [align=left] {{\scriptsize direct dependency}};
\draw (95,98.9) node [scale=0.7] [align=left] {{\scriptsize transitive dependency }};
\draw (237,67) node [scale=0.5] [align=left] {org.jxls};
\draw (237.07,76.47) node [scale=0.7] [align=left] {\textbf{jxls}};
\draw (237.07,85.47) node [scale=0.5] [align=left] {2.6.0};
\draw (469,95) node [scale=0.5] [align=left] {org.jxls};
\draw (469.07,104.47) node [scale=0.7] [align=left] {\textbf{jxls-examples}};
\draw (469.07,113.47) node [scale=0.5] [align=left] {2.6.0};
\draw (372,34) node [scale=0.5] [align=left] {org.jxls};
\draw (371.92,43.4) node [scale=0.7] [align=left] {\textbf{jxls-project}};
\draw (372.07,51.47) node [scale=0.5] [align=left] {2.6.0};
\draw (51,23.61) node  [align=left] {\textbf{Legend}};
\draw (101,42.9) node [scale=0.7] [align=left] {{\scriptsize parent module declaration}};
\draw (93,112.9) node [scale=0.7] [align=left] {{\scriptsize inherited dependency}};
\draw (110,68.9) node [scale=0.7] [align=left] {{\scriptsize dependency omitted by Maven}};
\draw (204,277.9) node [color={rgb, 255:red, 128; green, 128; blue, 128 }  ,opacity=1 ] [align=left] {{\tiny 1.2}};
\draw (266,281.9) node [color={rgb, 255:red, 128; green, 128; blue, 128 }  ,opacity=1 ] [align=left] {{\tiny 1.2}};
\draw (451,281.9) node [color={rgb, 255:red, 128; green, 128; blue, 128 }  ,opacity=1 ] [align=left] {{\tiny 1.7.26}};
\draw (428,181.9) node [color={rgb, 255:red, 128; green, 128; blue, 128 }  ,opacity=1 ] [align=left] {{\tiny 1.7.26}};
\draw (484,256.9) node [color={rgb, 255:red, 128; green, 128; blue, 128 }  ,opacity=1 ] [align=left] {{\tiny 1.7.12}};
\draw (574,73.53) node [rotate=-90] [align=left] {Project};
\draw (574,225.53) node [rotate=-90] [align=left] {Dependencies};

\end{tikzpicture}

%% file: figures/jxls_dt_labelled.tex
\tikzset{every picture/.style={line width=0.75pt}} %

\begin{tikzpicture}[x=0.75pt,y=0.75pt,yscale=-0.85,xscale=0.85]

\draw    (344.17,49.55) .. controls (390.04,97.22) and (399.48,187.83) .. (364.73,268.23) ;
\draw [shift={(363.67,270.67)}, rotate = 293.96] [fill={rgb, 255:red, 0; green, 0; blue, 0 }  ][line width=0.08]  [draw opacity=0] (8.93,-4.29) -- (0,0) -- (8.93,4.29) -- cycle    ;

\draw    (323.93,48.87) .. controls (321.93,69.87) and (255.93,85.87) .. (220.93,100.87) .. controls (186.28,115.72) and (145.75,180.55) .. (138.15,267.23) ;
\draw [shift={(137.93,269.87)}, rotate = 274.55] [fill={rgb, 255:red, 0; green, 0; blue, 0 }  ][line width=0.08]  [draw opacity=0] (8.93,-4.29) -- (0,0) -- (8.93,4.29) -- cycle    ;

\draw    (320.93,25.87) .. controls (297.93,12.87) and (360.93,62.87) .. (267.93,70.87) .. controls (177.26,78.67) and (107.5,176.78) .. (101.29,213.19) ;
\draw [shift={(100.93,215.87)}, rotate = 275.04] [fill={rgb, 255:red, 0; green, 0; blue, 0 }  ][line width=0.08]  [draw opacity=0] (8.93,-4.29) -- (0,0) -- (8.93,4.29) -- cycle    ;

\draw    (338.93,45.87) .. controls (305.78,169.69) and (320.16,171.84) .. (319.98,208.94) ;
\draw [shift={(319.93,211.87)}, rotate = 271.43] [fill={rgb, 255:red, 0; green, 0; blue, 0 }  ][line width=0.08]  [draw opacity=0] (8.93,-4.29) -- (0,0) -- (8.93,4.29) -- cycle    ;

\draw    (353.33,50.4) .. controls (374.12,67.23) and (478.81,82.56) .. (492.55,174.07) ;
\draw [shift={(492.93,176.87)}, rotate = 262.72] [fill={rgb, 255:red, 0; green, 0; blue, 0 }  ][line width=0.08]  [draw opacity=0] (8.93,-4.29) -- (0,0) -- (8.93,4.29) -- cycle    ;

\draw    (332.93,49.87) .. controls (283.93,237.05) and (233.97,167.8) .. (221.64,211.07) ;
\draw [shift={(220.93,213.87)}, rotate = 282.65] [fill={rgb, 255:red, 0; green, 0; blue, 0 }  ][line width=0.08]  [draw opacity=0] (8.93,-4.29) -- (0,0) -- (8.93,4.29) -- cycle    ;

\draw  [dash pattern={on 4.5pt off 4.5pt}]  (44.93,112) -- (584.93,112) ;

\draw    (355.67,48.55) .. controls (366.45,46.21) and (602.55,88.44) .. (525.52,267.69) ;
\draw [shift={(524.33,270.4)}, rotate = 293.94] [fill={rgb, 255:red, 0; green, 0; blue, 0 }  ][line width=0.08]  [draw opacity=0] (8.93,-4.29) -- (0,0) -- (8.93,4.29) -- cycle    ;

\draw    (328.17,50.05) .. controls (299.17,98.05) and (239.67,100.67) .. (218.93,112.87) .. controls (198.51,124.88) and (131.71,221.7) .. (173.16,268.31) ;
\draw [shift={(175.13,270.4)}, rotate = 225.17000000000002] [fill={rgb, 255:red, 0; green, 0; blue, 0 }  ][line width=0.08]  [draw opacity=0] (8.93,-4.29) -- (0,0) -- (8.93,4.29) -- cycle    ;

\draw    (341.67,45.55) .. controls (343.75,69.47) and (355.65,92.74) .. (357.6,129.94) ;
\draw [shift={(357.73,132.83)}, rotate = 267.84000000000003] [fill={rgb, 255:red, 0; green, 0; blue, 0 }  ][line width=0.08]  [draw opacity=0] (8.93,-4.29) -- (0,0) -- (8.93,4.29) -- cycle    ;

\draw    (332.67,46.55) .. controls (307.32,110.41) and (255.94,99.57) .. (245.64,132.26) ;
\draw [shift={(244.93,134.87)}, rotate = 283.09000000000003] [fill={rgb, 255:red, 0; green, 0; blue, 0 }  ][line width=0.08]  [draw opacity=0] (8.93,-4.29) -- (0,0) -- (8.93,4.29) -- cycle    ;

\draw    (320.93,25.87) .. controls (300.93,32.87) and (334.93,58.87) .. (281.93,62.87) .. controls (267.23,60.91) and (109.43,89.68) .. (93.72,131.3) ;
\draw [shift={(92.93,133.87)}, rotate = 283.09000000000003] [fill={rgb, 255:red, 0; green, 0; blue, 0 }  ][line width=0.08]  [draw opacity=0] (8.93,-4.29) -- (0,0) -- (8.93,4.29) -- cycle    ;

\draw  [fill={rgb, 255:red, 213; green, 255; blue, 197 }  ,fill opacity=1 ] (310.73,24.8) .. controls (310.73,21.27) and (313.6,18.4) .. (317.13,18.4) -- (353.33,18.4) .. controls (356.87,18.4) and (359.73,21.27) .. (359.73,24.8) -- (359.73,44) .. controls (359.73,47.53) and (356.87,50.4) .. (353.33,50.4) -- (317.13,50.4) .. controls (313.6,50.4) and (310.73,47.53) .. (310.73,44) -- cycle ;
\draw  [color={rgb, 255:red, 0; green, 0; blue, 0 }  ,draw opacity=1 ][fill={rgb, 255:red, 197; green, 234; blue, 255 }  ,fill opacity=1 ] (332.2,138.8) .. controls (332.2,135.27) and (335.07,132.4) .. (338.6,132.4) -- (369.8,132.4) .. controls (373.33,132.4) and (376.2,135.27) .. (376.2,138.8) -- (376.2,158) .. controls (376.2,161.53) and (373.33,164.4) .. (369.8,164.4) -- (338.6,164.4) .. controls (335.07,164.4) and (332.2,161.53) .. (332.2,158) -- cycle ;
\draw  [color={rgb, 255:red, 0; green, 0; blue, 0 }  ,draw opacity=1 ][fill={rgb, 255:red, 197; green, 234; blue, 255 }  ,fill opacity=1 ] (200.8,141.8) .. controls (200.8,138.27) and (203.67,135.4) .. (207.2,135.4) -- (276.4,135.4) .. controls (279.93,135.4) and (282.8,138.27) .. (282.8,141.8) -- (282.8,161) .. controls (282.8,164.53) and (279.93,167.4) .. (276.4,167.4) -- (207.2,167.4) .. controls (203.67,167.4) and (200.8,164.53) .. (200.8,161) -- cycle ;
\draw  [fill={rgb, 255:red, 255; green, 254; blue, 171 }  ,fill opacity=1 ] (168.73,219.8) .. controls (168.73,216.27) and (171.6,213.4) .. (175.13,213.4) -- (248.8,213.4) .. controls (252.33,213.4) and (255.2,216.27) .. (255.2,219.8) -- (255.2,239) .. controls (255.2,242.53) and (252.33,245.4) .. (248.8,245.4) -- (175.13,245.4) .. controls (171.6,245.4) and (168.73,242.53) .. (168.73,239) -- cycle ;
\draw  [fill={rgb, 255:red, 255; green, 254; blue, 171 }  ,fill opacity=1 ] (264.73,219.8) .. controls (264.73,216.27) and (267.6,213.4) .. (271.13,213.4) -- (362.33,213.4) .. controls (365.87,213.4) and (368.73,216.27) .. (368.73,219.8) -- (368.73,239) .. controls (368.73,242.53) and (365.87,245.4) .. (362.33,245.4) -- (271.13,245.4) .. controls (267.6,245.4) and (264.73,242.53) .. (264.73,239) -- cycle ;
\draw  [fill={rgb, 255:red, 255; green, 254; blue, 171 }  ,fill opacity=1 ] (386.73,218.8) .. controls (386.73,215.27) and (389.6,212.4) .. (393.13,212.4) -- (452.33,212.4) .. controls (455.87,212.4) and (458.73,215.27) .. (458.73,218.8) -- (458.73,238) .. controls (458.73,241.53) and (455.87,244.4) .. (452.33,244.4) -- (393.13,244.4) .. controls (389.6,244.4) and (386.73,241.53) .. (386.73,238) -- cycle ;
\draw  [color={rgb, 255:red, 0; green, 0; blue, 0 }  ,draw opacity=1 ][fill={rgb, 255:red, 250; green, 208; blue, 255 }  ,fill opacity=1 ] (463.73,182.8) .. controls (463.73,179.27) and (466.6,176.4) .. (470.13,176.4) -- (526.33,176.4) .. controls (529.87,176.4) and (532.73,179.27) .. (532.73,182.8) -- (532.73,202) .. controls (532.73,205.53) and (529.87,208.4) .. (526.33,208.4) -- (470.13,208.4) .. controls (466.6,208.4) and (463.73,205.53) .. (463.73,202) -- cycle ;
\draw  [color={rgb, 255:red, 0; green, 0; blue, 0 }  ,draw opacity=1 ][fill={rgb, 255:red, 197; green, 234; blue, 255 }  ,fill opacity=1 ] (60.8,140.8) .. controls (60.8,137.27) and (63.67,134.4) .. (67.2,134.4) -- (110.4,134.4) .. controls (113.93,134.4) and (116.8,137.27) .. (116.8,140.8) -- (116.8,160) .. controls (116.8,163.53) and (113.93,166.4) .. (110.4,166.4) -- (67.2,166.4) .. controls (63.67,166.4) and (60.8,163.53) .. (60.8,160) -- cycle ;
\draw  [fill={rgb, 255:red, 255; green, 254; blue, 171 }  ,fill opacity=1 ] (45.73,222.8) .. controls (45.73,219.27) and (48.6,216.4) .. (52.13,216.4) -- (130.33,216.4) .. controls (133.87,216.4) and (136.73,219.27) .. (136.73,222.8) -- (136.73,242) .. controls (136.73,245.53) and (133.87,248.4) .. (130.33,248.4) -- (52.13,248.4) .. controls (48.6,248.4) and (45.73,245.53) .. (45.73,242) -- cycle ;
\draw  [fill={rgb, 255:red, 255; green, 254; blue, 171 }  ,fill opacity=1 ] (40.73,278.8) .. controls (40.73,275.27) and (43.6,272.4) .. (47.13,272.4) -- (154.33,272.4) .. controls (157.87,272.4) and (160.73,275.27) .. (160.73,278.8) -- (160.73,298) .. controls (160.73,301.53) and (157.87,304.4) .. (154.33,304.4) -- (47.13,304.4) .. controls (43.6,304.4) and (40.73,301.53) .. (40.73,298) -- cycle ;
\draw  [fill={rgb, 255:red, 255; green, 254; blue, 171 }  ,fill opacity=1 ] (168.73,276.8) .. controls (168.73,273.27) and (171.6,270.4) .. (175.13,270.4) -- (262.33,270.4) .. controls (265.87,270.4) and (268.73,273.27) .. (268.73,276.8) -- (268.73,296) .. controls (268.73,299.53) and (265.87,302.4) .. (262.33,302.4) -- (175.13,302.4) .. controls (171.6,302.4) and (168.73,299.53) .. (168.73,296) -- cycle ;
\draw  [fill={rgb, 255:red, 255; green, 254; blue, 171 }  ,fill opacity=1 ] (276.73,276.8) .. controls (276.73,273.27) and (279.6,270.4) .. (283.13,270.4) -- (383.33,270.4) .. controls (386.87,270.4) and (389.73,273.27) .. (389.73,276.8) -- (389.73,296) .. controls (389.73,299.53) and (386.87,302.4) .. (383.33,302.4) -- (283.13,302.4) .. controls (279.6,302.4) and (276.73,299.53) .. (276.73,296) -- cycle ;
\draw  [color={rgb, 255:red, 0; green, 0; blue, 0 }  ,draw opacity=1 ][fill={rgb, 255:red, 250; green, 208; blue, 255 }  ,fill opacity=1 ] (479.73,276.8) .. controls (479.73,273.27) and (482.6,270.4) .. (486.13,270.4) -- (524.33,270.4) .. controls (527.87,270.4) and (530.73,273.27) .. (530.73,276.8) -- (530.73,296) .. controls (530.73,299.53) and (527.87,302.4) .. (524.33,302.4) -- (486.13,302.4) .. controls (482.6,302.4) and (479.73,299.53) .. (479.73,296) -- cycle ;
\draw [color={rgb, 255:red, 0; green, 0; blue, 0 }  ,draw opacity=1 ][line width=0.75]    (26.93,32.87) -- (45.73,33.01) ;
\draw [shift={(48.73,33.03)}, rotate = 180.44] [fill={rgb, 255:red, 0; green, 0; blue, 0 }  ,fill opacity=1 ][line width=0.08]  [draw opacity=0] (8.93,-4.29) -- (0,0) -- (8.93,4.29) -- cycle    ;

\draw  [color={rgb, 255:red, 0; green, 0; blue, 0 }  ,draw opacity=1 ][fill={rgb, 255:red, 197; green, 234; blue, 255 }  ,fill opacity=1 ] (36.8,46.62) .. controls (36.8,45.63) and (37.6,44.83) .. (38.59,44.83) -- (43.95,44.83) .. controls (44.93,44.83) and (45.73,45.63) .. (45.73,46.62) -- (45.73,52.61) .. controls (45.73,53.6) and (44.93,54.4) .. (43.95,54.4) -- (38.59,54.4) .. controls (37.6,54.4) and (36.8,53.6) .. (36.8,52.61) -- cycle ;
\draw  [color={rgb, 255:red, 0; green, 0; blue, 0 }  ,draw opacity=1 ][fill={rgb, 255:red, 255; green, 254; blue, 171 }  ,fill opacity=1 ] (36.8,61.62) .. controls (36.8,60.63) and (37.6,59.83) .. (38.59,59.83) -- (43.95,59.83) .. controls (44.93,59.83) and (45.73,60.63) .. (45.73,61.62) -- (45.73,67.61) .. controls (45.73,68.6) and (44.93,69.4) .. (43.95,69.4) -- (38.59,69.4) .. controls (37.6,69.4) and (36.8,68.6) .. (36.8,67.61) -- cycle ;
\draw  [color={rgb, 255:red, 0; green, 0; blue, 0 }  ,draw opacity=1 ][fill={rgb, 255:red, 250; green, 208; blue, 255 }  ,fill opacity=1 ] (36.8,75.62) .. controls (36.8,74.63) and (37.6,73.83) .. (38.59,73.83) -- (43.95,73.83) .. controls (44.93,73.83) and (45.73,74.63) .. (45.73,75.62) -- (45.73,81.61) .. controls (45.73,82.6) and (44.93,83.4) .. (43.95,83.4) -- (38.59,83.4) .. controls (37.6,83.4) and (36.8,82.6) .. (36.8,81.61) -- cycle ;
\draw    (349.67,50.55) .. controls (365.59,72.44) and (452.06,101.57) .. (417.46,210.22) ;
\draw [shift={(416.93,211.87)}, rotate = 288.12] [fill={rgb, 255:red, 0; green, 0; blue, 0 }  ][line width=0.08]  [draw opacity=0] (8.93,-4.29) -- (0,0) -- (8.93,4.29) -- cycle    ;

\draw (335,24) node [scale=0.5] [align=left] {org.jxls};
\draw (335.07,33.47) node [scale=0.7] [align=left] {\textbf{jxls-poi}};
\draw (335.07,42.47) node [scale=0.5] [align=left] {1.0.15};
\draw (354,138) node [scale=0.5] [align=left] {org.jxls};
\draw (354.07,147.47) node [scale=0.7] [align=left] {\textbf{jxls}};
\draw (354.07,157.47) node [scale=0.5] [align=left] {2.6.0};
\draw (242,141) node [scale=0.5] [align=left] {org.apache.commons};
\draw (242.07,150.47) node [scale=0.7] [align=left] {\textbf{commons-jexl}};
\draw (242.07,160.47) node [scale=0.5] [align=left] {2.1.1};
\draw (212,218) node [scale=0.5] [align=left] {org.apache.commons};
\draw (212.07,228.47) node [scale=0.7] [align=left] {\textbf{commons-jexl3}};
\draw (212.07,238.47) node [scale=0.5] [align=left] {3.1};
\draw (316,219) node [scale=0.5] [align=left] {commons-beanutils};
\draw (316.07,228.47) node [scale=0.7] [align=left] {\textbf{commons-beanutils}};
\draw (316.07,238.47) node [scale=0.5] [align=left] {1.9.3};
\draw (423,218) node [scale=0.5] [align=left] {ch.qos.logback};
\draw (423.07,227.47) node [scale=0.7] [align=left] {\textbf{logback-core}};
\draw (423.07,236.47) node [scale=0.5] [align=left] {1.2.3};
\draw (497,182) node [scale=0.5] [align=left] {org.slf4j};
\draw (497.07,191.47) node [scale=0.7] [align=left] {\textbf{jcl-over-slf4j}};
\draw (497.07,200.47) node [scale=0.5] [align=left] {1.7.12};
\draw (89,141) node [scale=0.5] [align=left] {org.apache.poi};
\draw (89.07,150.47) node [scale=0.7] [align=left] {\textbf{poi}};
\draw (89.07,160.47) node [scale=0.5] [align=left] {3.17};
\draw (91,221) node [scale=0.5] [align=left] {commons-codec};
\draw (91.07,230.47) node [scale=0.7] [align=left] {\textbf{commons-codec}};
\draw (91.07,240.47) node [scale=0.5] [align=left] {1.10};
\draw (101,277) node [scale=0.5] [align=left] {org.apache.commons};
\draw (101.07,286.47) node [scale=0.7] [align=left] {\textbf{commons-collections4}};
\draw (101.07,295.47) node [scale=0.5] [align=left] {4.1};
\draw (219,276) node [scale=0.5] [align=left] {commons-logging};
\draw (218.7,286.4) node [scale=0.7] [align=left] {\textbf{commons-logging}};
\draw (219.07,295.47) node [scale=0.5] [align=left] {1.1.1};
\draw (333,277) node [scale=0.5] [align=left] {commons-collections};
\draw (332.7,287.4) node [scale=0.7] [align=left] {\textbf{commons-collections}};
\draw (333.07,296.47) node [scale=0.5] [align=left] {3.2.2};
\draw (504,277) node [scale=0.5] [align=left] {org.slf4j};
\draw (504.07,286.47) node [scale=0.7] [align=left] {\textbf{slf4j-api}};
\draw (504.07,296.47) node [scale=0.5] [align=left] {1.7.12};
\draw (99,32.9) node [scale=0.7] [align=left] {{\scriptsize dependency usage status}};
\draw (87,47.9) node [scale=0.7] [align=left] {{\scriptsize direct dependency}};
\draw (95,61.9) node [scale=0.7] [align=left] {{\scriptsize transitive dependency }};
\draw (51,14.61) node  [align=left] {\textbf{Legend}};
\draw (93,75.9) node [scale=0.7] [align=left] {{\scriptsize inherited dependency}};
\draw (574,50.53) node [rotate=-90] [align=left] {Project};
\draw (574,199.53) node [rotate=-90] [align=left] {Dependencies};
\draw (531,124.9) node  [color={rgb, 255:red, 0; green, 0; blue, 0 }  ,opacity=1 ] [align=left] {{\scriptsize bi}};
\draw (117,93.9) node  [color={rgb, 255:red, 0; green, 0; blue, 0 }  ,opacity=1 ] [align=left] {{\scriptsize ud}};
\draw (343,96.9) node  [color={rgb, 255:red, 0; green, 0; blue, 0 }  ,opacity=1 ] [align=left] {{\scriptsize ud}};
\draw (298,99.9) node  [color={rgb, 255:red, 0; green, 0; blue, 0 }  ,opacity=1 ] [align=left] {{\scriptsize bd}};
\draw (326,175.9) node  [color={rgb, 255:red, 0; green, 0; blue, 0 }  ,opacity=1 ] [align=left] {{\scriptsize ut}};
\draw (152,172.9) node  [color={rgb, 255:red, 0; green, 0; blue, 0 }  ,opacity=1 ] [align=left] {{\scriptsize ut}};
\draw (432,128.9) node  [color={rgb, 255:red, 0; green, 0; blue, 0 }  ,opacity=1 ] [align=left] {{\scriptsize ut}};
\draw (456,94.9) node  [color={rgb, 255:red, 0; green, 0; blue, 0 }  ,opacity=1 ] [align=left] {{\scriptsize ui}};
\draw (179,183.9) node  [color={rgb, 255:red, 0; green, 0; blue, 0 }  ,opacity=1 ] [align=left] {{\scriptsize bt}};
\draw (396,160.9) node  [color={rgb, 255:red, 0; green, 0; blue, 0 }  ,opacity=1 ] [align=left] {{\scriptsize bt}};
\draw (300,154.9) node  [color={rgb, 255:red, 0; green, 0; blue, 0 }  ,opacity=1 ] [align=left] {{\scriptsize bt}};
\draw (135,136.9) node  [color={rgb, 255:red, 0; green, 0; blue, 0 }  ,opacity=1 ] [align=left] {{\scriptsize bt}};

\end{tikzpicture}

%% file: figures/depclean.tex
\tikzset{every picture/.style={line width=0.75pt}} %

\begin{tikzpicture}[x=0.75pt,y=0.75pt,yscale=-0.85,xscale=0.85]

\draw  [fill={rgb, 255:red, 236; green, 236; blue, 236 }  ,fill opacity=1 ] (158.67,868.67) -- (617.36,868.67) -- (617.36,1111.33) -- (158.67,1111.33) -- cycle ;
\draw  [fill={rgb, 255:red, 255; green, 246; blue, 136 }  ,fill opacity=0.3 ] (327.33,882.33) -- (460.93,882.33) -- (460.93,1073.33) -- (327.33,1073.33) -- cycle ;
\draw  [fill={rgb, 255:red, 243; green, 243; blue, 243 }  ,fill opacity=1 ][dash pattern={on 0.84pt off 2.51pt}] (259.81,1137.33) .. controls (259.81,1130.89) and (265.04,1125.67) .. (271.48,1125.67) -- (505,1125.67) .. controls (511.44,1125.67) and (516.67,1130.89) .. (516.67,1137.33) -- (516.67,1195.67) .. controls (516.67,1202.11) and (511.44,1207.33) .. (505,1207.33) -- (271.48,1207.33) .. controls (265.04,1207.33) and (259.81,1202.11) .. (259.81,1195.67) -- cycle ;
\draw  [fill={rgb, 255:red, 243; green, 243; blue, 243 }  ,fill opacity=1 ][dash pattern={on 0.84pt off 2.51pt}] (258.67,780.15) .. controls (258.67,773.07) and (264.41,767.33) .. (271.49,767.33) -- (502.7,767.33) .. controls (509.78,767.33) and (515.52,773.07) .. (515.52,780.15) -- (515.52,844.25) .. controls (515.52,851.33) and (509.78,857.07) .. (502.7,857.07) -- (271.49,857.07) .. controls (264.41,857.07) and (258.67,851.33) .. (258.67,844.25) -- cycle ;
\draw  [fill={rgb, 255:red, 255; green, 246; blue, 136 }  ,fill opacity=0.3 ] (171.6,893.3) -- (298.6,893.3) -- (298.6,1072.3) -- (171.6,1072.3) -- cycle ;
\draw    (207.67,1003.67) -- (193.55,1027.1) ;
\draw [shift={(192,1029.67)}, rotate = 301.07] [fill={rgb, 255:red, 0; green, 0; blue, 0 }  ][line width=0.08]  [draw opacity=0] (8.93,-4.29) -- (0,0) -- (8.93,4.29) -- cycle    ;

\draw  [fill={rgb, 255:red, 208; green, 2; blue, 27 }  ,fill opacity=0.3 ] (479,784.67) -- (479,830.13) .. controls (479,835.1) and (465.57,839.13) .. (449,839.13) .. controls (432.43,839.13) and (419,835.1) .. (419,830.13) -- (419,784.67) .. controls (419,779.7) and (432.43,775.67) .. (449,775.67) .. controls (465.57,775.67) and (479,779.7) .. (479,784.67) .. controls (479,789.64) and (465.57,793.67) .. (449,793.67) .. controls (432.43,793.67) and (419,789.64) .. (419,784.67) ;
\draw  [fill={rgb, 255:red, 241; green, 182; blue, 250 }  ,fill opacity=1 ] (290.67,774.67) -- (389.67,774.67) -- (389.67,849.78) -- (290.67,849.78) -- cycle ;
\draw    (291.67,791.67) -- (389.67,791.67) ;

\draw  [fill={rgb, 255:red, 255; green, 255; blue, 255 }  ,fill opacity=1 ] (300.67,807.5) .. controls (300.67,802.07) and (305.07,797.67) .. (310.5,797.67) -- (369.19,797.67) .. controls (374.62,797.67) and (379.02,802.07) .. (379.02,807.5) -- (379.02,807.5) .. controls (379.02,812.93) and (374.62,817.33) .. (369.19,817.33) -- (310.5,817.33) .. controls (305.07,817.33) and (300.67,812.93) .. (300.67,807.5) -- cycle ;
\draw  [fill={rgb, 255:red, 255; green, 255; blue, 255 }  ,fill opacity=1 ] (301.31,830.98) .. controls (301.31,825.63) and (305.65,821.3) .. (311,821.3) -- (371.98,821.3) .. controls (377.33,821.3) and (381.67,825.63) .. (381.67,830.98) -- (381.67,830.98) .. controls (381.67,836.33) and (377.33,840.67) .. (371.98,840.67) -- (311,840.67) .. controls (305.65,840.67) and (301.31,836.33) .. (301.31,830.98) -- cycle ;
\draw    (171.67,914.67) -- (298.67,914.67) ;

\draw    (226.92,1005.83) -- (240.96,1028.13) ;
\draw [shift={(242.56,1030.67)}, rotate = 237.8] [fill={rgb, 255:red, 0; green, 0; blue, 0 }  ][line width=0.08]  [draw opacity=0] (8.93,-4.29) -- (0,0) -- (8.93,4.29) -- cycle    ;

\draw    (243.58,995.75) -- (271.03,1027.4) ;
\draw [shift={(273,1029.67)}, rotate = 229.06] [fill={rgb, 255:red, 0; green, 0; blue, 0 }  ][line width=0.08]  [draw opacity=0] (8.93,-4.29) -- (0,0) -- (8.93,4.29) -- cycle    ;

\draw  [color={rgb, 255:red, 0; green, 0; blue, 0 }  ,draw opacity=1 ][fill={rgb, 255:red, 197; green, 234; blue, 255 }  ,fill opacity=1 ] (198.37,982.8) .. controls (198.37,979.27) and (201.24,976.4) .. (204.77,976.4) -- (224.37,976.4) .. controls (227.91,976.4) and (230.77,979.27) .. (230.77,982.8) -- (230.77,1002) .. controls (230.77,1005.53) and (227.91,1008.4) .. (224.37,1008.4) -- (204.77,1008.4) .. controls (201.24,1008.4) and (198.37,1005.53) .. (198.37,1002) -- cycle ;
\draw  [color={rgb, 255:red, 0; green, 0; blue, 0 }  ,draw opacity=1 ][fill={rgb, 255:red, 250; green, 240; blue, 122 }  ,fill opacity=1 ] (180.18,1036.47) .. controls (180.18,1032.93) and (183.05,1030.07) .. (186.58,1030.07) -- (206.18,1030.07) .. controls (209.71,1030.07) and (212.58,1032.93) .. (212.58,1036.47) -- (212.58,1055.67) .. controls (212.58,1059.2) and (209.71,1062.07) .. (206.18,1062.07) -- (186.58,1062.07) .. controls (183.05,1062.07) and (180.18,1059.2) .. (180.18,1055.67) -- cycle ;
\draw    (226.67,950.33) -- (211.56,975.11) ;
\draw [shift={(210,977.67)}, rotate = 301.37] [fill={rgb, 255:red, 0; green, 0; blue, 0 }  ][line width=0.08]  [draw opacity=0] (8.93,-4.29) -- (0,0) -- (8.93,4.29) -- cycle    ;

\draw  [color={rgb, 255:red, 0; green, 0; blue, 0 }  ,draw opacity=1 ][fill={rgb, 255:red, 197; green, 234; blue, 255 }  ,fill opacity=1 ] (240.1,982.8) .. controls (240.1,979.27) and (242.96,976.4) .. (246.5,976.4) -- (266.1,976.4) .. controls (269.63,976.4) and (272.5,979.27) .. (272.5,982.8) -- (272.5,1002) .. controls (272.5,1005.53) and (269.63,1008.4) .. (266.1,1008.4) -- (246.5,1008.4) .. controls (242.96,1008.4) and (240.1,1005.53) .. (240.1,1002) -- cycle ;
\draw    (240.56,952) -- (260.08,975.36) ;
\draw [shift={(262,977.67)}, rotate = 230.12] [fill={rgb, 255:red, 0; green, 0; blue, 0 }  ][line width=0.08]  [draw opacity=0] (8.93,-4.29) -- (0,0) -- (8.93,4.29) -- cycle    ;

\draw  [color={rgb, 255:red, 0; green, 0; blue, 0 }  ,draw opacity=1 ][fill={rgb, 255:red, 250; green, 240; blue, 122 }  ,fill opacity=1 ] (221.13,1036.47) .. controls (221.13,1032.93) and (223.99,1030.07) .. (227.53,1030.07) -- (247.13,1030.07) .. controls (250.66,1030.07) and (253.53,1032.93) .. (253.53,1036.47) -- (253.53,1055.67) .. controls (253.53,1059.2) and (250.66,1062.07) .. (247.13,1062.07) -- (227.53,1062.07) .. controls (223.99,1062.07) and (221.13,1059.2) .. (221.13,1055.67) -- cycle ;
\draw  [color={rgb, 255:red, 0; green, 0; blue, 0 }  ,draw opacity=1 ][fill={rgb, 255:red, 250; green, 240; blue, 122 }  ,fill opacity=1 ] (259.75,1036.47) .. controls (259.75,1032.93) and (262.61,1030.07) .. (266.15,1030.07) -- (285.75,1030.07) .. controls (289.28,1030.07) and (292.15,1032.93) .. (292.15,1036.47) -- (292.15,1055.67) .. controls (292.15,1059.2) and (289.28,1062.07) .. (285.75,1062.07) -- (266.15,1062.07) .. controls (262.61,1062.07) and (259.75,1059.2) .. (259.75,1055.67) -- cycle ;
\draw  [color={rgb, 255:red, 0; green, 0; blue, 0 }  ,draw opacity=1 ][fill={rgb, 255:red, 241; green, 182; blue, 250 }  ,fill opacity=1 ] (210.67,933.13) .. controls (210.67,929.6) and (213.53,926.73) .. (217.07,926.73) -- (250.84,926.73) .. controls (254.37,926.73) and (257.24,929.6) .. (257.24,933.13) -- (257.24,952.33) .. controls (257.24,955.87) and (254.37,958.73) .. (250.84,958.73) -- (217.07,958.73) .. controls (213.53,958.73) and (210.67,955.87) .. (210.67,952.33) -- cycle ;
\draw  [color={rgb, 255:red, 0; green, 0; blue, 0 }  ,draw opacity=1 ][fill={rgb, 255:red, 197; green, 234; blue, 255 }  ,fill opacity=1 ] (355.62,983.8) .. controls (355.62,980.27) and (358.49,977.4) .. (362.02,977.4) -- (382.06,977.4) .. controls (385.6,977.4) and (388.46,980.27) .. (388.46,983.8) -- (388.46,1003) .. controls (388.46,1006.53) and (385.6,1009.4) .. (382.06,1009.4) -- (362.02,1009.4) .. controls (358.49,1009.4) and (355.62,1006.53) .. (355.62,1003) -- cycle ;
\draw  [color={rgb, 255:red, 0; green, 0; blue, 0 }  ,draw opacity=1 ][fill={rgb, 255:red, 250; green, 240; blue, 122 }  ,fill opacity=1 ] (337.87,1037.47) .. controls (337.87,1033.93) and (340.74,1031.07) .. (344.27,1031.07) -- (363.87,1031.07) .. controls (367.41,1031.07) and (370.27,1033.93) .. (370.27,1037.47) -- (370.27,1056.67) .. controls (370.27,1060.2) and (367.41,1063.07) .. (363.87,1063.07) -- (344.27,1063.07) .. controls (340.74,1063.07) and (337.87,1060.2) .. (337.87,1056.67) -- cycle ;
\draw  [fill={rgb, 255:red, 250; green, 240; blue, 122 }  ,fill opacity=1 ] (343.13,1051.07) .. controls (343.13,1047.07) and (346.37,1043.83) .. (350.37,1043.83) .. controls (354.37,1043.83) and (357.61,1047.07) .. (357.61,1051.07) .. controls (357.61,1055.06) and (354.37,1058.31) .. (350.37,1058.31) .. controls (346.37,1058.31) and (343.13,1055.06) .. (343.13,1051.07) -- cycle ;
\draw    (364.17,1003.67) -- (351.35,1040.99) ;
\draw [shift={(350.37,1043.83)}, rotate = 288.96] [fill={rgb, 255:red, 0; green, 0; blue, 0 }  ][line width=0.08]  [draw opacity=0] (8.93,-4.29) -- (0,0) -- (8.93,4.29) -- cycle    ;

\draw  [color={rgb, 255:red, 0; green, 0; blue, 0 }  ,draw opacity=1 ][fill={rgb, 255:red, 250; green, 240; blue, 122 }  ,fill opacity=1 ] (379.82,1037.47) .. controls (379.82,1033.93) and (382.68,1031.07) .. (386.22,1031.07) -- (405.82,1031.07) .. controls (409.35,1031.07) and (412.22,1033.93) .. (412.22,1037.47) -- (412.22,1056.67) .. controls (412.22,1060.2) and (409.35,1063.07) .. (405.82,1063.07) -- (386.22,1063.07) .. controls (382.68,1063.07) and (379.82,1060.2) .. (379.82,1056.67) -- cycle ;
\draw    (328.67,914.67) -- (460.69,914.83) ;

\draw  [fill={rgb, 255:red, 171; green, 214; blue, 127 }  ,fill opacity=1 ] (322.33,1134.67) -- (367.67,1134.67) -- (367.67,1162.94) .. controls (339.33,1162.94) and (345,1173.14) .. (322.33,1166.54) -- cycle ; \draw  [fill={rgb, 255:red, 171; green, 214; blue, 127 }  ,fill opacity=1 ] (316.67,1138.95) -- (362,1138.95) -- (362,1167.23) .. controls (333.67,1167.23) and (339.33,1177.42) .. (316.67,1170.83) -- cycle ; \draw  [fill={rgb, 255:red, 171; green, 214; blue, 127 }  ,fill opacity=1 ] (311,1143.24) -- (356.33,1143.24) -- (356.33,1171.51) .. controls (328,1171.51) and (333.67,1181.71) .. (311,1175.11) -- cycle ;
\draw [fill={rgb, 255:red, 126; green, 211; blue, 33 }  ,fill opacity=1 ]   (347.05,1149.93) -- (319.66,1149.93) ;

\draw [fill={rgb, 255:red, 126; green, 211; blue, 33 }  ,fill opacity=1 ]   (347.05,1155.02) -- (319.66,1155.02) ;

\draw [fill={rgb, 255:red, 126; green, 211; blue, 33 }  ,fill opacity=1 ]   (347.05,1160.11) -- (319.66,1160.11) ;

\draw [fill={rgb, 255:red, 126; green, 211; blue, 33 }  ,fill opacity=1 ]   (347.05,1165.2) -- (319.66,1165.2) ;

\draw  [fill={rgb, 255:red, 255; green, 246; blue, 136 }  ,fill opacity=0.3 ] (489.95,881.67) -- (605,881.67) -- (605,1073.19) -- (489.95,1073.19) -- cycle ;
\draw    (533.67,1004.67) -- (519.55,1028.1) ;
\draw [shift={(518,1030.67)}, rotate = 301.07] [fill={rgb, 255:red, 0; green, 0; blue, 0 }  ][line width=0.08]  [draw opacity=0] (8.93,-4.29) -- (0,0) -- (8.93,4.29) -- cycle    ;

\draw    (490.33,914.67) -- (604.33,914.67) ;

\draw  [color={rgb, 255:red, 0; green, 0; blue, 0 }  ,draw opacity=1 ][fill={rgb, 255:red, 197; green, 234; blue, 255 }  ,fill opacity=1 ] (517.88,983.8) .. controls (517.88,980.27) and (520.74,977.4) .. (524.28,977.4) -- (543.88,977.4) .. controls (547.41,977.4) and (550.28,980.27) .. (550.28,983.8) -- (550.28,1003) .. controls (550.28,1006.53) and (547.41,1009.4) .. (543.88,1009.4) -- (524.28,1009.4) .. controls (520.74,1009.4) and (517.88,1006.53) .. (517.88,1003) -- cycle ;
\draw  [color={rgb, 255:red, 0; green, 0; blue, 0 }  ,draw opacity=1 ][fill={rgb, 255:red, 250; green, 240; blue, 122 }  ,fill opacity=1 ] (499.02,1037.47) .. controls (499.02,1033.93) and (501.89,1031.07) .. (505.42,1031.07) -- (525.02,1031.07) .. controls (528.55,1031.07) and (531.42,1033.93) .. (531.42,1037.47) -- (531.42,1056.67) .. controls (531.42,1060.2) and (528.55,1063.07) .. (525.02,1063.07) -- (505.42,1063.07) .. controls (501.89,1063.07) and (499.02,1060.2) .. (499.02,1056.67) -- cycle ;
\draw    (543.67,949.33) -- (528.56,974.11) ;
\draw [shift={(527,976.67)}, rotate = 301.37] [fill={rgb, 255:red, 0; green, 0; blue, 0 }  ][line width=0.08]  [draw opacity=0] (8.93,-4.29) -- (0,0) -- (8.93,4.29) -- cycle    ;

\draw  [color={rgb, 255:red, 0; green, 0; blue, 0 }  ,draw opacity=1 ][fill={rgb, 255:red, 197; green, 234; blue, 255 }  ,fill opacity=1 ] (559.6,983.8) .. controls (559.6,980.27) and (562.47,977.4) .. (566,977.4) -- (585.6,977.4) .. controls (589.13,977.4) and (592,980.27) .. (592,983.8) -- (592,1003) .. controls (592,1006.53) and (589.13,1009.4) .. (585.6,1009.4) -- (566,1009.4) .. controls (562.47,1009.4) and (559.6,1006.53) .. (559.6,1003) -- cycle ;
\draw    (556.56,952) -- (576.08,975.36) ;
\draw [shift={(578,977.67)}, rotate = 230.12] [fill={rgb, 255:red, 0; green, 0; blue, 0 }  ][line width=0.08]  [draw opacity=0] (8.93,-4.29) -- (0,0) -- (8.93,4.29) -- cycle    ;

\draw  [color={rgb, 255:red, 0; green, 0; blue, 0 }  ,draw opacity=1 ][fill={rgb, 255:red, 241; green, 182; blue, 250 }  ,fill opacity=1 ] (528.76,933.13) .. controls (528.76,929.6) and (531.63,926.73) .. (535.16,926.73) -- (568.93,926.73) .. controls (572.47,926.73) and (575.33,929.6) .. (575.33,933.13) -- (575.33,952.33) .. controls (575.33,955.87) and (572.47,958.73) .. (568.93,958.73) -- (535.16,958.73) .. controls (531.63,958.73) and (528.76,955.87) .. (528.76,952.33) -- cycle ;
\draw  [fill={rgb, 255:red, 255; green, 255; blue, 255 }  ,fill opacity=1 ] (295.67,975.63) -- (315.87,975.63) -- (315.87,967.47) -- (329.33,983.8) -- (315.87,1000.13) -- (315.87,991.97) -- (295.67,991.97) -- cycle ;
\draw    (448,839.13) -- (448.6,864.67) ;
\draw [shift={(448.67,867.67)}, rotate = 268.65999999999997] [fill={rgb, 255:red, 0; green, 0; blue, 0 }  ][line width=0.08]  [draw opacity=0] (10.72,-5.15) -- (0,0) -- (10.72,5.15) -- (7.12,0) -- cycle    ;

\draw    (340.67,1111.17) -- (340.67,1131.67) ;
\draw [shift={(340.67,1134.67)}, rotate = 270] [fill={rgb, 255:red, 0; green, 0; blue, 0 }  ][line width=0.08]  [draw opacity=0] (10.72,-5.15) -- (0,0) -- (10.72,5.15) -- (7.12,0) -- cycle    ;

\draw    (447.67,1111.67) -- (447.67,1135.33) ;
\draw [shift={(447.67,1138.33)}, rotate = 270] [fill={rgb, 255:red, 0; green, 0; blue, 0 }  ][line width=0.08]  [draw opacity=0] (10.72,-5.15) -- (0,0) -- (10.72,5.15) -- (7.12,0) -- cycle    ;

\draw  [fill={rgb, 255:red, 255; green, 221; blue, 164 }  ,fill opacity=1 ] (411.25,1150.23) .. controls (411.25,1143.91) and (416.37,1138.78) .. (422.69,1138.78) -- (469.81,1138.78) .. controls (476.13,1138.78) and (481.25,1143.91) .. (481.25,1150.23) -- (481.25,1150.23) .. controls (481.25,1156.54) and (476.13,1161.67) .. (469.81,1161.67) -- (422.69,1161.67) .. controls (416.37,1161.67) and (411.25,1156.54) .. (411.25,1150.23) -- cycle ;
\draw  [fill={rgb, 255:red, 191; green, 227; blue, 249 }  ,fill opacity=1 ] (360.8,999.07) .. controls (360.8,995.07) and (364.04,991.83) .. (368.04,991.83) .. controls (372.04,991.83) and (375.28,995.07) .. (375.28,999.07) .. controls (375.28,1003.06) and (372.04,1006.31) .. (368.04,1006.31) .. controls (364.04,1006.31) and (360.8,1003.06) .. (360.8,999.07) -- cycle ;
\draw  [fill={rgb, 255:red, 255; green, 255; blue, 255 }  ,fill opacity=1 ] (457.67,975.63) -- (477.87,975.63) -- (477.87,967.47) -- (491.33,983.8) -- (477.87,1000.13) -- (477.87,991.97) -- (457.67,991.97) -- cycle ;
\draw    (339.67,848.67) -- (339.67,864.67) ;
\draw [shift={(339.67,867.67)}, rotate = 270] [fill={rgb, 255:red, 0; green, 0; blue, 0 }  ][line width=0.08]  [draw opacity=0] (10.72,-5.15) -- (0,0) -- (10.72,5.15) -- (7.12,0) -- cycle    ;

\draw  [color={rgb, 255:red, 0; green, 0; blue, 0 }  ,draw opacity=1 ][fill={rgb, 255:red, 250; green, 240; blue, 122 }  ,fill opacity=1 ] (417.75,1036.47) .. controls (417.75,1032.93) and (420.61,1030.07) .. (424.15,1030.07) -- (443.75,1030.07) .. controls (447.28,1030.07) and (450.15,1032.93) .. (450.15,1036.47) -- (450.15,1055.67) .. controls (450.15,1059.2) and (447.28,1062.07) .. (443.75,1062.07) -- (424.15,1062.07) .. controls (420.61,1062.07) and (417.75,1059.2) .. (417.75,1055.67) -- cycle ;
\draw    (401.58,995.75) -- (418.98,1028.24) ;
\draw [shift={(420.4,1030.89)}, rotate = 241.82999999999998] [fill={rgb, 255:red, 0; green, 0; blue, 0 }  ][line width=0.08]  [draw opacity=0] (8.93,-4.29) -- (0,0) -- (8.93,4.29) -- cycle    ;

\draw  [color={rgb, 255:red, 0; green, 0; blue, 0 }  ,draw opacity=1 ][fill={rgb, 255:red, 197; green, 234; blue, 255 }  ,fill opacity=1 ] (397.79,983.8) .. controls (397.79,980.27) and (400.65,977.4) .. (404.19,977.4) -- (423.79,977.4) .. controls (427.32,977.4) and (430.19,980.27) .. (430.19,983.8) -- (430.19,1003) .. controls (430.19,1006.53) and (427.32,1009.4) .. (423.79,1009.4) -- (404.19,1009.4) .. controls (400.65,1009.4) and (397.79,1006.53) .. (397.79,1003) -- cycle ;
\draw  [fill={rgb, 255:red, 191; green, 227; blue, 249 }  ,fill opacity=1 ] (404.9,996.04) .. controls (404.9,990.59) and (409.33,986.16) .. (414.79,986.16) .. controls (420.24,986.16) and (424.67,990.59) .. (424.67,996.04) .. controls (424.67,1001.5) and (420.24,1005.92) .. (414.79,1005.92) .. controls (409.33,1005.92) and (404.9,1001.5) .. (404.9,996.04) -- cycle ;
\draw  [color={rgb, 255:red, 0; green, 0; blue, 0 }  ,draw opacity=1 ][fill={rgb, 255:red, 241; green, 182; blue, 250 }  ,fill opacity=1 ] (368.51,933.8) .. controls (368.51,930.27) and (371.37,927.4) .. (374.91,927.4) -- (408.68,927.4) .. controls (412.21,927.4) and (415.08,930.27) .. (415.08,933.8) -- (415.08,953) .. controls (415.08,956.53) and (412.21,959.4) .. (408.68,959.4) -- (374.91,959.4) .. controls (371.37,959.4) and (368.51,956.53) .. (368.51,953) -- cycle ;
\draw    (385.67,950) -- (369.2,989.06) ;
\draw [shift={(368.04,991.83)}, rotate = 292.85] [fill={rgb, 255:red, 0; green, 0; blue, 0 }  ][line width=0.08]  [draw opacity=0] (8.93,-4.29) -- (0,0) -- (8.93,4.29) -- cycle    ;

\draw    (399.56,951.67) -- (413.59,984.07) ;
\draw [shift={(414.79,986.83)}, rotate = 246.57999999999998] [fill={rgb, 255:red, 0; green, 0; blue, 0 }  ][line width=0.08]  [draw opacity=0] (8.93,-4.29) -- (0,0) -- (8.93,4.29) -- cycle    ;

\draw  [fill={rgb, 255:red, 241; green, 182; blue, 250 }  ,fill opacity=1 ] (380.57,952.42) .. controls (380.57,949.92) and (382.59,947.9) .. (385.08,947.9) .. controls (387.58,947.9) and (389.6,949.92) .. (389.6,952.42) .. controls (389.6,954.91) and (387.58,956.93) .. (385.08,956.93) .. controls (382.59,956.93) and (380.57,954.91) .. (380.57,952.42) -- cycle ;
\draw  [fill={rgb, 255:red, 241; green, 182; blue, 250 }  ,fill opacity=1 ] (396.57,952.02) .. controls (396.57,950.3) and (397.96,948.9) .. (399.68,948.9) .. controls (401.4,948.9) and (402.8,950.3) .. (402.8,952.02) .. controls (402.8,953.74) and (401.4,955.13) .. (399.68,955.13) .. controls (397.96,955.13) and (396.57,953.74) .. (396.57,952.02) -- cycle ;
\draw  [fill={rgb, 255:red, 250; green, 240; blue, 122 }  ,fill opacity=1 ] (391.75,1051.33) .. controls (391.75,1048.98) and (393.66,1047.07) .. (396.02,1047.07) .. controls (398.38,1047.07) and (400.29,1048.98) .. (400.29,1051.33) .. controls (400.29,1053.69) and (398.38,1055.6) .. (396.02,1055.6) .. controls (393.66,1055.6) and (391.75,1053.69) .. (391.75,1051.33) -- cycle ;
\draw  [fill={rgb, 255:red, 191; green, 227; blue, 249 }  ,fill opacity=1 ] (377.6,1003.26) .. controls (377.6,1001.14) and (379.32,999.43) .. (381.43,999.43) .. controls (383.55,999.43) and (385.27,1001.14) .. (385.27,1003.26) .. controls (385.27,1005.38) and (383.55,1007.09) .. (381.43,1007.09) .. controls (379.32,1007.09) and (377.6,1005.38) .. (377.6,1003.26) -- cycle ;
\draw    (381.43,1007.09) -- (392.79,1044.45) ;
\draw [shift={(393.67,1047.32)}, rotate = 253.07999999999998] [fill={rgb, 255:red, 0; green, 0; blue, 0 }  ][line width=0.08]  [draw opacity=0] (8.93,-4.29) -- (0,0) -- (8.93,4.29) -- cycle    ;

\draw (197,880) node   [align=left] {\textsc{DepClean}};
\draw (548,1165) node  [rotate=-90] [align=left] {\textbf{Output}};
\draw (549,814) node  [rotate=-90] [align=left] {\textbf{Input}};
\draw (448.33,810.74) node   [align=left] {{\scriptsize \textbf{Maven }}\\{\scriptsize \textbf{Central}}};
\draw (340,782.02) node  [font=\normalsize] [align=left] {\textbf{{\scriptsize Project}}};
\draw (340.83,830.98) node  [font=\small] [align=left] {{\footnotesize POM}};
\draw (338.96,807.34) node  [font=\small] [align=left] {{\scriptsize Bytecode}};
\draw (236.5,903.52) node  [font=\tiny] [align=left] {{\scriptsize \textbf{Dependency tree}}};
\draw (386.36,899.18) node  [font=\tiny] [align=left] {{\scriptsize \textbf{Dependency }}\\{\scriptsize \textbf{usage analysis}}};
\draw (345,1191.76) node  [font=\scriptsize] [align=left] {Dependency \\usage report};
\draw (551.06,899.52) node  [font=\tiny] [align=left] {{\scriptsize \textbf{Debloated }}\\{\scriptsize \textbf{dependency tree}}};
\draw (448,1178.76) node  [font=\scriptsize] [align=left] {Debloated\\POM file};
\draw (233.95,942.73) node  [font=\tiny] [align=left] {\textbf{{\scriptsize Project}}};
\draw (446.25,1150.23) node  [font=\small] [align=left] {{$\text{\large{POM}}_d$}};
\draw (238,1090.28) node   [align=left] {{\scriptsize Maven dependency }\\{\scriptsize resolution}};
\draw (384,1092.28) node   [align=left] {{\scriptsize Bytecode analysis of }\\{\scriptsize API members calls}};
\draw (551,1092.28) node   [align=left] {{\scriptsize Actually used }\\{\scriptsize dependencies}};
\draw (391.79,942.73) node  [font=\tiny] [align=left] {\textbf{{\scriptsize Project}}};
\draw (551.38,942.73) node  [font=\tiny] [align=left] {\textbf{{\scriptsize Project}}};

\end{tikzpicture}

%% file: figures/exp_framework.tex
\tikzset{every picture/.style={line width=0.75pt}} %

\begin{tikzpicture}[x=0.75pt,y=0.75pt,yscale=-0.8,xscale=0.8]

\draw [line width=1.5]    (257.75,254.58) -- (296.73,254.54) ;
\draw [shift={(300.73,254.53)}, rotate = 539.9300000000001] [fill={rgb, 255:red, 0; green, 0; blue, 0 }  ][line width=0.08]  [draw opacity=0] (13.4,-6.43) -- (0,0) -- (13.4,6.44) -- (8.9,0) -- cycle    ;

\draw  [fill={rgb, 255:red, 243; green, 243; blue, 243 }  ,fill opacity=1 ][dash pattern={on 0.84pt off 2.51pt}] (33.73,213.96) .. controls (33.73,204.85) and (41.12,197.47) .. (50.23,197.47) -- (253.24,197.47) .. controls (262.35,197.47) and (269.73,204.85) .. (269.73,213.96) -- (269.73,296.44) .. controls (269.73,305.55) and (262.35,312.93) .. (253.24,312.93) -- (50.23,312.93) .. controls (41.12,312.93) and (33.73,305.55) .. (33.73,296.44) -- cycle ;
\draw  [fill={rgb, 255:red, 243; green, 243; blue, 243 }  ,fill opacity=1 ][dash pattern={on 0.84pt off 2.51pt}] (302.17,211.19) .. controls (302.17,202.01) and (309.61,194.57) .. (318.79,194.57) -- (448.14,194.57) .. controls (457.32,194.57) and (464.77,202.01) .. (464.77,211.19) -- (464.77,294.31) .. controls (464.77,303.49) and (457.32,310.93) .. (448.14,310.93) -- (318.79,310.93) .. controls (309.61,310.93) and (302.17,303.49) .. (302.17,294.31) -- cycle ;
\draw [line width=1.5]    (185.75,99.58) -- (224.73,99.54) ;
\draw [shift={(228.73,99.53)}, rotate = 539.9300000000001] [fill={rgb, 255:red, 0; green, 0; blue, 0 }  ][line width=0.08]  [draw opacity=0] (13.4,-6.43) -- (0,0) -- (13.4,6.44) -- (8.9,0) -- cycle    ;

\draw  [fill={rgb, 255:red, 243; green, 243; blue, 243 }  ,fill opacity=1 ][dash pattern={on 0.84pt off 2.51pt}] (32.25,57.63) .. controls (32.25,48.45) and (39.7,41) .. (48.88,41) -- (178.23,41) .. controls (187.41,41) and (194.85,48.45) .. (194.85,57.63) -- (194.85,140.75) .. controls (194.85,149.93) and (187.41,157.37) .. (178.23,157.37) -- (48.88,157.37) .. controls (39.7,157.37) and (32.25,149.93) .. (32.25,140.75) -- cycle ;
\draw (114.43,100.02) node  {\includegraphics[width=43.96pt,height=43.96pt]{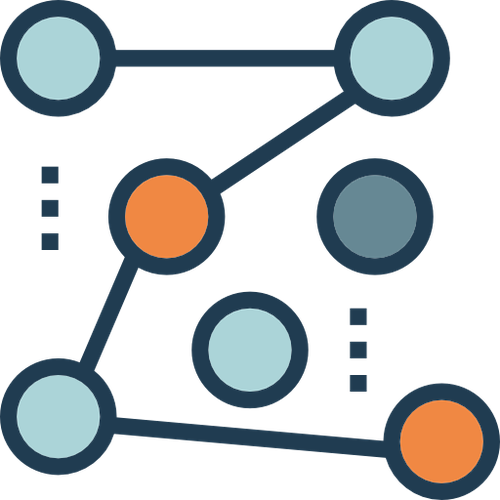}};
\draw  [fill={rgb, 255:red, 243; green, 243; blue, 243 }  ,fill opacity=1 ][dash pattern={on 0.84pt off 2.51pt}] (228.73,57.96) .. controls (228.73,48.85) and (236.12,41.47) .. (245.23,41.47) -- (448.24,41.47) .. controls (457.35,41.47) and (464.73,48.85) .. (464.73,57.96) -- (464.73,140.44) .. controls (464.73,149.55) and (457.35,156.93) .. (448.24,156.93) -- (245.23,156.93) .. controls (236.12,156.93) and (228.73,149.55) .. (228.73,140.44) -- cycle ;
\draw  [fill={rgb, 255:red, 240; green, 225; blue, 100 }  ,fill opacity=1 ] (15.76,42.2) .. controls (15.76,33.68) and (23.14,26.77) .. (32.25,26.77) .. controls (41.36,26.77) and (48.75,33.68) .. (48.75,42.2) .. controls (48.75,50.72) and (41.36,57.63) .. (32.25,57.63) .. controls (23.14,57.63) and (15.76,50.72) .. (15.76,42.2) -- cycle ;
\draw  [fill={rgb, 255:red, 240; green, 225; blue, 100 }  ,fill opacity=1 ] (210.76,44.2) .. controls (210.76,35.68) and (218.14,28.77) .. (227.25,28.77) .. controls (236.36,28.77) and (243.75,35.68) .. (243.75,44.2) .. controls (243.75,52.72) and (236.36,59.63) .. (227.25,59.63) .. controls (218.14,59.63) and (210.76,52.72) .. (210.76,44.2) -- cycle ;
\draw  [fill={rgb, 255:red, 240; green, 225; blue, 100 }  ,fill opacity=1 ] (17.92,198.17) .. controls (17.92,189.65) and (25.31,182.75) .. (34.42,182.75) .. controls (43.53,182.75) and (50.91,189.65) .. (50.91,198.17) .. controls (50.91,206.69) and (43.53,213.6) .. (34.42,213.6) .. controls (25.31,213.6) and (17.92,206.69) .. (17.92,198.17) -- cycle ;
\draw  [fill={rgb, 255:red, 240; green, 225; blue, 100 }  ,fill opacity=1 ] (282.76,198.2) .. controls (282.76,189.68) and (290.14,182.77) .. (299.25,182.77) .. controls (308.36,182.77) and (315.75,189.68) .. (315.75,198.2) .. controls (315.75,206.72) and (308.36,213.63) .. (299.25,213.63) .. controls (290.14,213.63) and (282.76,206.72) .. (282.76,198.2) -- cycle ;
\draw (427.29,93.94) node  {\includegraphics[width=24.17pt,height=24.17pt]{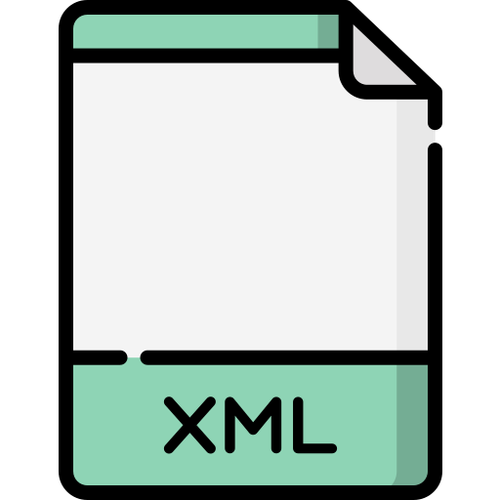}};

\draw (387.83,130.61) node  {\includegraphics[width=23.53pt,height=24.22pt]{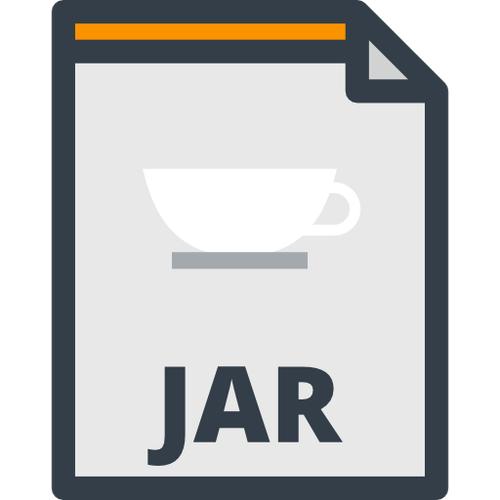}};
\draw  [dash pattern={on 4.5pt off 4.5pt}]  (305.73,102.53) -- (412.47,95.2) ;
\draw [shift={(414.47,95.07)}, rotate = 536.0699999999999] [color={rgb, 255:red, 0; green, 0; blue, 0 }  ][line width=0.75]    (10.93,-3.29) .. controls (6.95,-1.4) and (3.31,-0.3) .. (0,0) .. controls (3.31,0.3) and (6.95,1.4) .. (10.93,3.29)   ;

\draw  [dash pattern={on 4.5pt off 4.5pt}]  (305.73,112.53) -- (372.79,128.08) ;
\draw [shift={(374.73,128.53)}, rotate = 193.06] [color={rgb, 255:red, 0; green, 0; blue, 0 }  ][line width=0.75]    (10.93,-3.29) .. controls (6.95,-1.4) and (3.31,-0.3) .. (0,0) .. controls (3.31,0.3) and (6.95,1.4) .. (10.93,3.29)   ;

\draw (156.23,275.34) node  {\includegraphics[width=115.14pt,height=36.3pt]{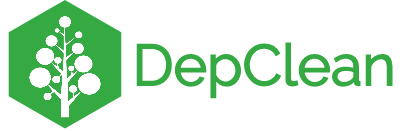}};
\draw  [fill={rgb, 255:red, 208; green, 2; blue, 27 }  ,fill opacity=0.3 ] (303.93,81.52) -- (303.93,123.85) .. controls (303.93,128.48) and (291.43,132.23) .. (276,132.23) .. controls (260.57,132.23) and (248.07,128.48) .. (248.07,123.85) -- (248.07,81.52) .. controls (248.07,76.89) and (260.57,73.14) .. (276,73.14) .. controls (291.43,73.14) and (303.93,76.89) .. (303.93,81.52) .. controls (303.93,86.14) and (291.43,89.9) .. (276,89.9) .. controls (260.57,89.9) and (248.07,86.14) .. (248.07,81.52) ;
\draw  [fill={rgb, 255:red, 171; green, 214; blue, 127 }  ,fill opacity=1 ] (363.93,248.13) -- (411.67,248.13) -- (411.67,277.91) .. controls (381.84,277.91) and (387.8,288.65) .. (363.93,281.7) -- cycle ; \draw  [fill={rgb, 255:red, 171; green, 214; blue, 127 }  ,fill opacity=1 ] (357.97,252.65) -- (405.7,252.65) -- (405.7,282.42) .. controls (375.87,282.42) and (381.84,293.16) .. (357.97,286.21) -- cycle ; \draw  [fill={rgb, 255:red, 171; green, 214; blue, 127 }  ,fill opacity=1 ] (352,257.16) -- (399.74,257.16) -- (399.74,286.94) .. controls (369.9,286.94) and (375.87,297.67) .. (352,290.73) -- cycle ;
\draw [fill={rgb, 255:red, 126; green, 211; blue, 33 }  ,fill opacity=1 ]   (389.96,265.25) -- (361.12,265.25) ;

\draw [fill={rgb, 255:red, 126; green, 211; blue, 33 }  ,fill opacity=1 ]   (389.96,270.61) -- (361.12,270.61) ;

\draw [fill={rgb, 255:red, 126; green, 211; blue, 33 }  ,fill opacity=1 ]   (389.96,275.97) -- (361.12,275.97) ;

\draw [fill={rgb, 255:red, 126; green, 211; blue, 33 }  ,fill opacity=1 ]   (389.96,281.33) -- (361.12,281.33) ;

\draw  [dash pattern={on 4.5pt off 4.5pt}]  (15.47,171) -- (504.47,171.07) ;

\draw (155,223) node   [align=left] {\textbf{Detect and analyze }\\\textbf{bloated dependencies}};
\draw (111.47,54.8) node   [align=left] {\textbf{Filter artifacts}};
\draw (352.97,55.8) node   [align=left] {\textbf{Resolve dependencies }};
\draw (32.25,42.2) node  [font=\large]  {$1$};
\draw (227.25,43.2) node  [font=\large]  {$2$};
\draw (34.42,198.17) node  [font=\large]  {$3$};
\draw (299.25,197.2) node  [font=\large]  {$4$};
\draw (111.77,141.14) node   [align=left] {MDG};
\draw (388,221) node   [align=left] {\textbf{Report dependency }\\\textbf{usage metrics}};
\draw (356,90.17) node  [rotate=-356.04] [align=left] {{\tiny donwload}};
\draw (344,114.17) node  [rotate=-12.28] [align=left] {{\tiny donwload}};
\draw (275.38,105.79) node   [align=left] {{\scriptsize \textbf{Maven }}\\{\scriptsize \textbf{Central}}};
\draw (495.23,101.53) node  [rotate=-90] [align=left] {\textbf{Data collection}};
\draw (495.23,247.53) node  [rotate=-90] [align=left] {\textbf{Data analysis}};
\draw (426.39,91.26) node  [font=\tiny]  {$POM$};

\end{tikzpicture}

%% file: main.bbl
\begin{thebibliography}{43}
\providecommand{\natexlab}[1]{#1}
\providecommand{\url}[1]{\texttt{#1}}
\expandafter\ifx\csname urlstyle\endcsname\relax
  \providecommand{\doi}[1]{doi: #1}\else
  \providecommand{\doi}{doi: \begingroup \urlstyle{rm}\Url}\fi

\bibitem[Azad et~al.(2019)Azad, Laperdrix, and Nikiforakis]{Babak2019}
B.~A. Azad, P.~Laperdrix, and N.~Nikiforakis.
\newblock Less is more: Quantifying the security benefits of debloating web
  applications.
\newblock In \emph{Proceedings of the 28th USENIX Conference on Security
  Symposium}, SEC, page 1697–1714, USA, 2019. USENIX Association.
\newblock ISBN 9781939133069.

\bibitem[Bauer et~al.(2014)Bauer, Eckhardt, Hauptmann, and Klimek]{Bauer2014}
V.~Bauer, J.~Eckhardt, B.~Hauptmann, and M.~Klimek.
\newblock {An Exploratory Study on Reuse at Google}.
\newblock In \emph{Proceedings of the 1st International Workshop on Software
  Engineering Research and Industrial Practices}, SERIP, pages 14--23, New
  York, NY, USA, 2014. ACM.
\newblock ISBN 978-1-4503-2859-3.
\newblock \doi{10.1145/2593850.2593854}.
\newblock URL \url{http://doi.acm.org/10.1145/2593850.2593854}.

\bibitem[Bavota et~al.(2015)Bavota, Canfora, Di~Penta, Oliveto, and
  Panichella]{Bavota2015}
G.~Bavota, G.~Canfora, M.~Di~Penta, R.~Oliveto, and S.~Panichella.
\newblock How the apache community upgrades dependencies: an evolutionary
  study.
\newblock \emph{Empirical Software Engineering}, 20\penalty0 (5):\penalty0
  1275--1317, Oct 2015.
\newblock ISSN 1573-7616.
\newblock \doi{10.1007/s10664-014-9325-9}.
\newblock URL \url{https://doi.org/10.1007/s10664-014-9325-9}.

\bibitem[{Benelallam} et~al.(2019){Benelallam}, {Harrand}, {Soto-Valero},
  {Baudry}, and {Barais}]{Benelallam2019}
A.~{Benelallam}, N.~{Harrand}, C.~{Soto-Valero}, B.~{Baudry}, and O.~{Barais}.
\newblock {The Maven Dependency Graph: a Temporal Graph-based Representation of
  Maven Central}.
\newblock In \emph{16th International Conference on Mining Software
  Repositories (MSR)}, Montreal, Canada, 2019. IEEE/ACM.

\bibitem[Bezemer et~al.(2017)Bezemer, McIntosh, Adams, German, and
  Hassan]{Bezemer2017}
C.-P. Bezemer, S.~McIntosh, B.~Adams, D.~M. German, and A.~E. Hassan.
\newblock An empirical study of unspecified dependencies in make-based build
  systems.
\newblock \emph{Empirical Software Engineering}, 22\penalty0 (6):\penalty0
  3117--3148, Dec 2017.
\newblock ISSN 1573-7616.
\newblock \doi{10.1007/s10664-017-9510-8}.
\newblock URL \url{https://doi.org/10.1007/s10664-017-9510-8}.

\bibitem[Brooks(1987)]{Brooks1987}
F.~P. Brooks.
\newblock No silver bullet essence and accidents of software engineering.
\newblock \emph{Computer}, 20\penalty0 (4):\penalty0 10–19, Apr. 1987.
\newblock ISSN 0018-9162.
\newblock \doi{10.1109/MC.1987.1663532}.
\newblock URL \url{https://doi.org/10.1109/MC.1987.1663532}.

\bibitem[Callo~Arias et~al.(2011)Callo~Arias, van~der Spek, and
  Avgeriou]{CalloArias2011}
T.~B. Callo~Arias, P.~van~der Spek, and P.~Avgeriou.
\newblock A practice-driven systematic review of dependency analysis solutions.
\newblock \emph{Empirical Software Engineering}, 16\penalty0 (5):\penalty0
  544--586, Oct 2011.
\newblock \doi{10.1007/s10664-011-9158-8}.
\newblock URL \url{https://doi.org/10.1007/s10664-011-9158-8}.

\bibitem[Celik et~al.(2016)Celik, Knaust, Milicevic, and Gligoric]{Celik2016}
A.~Celik, A.~Knaust, A.~Milicevic, and M.~Gligoric.
\newblock Build system with lazy retrieval for java projects.
\newblock In \emph{Proceedings of the 2016 24th ACM SIGSOFT International
  Symposium on Foundations of Software Engineering}, FSE, pages 643--654, New
  York, NY, USA, 2016. ACM.
\newblock ISBN 978-1-4503-4218-6.
\newblock \doi{10.1145/2950290.2950358}.
\newblock URL \url{http://doi.acm.org/10.1145/2950290.2950358}.

\bibitem[Cox(2019)]{cox2019surviving}
R.~Cox.
\newblock Surviving software dependencies.
\newblock \emph{Communications of the ACM}, 62\penalty0 (9):\penalty0 36--43,
  2019.
\newblock URL
  \url{http://delivery.acm.org/10.1145/3350000/3347446/p36-cox.pdf?ip=130.237.224.202&id=3347446&acc=OA&key=74F7687761D7AE37\%2EE53E9A92DC589BF3\%2E4D4702B0C3E38B35\%2E5945DC2EABF3343C&__acm__=1575959910_051299ecdd4bf0a7442ae3476bd0e750}.

\bibitem[Decan et~al.(2017)Decan, Mens, and Claes]{Decan2017}
A.~Decan, T.~Mens, and M.~Claes.
\newblock {An Empirical Comparison of Dependency Issues in OSS Packaging
  Ecosystems}.
\newblock In \emph{2017 IEEE 24th International Conference on Software
  Analysis, Evolution and Reengineering}, SANER, pages 2--12, Feb 2017.
\newblock \doi{10.1109/SANER.2017.7884604}.

\bibitem[Decan et~al.(2019)Decan, Mens, and Grosjean]{Decan2019}
A.~Decan, T.~Mens, and P.~Grosjean.
\newblock An empirical comparison of dependency network evolution in seven
  software packaging ecosystems.
\newblock \emph{Empirical Software Engineering}, 24\penalty0 (1):\penalty0
  381--416, Feb 2019.
\newblock ISSN 1573-7616.
\newblock \doi{10.1007/s10664-017-9589-y}.
\newblock URL \url{https://doi.org/10.1007/s10664-017-9589-y}.

\bibitem[Gkortzis et~al.(2019)Gkortzis, Feitosa, and Spinellis]{Gkortzis2019}
A.~Gkortzis, D.~Feitosa, and D.~Spinellis.
\newblock A double-edged sword? software reuse and potential security
  vulnerabilities.
\newblock In X.~Peng, A.~Ampatzoglou, and T.~Bhowmik, editors, \emph{Reuse in
  the Big Data Era}, pages 187--203, Cham, 2019. Springer International
  Publishing.
\newblock ISBN 978-3-030-22888-0.

\bibitem[Harrand et~al.(2019)Harrand, Benelallam, Soto-Valero, Barais, and
  Baudry]{Harrand2019}
N.~Harrand, A.~Benelallam, C.~Soto-Valero, O.~Barais, and B.~Baudry.
\newblock Analyzing 2.3 million maven dependencies to reveal an essential core
  in apis.
\newblock \emph{arXiv preprint arXiv:1908.09757}, 2019.

\bibitem[{Heath} et~al.(2019){Heath}, {Velingker}, {Bastani}, and
  {Naik}]{Heath2019}
B.~{Heath}, N.~{Velingker}, O.~{Bastani}, and M.~{Naik}.
\newblock {PolyDroid: Learning-Driven Specialization of Mobile Applications}.
\newblock \emph{arXiv e-prints}, art. arXiv:1902.09589, Feb 2019.

\bibitem[Holzmann(2015)]{Holzmann2015}
G.~J. Holzmann.
\newblock Code inflation.
\newblock \emph{IEEE Software}, 1\penalty0 (2):\penalty0 10--13, 2015.

\bibitem[{Jezek} and {Dietrich}(2014)]{Jezek2014}
K.~{Jezek} and J.~{Dietrich}.
\newblock On the use of static analysis to safeguard recursive dependency
  resolution.
\newblock In \emph{2014 40th EUROMICRO Conference on Software Engineering and
  Advanced Applications}, pages 166--173, Aug 2014.
\newblock \doi{10.1109/SEAA.2014.35}.

\bibitem[{Jiang} et~al.(2016){Jiang}, {Wu}, and {Liu}]{Jiang2016}
Y.~{Jiang}, D.~{Wu}, and P.~{Liu}.
\newblock Jred: Program customization and bloatware mitigation based on static
  analysis.
\newblock In \emph{2016 IEEE 40th Annual Computer Software and Applications
  Conference (COMPSAC)}, volume~1, pages 12--21, June 2016.
\newblock \doi{10.1109/COMPSAC.2016.146}.

\bibitem[Krueger(1992)]{Krueger1992}
C.~W. Krueger.
\newblock Software reuse.
\newblock \emph{ACM Comput. Surv.}, 24\penalty0 (2):\penalty0 131–183, June
  1992.
\newblock ISSN 0360-0300.
\newblock \doi{10.1145/130844.130856}.
\newblock URL \url{https://doi.org/10.1145/130844.130856}.

\bibitem[Kula et~al.(2018)Kula, German, Ouni, Ishio, and Inoue]{Kula2018}
R.~G. Kula, D.~M. German, A.~Ouni, T.~Ishio, and K.~Inoue.
\newblock Do developers update their library dependencies?
\newblock \emph{Empirical Software Engineering}, 23\penalty0 (1):\penalty0
  384--417, Feb 2018.
\newblock ISSN 1573-7616.
\newblock \doi{10.1007/s10664-017-9521-5}.
\newblock URL \url{https://doi.org/10.1007/s10664-017-9521-5}.

\bibitem[L\"{a}mmel et~al.(2011)L\"{a}mmel, Pek, and Starek]{Lammel2011}
R.~L\"{a}mmel, E.~Pek, and J.~Starek.
\newblock {Large-scale, AST-based API-usage Analysis of Open-source Java
  Projects}.
\newblock In \emph{Proceedings of the 2011 ACM Symposium on Applied Computing},
  SAC '11, pages 1317--1324, New York, NY, USA, 2011. ACM.
\newblock ISBN 978-1-4503-0113-8.
\newblock \doi{10.1145/1982185.1982471}.
\newblock URL \url{http://doi.acm.org/10.1145/1982185.1982471}.

\bibitem[{Landman} et~al.(2017){Landman}, {Serebrenik}, and
  {Vinju}]{Landman2017}
D.~{Landman}, A.~{Serebrenik}, and J.~J. {Vinju}.
\newblock {Challenges for Static Analysis of Java Reflection - Literature
  Review and Empirical Study}.
\newblock In \emph{2017 IEEE/ACM 39th International Conference on Software
  Engineering (ICSE)}, pages 507--518, May 2017.
\newblock \doi{10.1109/ICSE.2017.53}.

\bibitem[Lim(1994)]{Lim1994}
W.~C. Lim.
\newblock {Effects of reuse on quality, productivity, and economics}.
\newblock \emph{IEEE Software}, 11\penalty0 (5):\penalty0 23--30, 1994.
\newblock ISSN 0740-7459.
\newblock \doi{10.1109/52.311048}.

\bibitem[Lindholm et~al.(2014)Lindholm, Yellin, Bracha, and
  Buckley]{Lindholm2014}
T.~Lindholm, F.~Yellin, G.~Bracha, and A.~Buckley.
\newblock \emph{The Java virtual machine specification}.
\newblock Pearson Education, 2014.

\bibitem[McIntosh et~al.(2012)McIntosh, Adams, and Hassan]{McIntosh2012}
S.~McIntosh, B.~Adams, and A.~E. Hassan.
\newblock The evolution of java build systems.
\newblock \emph{Empirical Software Engineering}, 17\penalty0 (4):\penalty0
  578--608, Aug 2012.
\newblock ISSN 1573-7616.
\newblock \doi{10.1007/s10664-011-9169-5}.
\newblock URL \url{https://doi.org/10.1007/s10664-011-9169-5}.

\bibitem[McIntosh et~al.(2014)McIntosh, Poehlmann, Juergens, Mockus, Adams,
  Hassan, Haupt, and Wagner]{McIntoshicse2014}
S.~McIntosh, M.~Poehlmann, E.~Juergens, A.~Mockus, B.~Adams, A.~E. Hassan,
  B.~Haupt, and C.~Wagner.
\newblock Collecting and leveraging a benchmark of build system clones to aid
  in quality assessments.
\newblock In \emph{Companion Proceedings of the 36th International Conference
  on Software Engineering}, ICSE Companion 2014, page 145–154, New York, NY,
  USA, 2014. Association for Computing Machinery.
\newblock ISBN 9781450327688.
\newblock \doi{10.1145/2591062.2591181}.
\newblock URL \url{https://doi.org/10.1145/2591062.2591181}.

\bibitem[Mitropoulos et~al.(2014)Mitropoulos, Karakoidas, Louridas, Gousios,
  and Spinellis]{Mitropoulos2014}
D.~Mitropoulos, V.~Karakoidas, P.~Louridas, G.~Gousios, and D.~Spinellis.
\newblock The bug catalog of the maven ecosystem.
\newblock In \emph{Proceedings of the 11th Working Conference on Mining
  Software Repositories}, MSR 2014, pages 372--375, New York, NY, USA, 2014.
  ACM.
\newblock ISBN 978-1-4503-2863-0.
\newblock \doi{10.1145/2597073.2597123}.
\newblock URL \url{http://doi.acm.org/10.1145/2597073.2597123}.

\bibitem[Myers and Stylos(2016)]{Myers2016}
B.~A. Myers and J.~Stylos.
\newblock {Improving API usability}.
\newblock \emph{Communications of the ACM}, 59\penalty0 (6):\penalty0 62--69,
  2016.

\bibitem[Naur and Randell(1969)]{naur1969software}
P.~Naur and B.~Randell, editors.
\newblock \emph{Software Engineering: Report of a Conference Sponsored by the
  NATO Science Committee, Garmisch, Germany, 7-11 Oct. 1968, Brussels,
  Scientific Affairs Division, NATO}.
\newblock Newcastle University, 1969.

\bibitem[Nguyen et~al.(2020)Nguyen, Di~Rocco, Di~Ruscio, and
  Di~Penta]{Nguyen2020}
P.~T. Nguyen, J.~Di~Rocco, D.~Di~Ruscio, and M.~Di~Penta.
\newblock Crossrec: Supporting software developers by recommending third-party
  libraries.
\newblock \emph{Journal of Systems and Software}, 161:\penalty0 110460, 2020.

\bibitem[Pham et~al.(2016)Pham, Vu, Nguyen, et~al.]{Pham2016}
H.~V. Pham, P.~M. Vu, T.~T. Nguyen, et~al.
\newblock {Learning API usages from bytecode: a statistical approach}.
\newblock In \emph{Proceedings of the 38th International Conference on Software
  Engineering}, pages 416--427. ACM, 2016.

\bibitem[Qiu et~al.(2016)Qiu, Li, and Leung]{Qiu2016}
D.~Qiu, B.~Li, and H.~Leung.
\newblock {Understanding the API usage in Java}.
\newblock \emph{Information and software technology}, 73:\penalty0 81--100,
  2016.

\bibitem[Quach et~al.(2017)Quach, Erinfolami, Demicco, and Prakash]{Quach2017}
A.~Quach, R.~Erinfolami, D.~Demicco, and A.~Prakash.
\newblock A multi-os cross-layer study of bloating in user programs, kernel and
  managed execution environments.
\newblock In \emph{Proceedings of the 2017 Workshop on Forming an Ecosystem
  Around Software Transformation}, pages 65--70. ACM, 2017.

\bibitem[Rastogi et~al.(2017)Rastogi, Davidson, De~Carli, Jha, and
  McDaniel]{Rastogi2017}
V.~Rastogi, D.~Davidson, L.~De~Carli, S.~Jha, and P.~McDaniel.
\newblock Cimplifier: Automatically debloating containers.
\newblock In \emph{Proceedings of the 2017 11th Joint Meeting on Foundations of
  Software Engineering}, ESEC/FSE 2017, pages 476--486, New York, NY, USA,
  2017. ACM.
\newblock ISBN 978-1-4503-5105-8.
\newblock \doi{10.1145/3106237.3106271}.
\newblock URL \url{http://doi.acm.org/10.1145/3106237.3106271}.

\bibitem[Roover et~al.(2013)Roover, Lämmel, and Pek]{Roover2013}
C.~D. Roover, R.~Lämmel, and E.~Pek.
\newblock {Multi-dimensional exploration of API usage}.
\newblock In \emph{21st International Conference on Program Comprehension},
  ICPC, pages 152--161, 2013.
\newblock \doi{10.1109/ICPC.2013.6613843}.

\bibitem[Salza et~al.(2019)Salza, Palomba, Di~Nucci, De~Lucia, and
  Ferrucci]{Salza2019}
P.~Salza, F.~Palomba, D.~Di~Nucci, A.~De~Lucia, and F.~Ferrucci.
\newblock Third-party libraries in mobile apps.
\newblock \emph{Empirical Software Engineering}, Aug 2019.
\newblock ISSN 1573-7616.
\newblock \doi{10.1007/s10664-019-09754-1}.
\newblock URL \url{https://doi.org/10.1007/s10664-019-09754-1}.

\bibitem[Seo et~al.(2014)Seo, Sadowski, Elbaum, Aftandilian, and
  Bowdidge]{Seo2014}
H.~Seo, C.~Sadowski, S.~Elbaum, E.~Aftandilian, and R.~Bowdidge.
\newblock Programmers' build errors: A case study (at google).
\newblock In \emph{Proceedings of the 36th International Conference on Software
  Engineering}, ICSE 2014, pages 724--734, New York, NY, USA, 2014. ACM.
\newblock ISBN 978-1-4503-2756-5.
\newblock \doi{10.1145/2568225.2568255}.
\newblock URL \url{http://doi.acm.org/10.1145/2568225.2568255}.

\bibitem[Sharif et~al.(2018)Sharif, Abubakar, Gehani, and Zaffar]{Sharif2018}
H.~Sharif, M.~Abubakar, A.~Gehani, and F.~Zaffar.
\newblock Trimmer: Application specialization for code debloating.
\newblock In \emph{Proceedings of the 33rd ACM/IEEE International Conference on
  Automated Software Engineering}, ASE 2018, pages 329--339, New York, NY, USA,
  2018. ACM.
\newblock ISBN 978-1-4503-5937-5.
\newblock \doi{10.1145/3238147.3238160}.
\newblock URL \url{http://doi.acm.org/10.1145/3238147.3238160}.

\bibitem[Shull et~al.(2007)Shull, Singer, and Sj{\o}berg]{Shull2007}
F.~Shull, J.~Singer, and D.~I. Sj{\o}berg.
\newblock \emph{Guide to Advanced Empirical Software Engineering}.
\newblock Springer-Verlag, Berlin, Heidelberg, 2007.
\newblock ISBN 184800043X.

\bibitem[Soto-Valero et~al.(2019)Soto-Valero, Benelallam, Harrand, Barais, and
  Baudry]{SotoValero2019}
C.~Soto-Valero, A.~Benelallam, N.~Harrand, O.~Barais, and B.~Baudry.
\newblock {The Emergence of Software Diversity in Maven Central}.
\newblock In \emph{16th International Conference on Mining Software
  Repositories}, MSR 2019, pages 1--10, New York, NY, USA, 2019. ACM.
\newblock \doi{10.1145/2597073.2597097}.
\newblock URL \url{http://doi.acm.org/10.1145/2597073.2597097}.

\bibitem[Varanasi and Belida(2014)]{Varanasi2014}
B.~Varanasi and S.~Belida.
\newblock \emph{Maven Dependency Management}, pages 15--22.
\newblock Apress, Berkeley, CA, 2014.
\newblock ISBN 978-1-4842-0841-0.
\newblock \doi{10.1007/978-1-4842-0841-0_3}.
\newblock URL \url{https://doi.org/10.1007/978-1-4842-0841-0_3}.

\bibitem[V\'{a}zquez et~al.(2019)V\'{a}zquez, Bergel, Vidal, Pace, and
  Marcos]{Vazquez2019}
H.~V\'{a}zquez, A.~Bergel, S.~Vidal, J.~D. Pace, and C.~Marcos.
\newblock Slimming javascript applications: An approach for removing unused
  functions from javascript libraries.
\newblock \emph{Information and Software Technology}, 107:\penalty0 18--29,
  2019.
\newblock ISSN 0950-5849.
\newblock \doi{https://doi.org/10.1016/j.infsof.2018.10.009}.
\newblock URL
  \url{http://www.sciencedirect.com/science/article/pii/S0950584918302210}.

\bibitem[Wu et~al.(2017)Wu, Manabe, Kanda, German, and Inoue]{Wu2017}
Y.~Wu, Y.~Manabe, T.~Kanda, D.~M. German, and K.~Inoue.
\newblock Analysis of license inconsistency in large collections of open source
  projects.
\newblock \emph{Empirical Software Engineering}, 22\penalty0 (3):\penalty0
  1194--1222, Jun 2017.
\newblock ISSN 1573-7616.
\newblock \doi{10.1007/s10664-016-9487-8}.
\newblock URL \url{https://doi.org/10.1007/s10664-016-9487-8}.

\bibitem[Yu et~al.(2003)Yu, Dayani-Fard, and Mylopoulos]{Yu2003}
Y.~Yu, H.~Dayani-Fard, and J.~Mylopoulos.
\newblock Removing false code dependencies to speedup software build processes.
\newblock In \emph{Proceedings of the 2003 Conference of the Centre for
  Advanced Studies on Collaborative Research}, CASCON, pages 343--352. IBM
  Press, 2003.
\newblock URL \url{http://dl.acm.org/citation.cfm?id=961322.961375}.

\end{thebibliography}
